%% file: main.tex
\pdfoutput=1
\documentclass[12pt,a4paper]{article}
\usepackage{placeins}

\usepackage{ifthen} % for conditional statements
\newboolean{pdflatex}
\setboolean{pdflatex}{true} % False for eps figures 

\newboolean{articletitles}
\setboolean{articletitles}{true} % False removes titles in references

\newboolean{uprightparticles}
\setboolean{uprightparticles}{false} %True for upright particle symbols

\usepackage{longtable} % only for template; not usually to be used in PAPERs
\input{preamble}

\begin{document}

%%%%%%%%%%%%%%%%%%%%%%%%%
%%%%% Title     %%%%%%%%%
%%%%%%%%%%%%%%%%%%%%%%%%%
\renewcommand{\thefootnote}{\fnsymbol{footnote}}
\setcounter{footnote}{1}

% %%%%%%% CHOOSE TITLE PAGE--------
%\onecolumn
% \input{title-LHCb-ANA}
%\input{title-LHCb-CONF}
\input{title-LHCb-PAPER}

%\twocolumn
% %%%%%%%%%%%%% ---------

\renewcommand{\thefootnote}{\arabic{footnote}}
\setcounter{footnote}{0}

%%%%%%%%%%%%%%%%%%%%%%%%%%%%%%%%
%%%%%  Table of Content   %%%%%%
%%%%%%%%%%%%%%%%%%%%%%%%%%%%%%%%
%%%% Uncomment next 2 lines if desired
%\tableofcontents
%\cleardoublepage

%%%%%%%%%%%%%%%%%%%%%%%%%
%%%%% Main text %%%%%%%%%
%%%%%%%%%%%%%%%%%%%%%%%%%

\pagestyle{plain} % restore page numbers for the main text
\setcounter{page}{1}
\pagenumbering{arabic}

%% Uncomment during review phase. 
%% Comment before a final submission.
%\linenumbers

% You can include short sections directly in the main tex file.
% However, for larger papers it is desirable to split the text into
% several semiautonomous files, which can be revised independently.
% This is especially useful when developing a document in
% collaboration with several people, since then different parts can be
% edited independently.  This type of file organization is shown here.
% 

\newboolean{prl}
\setboolean{prl}{false} % False for eps figures 

\newlength{\figsize}
\setlength{\figsize}{0.9\hsize}
\input{paper-body}
%\input{justification}

\clearpage

% Author List ----------------------------
%  You need to get a new author list!
\input{LHCb_authorlist}

\end{document}

%% file: preamble.tex
\usepackage{microtype}
\usepackage{lineno}  % for line numbering during review
\usepackage{xspace} % To avoid problems with missing or double spaces after
\usepackage{caption} %these three command get the figure and table captions automatically small

\usepackage{graphicx}  % to include figures (can also use other packages)
\usepackage{color}
\usepackage{colortbl}
\graphicspath{{./figs/}} % Make Latex search fig subdir for figures

\usepackage{amsmath} % Adds a large collection of math symbols
\usepackage{amssymb}
\usepackage{amsfonts}
\usepackage{upgreek} % Adds in support for greek letters in roman typeset

\newcommand*\patchAmsMathEnvironmentForLineno[1]{%
\expandafter\let\csname old#1\expandafter\endcsname\csname #1\endcsname
\expandafter\let\csname oldend#1\expandafter\endcsname\csname
end#1\endcsname
 \renewenvironment{#1}%
   {\linenomath\csname old#1\endcsname}%
   {\csname oldend#1\endcsname\endlinenomath}%
}
\newcommand*\patchBothAmsMathEnvironmentsForLineno[1]{%
  \patchAmsMathEnvironmentForLineno{#1}%
  \patchAmsMathEnvironmentForLineno{#1*}%
}
\AtBeginDocument{%
\patchBothAmsMathEnvironmentsForLineno{equation}%
\patchBothAmsMathEnvironmentsForLineno{align}%
\patchBothAmsMathEnvironmentsForLineno{flalign}%
\patchBothAmsMathEnvironmentsForLineno{alignat}%
\patchBothAmsMathEnvironmentsForLineno{gather}%
\patchBothAmsMathEnvironmentsForLineno{multline}%
\patchBothAmsMathEnvironmentsForLineno{eqnarray}%
}

\usepackage{hyperref}    % Hyperlinks in references
\usepackage[all]{hypcap} % Internal hyperlinks to floats.

\input{lhcb-symbols-def} % Add in the predefined LHCb symbols

\usepackage{cite} % Allows for ranges in citations
\usepackage{mciteplus}

%% file: lhcb-symbols-def.tex
\usepackage{xspace} 
\usepackage{upgreek}

\def\lhcb {\mbox{LHCb}\xspace}

\def\MagUp {\mbox{\em Mag\kern -0.05em Up}\xspace}

\ifthenelse{\boolean{uprightparticles}}%
{

 \def\Pmu         {\ensuremath{\upmu}\xspace}

 \def\Pphi        {\ensuremath{\upphi}\xspace}

 \def\Ppsi        {\ensuremath{\uppsi}\xspace}

 \def\PDelta      {\ensuremath{\Delta}\xspace}                 
 \def\PXi      {\ensuremath{\Xi}\xspace}                 
 \def\PLambda      {\ensuremath{\Lambda}\xspace}                 
 \def\PSigma      {\ensuremath{\Sigma}\xspace}                 
 \def\POmega      {\ensuremath{\Omega}\xspace}                 
 \def\PUpsilon      {\ensuremath{\Upsilon}\xspace}                 
 
 \def\PB      {\ensuremath{\mathrm{B}}\xspace}                 
                  
 \def\PD      {\ensuremath{\mathrm{D}}\xspace}

 \def\PJ      {\ensuremath{\mathrm{J}}\xspace}                 
 \def\PK      {\ensuremath{\mathrm{K}}\xspace}

 \def\Pb      {\ensuremath{\mathrm{b}}\xspace}                 
 \def\Pc      {\ensuremath{\mathrm{c}}\xspace}

}
{

 \def\Pmu         {\ensuremath{\mu}\xspace}

 \def\Pphi        {\ensuremath{\phi}\xspace}

 \def\Ppsi        {\ensuremath{\psi}\xspace}                 
                  
 \mathchardef\PDelta="7101
 \mathchardef\PXi="7104
 \mathchardef\PLambda="7103
 \mathchardef\PSigma="7106
 \mathchardef\POmega="710A
 \mathchardef\PUpsilon="7107
                  
 \def\PB      {\ensuremath{B}\xspace}                 
                  
 \def\PD      {\ensuremath{D}\xspace}

 \def\PJ      {\ensuremath{J}\xspace}                 
 \def\PK      {\ensuremath{K}\xspace}

 \def\Pb      {\ensuremath{b}\xspace}                 
 \def\Pc      {\ensuremath{c}\xspace}

}

\makeatletter
  \newcommand{\miniscule}{\@setfontsize\miniscule{5}{6}}% \tiny: 6/7
\makeatother

\DeclareRobustCommand{\optbar}[1]{\shortstack{{\miniscule (\rule[.5ex]{1.25em}{.18mm})}
  \\ [-.7ex] $#1$}}

\def\mup        {{\ensuremath{\Pmu^+}}\xspace}
\def\cquark    {{\ensuremath{\Pc}}\xspace}

\def\bquark    {{\ensuremath{\Pb}}\xspace}

\def\kaon    {{\ensuremath{\PK}}\xspace}
  \def\Kbar    {{\kern 0.2em\overline{\kern -0.2em \PK}{}}\xspace}

\def\KorKbar    {\kern 0.18em\optbar{\kern -0.18em K}{}\xspace}

\def\Kstar   {{\ensuremath{\kaon^*}}\xspace}

\def\Kstarp  {{\ensuremath{\kaon^{*+}}}\xspace}

\newcommand{\phiz}{\ensuremath{\Pphi}\xspace}

  \def\Dbar    {{\kern 0.2em\overline{\kern -0.2em \PD}{}}\xspace}

\def\DorDbar    {\kern 0.18em\optbar{\kern -0.18em D}{}\xspace}

\def\Bbar    {{\ensuremath{\kern 0.18em\overline{\kern -0.18em \PB}{}}}\xspace}

\def\BorBbar    {\kern 0.18em\optbar{\kern -0.18em B}{}\xspace}

\def\jpsi     {{\ensuremath{{\PJ\mskip -3mu/\mskip -2mu\Ppsi\mskip 2mu}}}\xspace}

  \def\Y#1S{\ensuremath{\PUpsilon{(#1S)}}\xspace}% no space before {...}!

\def\Lbar        {{\ensuremath{\kern 0.1em\overline{\kern -0.1em\PLambda}}}\xspace}
\def\LorLbar    {\kern 0.18em\optbar{\kern -0.18em \PLambda}{}\xspace}

\def\BR         {\BF}
\def\to                 {\ensuremath{\rightarrow}\xspace}

\def\AT#1     {\ensuremath{A_{\mathrm{T}}^{#1}}\xspace}           % 2

\def\C#1      {\ensuremath{\mathcal{C}_{#1}}\xspace}                       % 9
\def\Cp#1     {\ensuremath{\mathcal{C}_{#1}^{'}}\xspace}                    % 7
\def\Ceff#1   {\ensuremath{\mathcal{C}_{#1}^{\mathrm{(eff)}}}\xspace}        % 9  
\def\Cpeff#1  {\ensuremath{\mathcal{C}_{#1}^{'\mathrm{(eff)}}}\xspace}       % 7
\def\Ope#1    {\ensuremath{\mathcal{O}_{#1}}\xspace}                       % 2
\def\Opep#1   {\ensuremath{\mathcal{O}_{#1}^{'}}\xspace}                    % 7

\newcommand{\tev}{\ensuremath{\mathrm{\,Te\kern -0.1em V}}\xspace}
\newcommand{\gev}{\ensuremath{\mathrm{\,Ge\kern -0.1em V}}\xspace}
\newcommand{\mev}{\ensuremath{\mathrm{\,Me\kern -0.1em V}}\xspace}
\newcommand{\kev}{\ensuremath{\mathrm{\,ke\kern -0.1em V}}\xspace}
\newcommand{\ev}{\ensuremath{\mathrm{\,e\kern -0.1em V}}\xspace}
\newcommand{\gevc}{\ensuremath{{\mathrm{\,Ge\kern -0.1em V\!/}c}}\xspace}
\newcommand{\mevc}{\ensuremath{{\mathrm{\,Me\kern -0.1em V\!/}c}}\xspace}
\newcommand{\gevcc}{\ensuremath{{\mathrm{\,Ge\kern -0.1em V\!/}c^2}}\xspace}
\newcommand{\gevgevcccc}{\ensuremath{{\mathrm{\,Ge\kern -0.1em V^2\!/}c^4}}\xspace}
\newcommand{\mevcc}{\ensuremath{{\mathrm{\,Me\kern -0.1em V\!/}c^2}}\xspace}

\def\km   {\ensuremath{\mathrm{ \,km}}\xspace}

\def\mum  {\ensuremath{{\,\upmu\mathrm{m}}}\xspace}

\newcommand{\chisq}{\ensuremath{\chi^2}\xspace}

\def\gsim{{~\raise.15em\hbox{$>$}\kern-.85em
          \lower.35em\hbox{$\sim$}~}\xspace}
\def\lsim{{~\raise.15em\hbox{$<$}\kern-.85em
          \lower.35em\hbox{$\sim$}~}\xspace}

\def\PDF {PDF\xspace}

\def\sPlot{\mbox{\em sPlot}\xspace}
\def\ptot       {\mbox{$p$}\xspace}
\def\pt         {\mbox{$p_{\mathrm{ T}}$}\xspace}

\def\pythia     {\mbox{\textsc{Pythia}}\xspace}

\def\tell1  {TELL1\xspace}
\def\ukl1   {UKL1\xspace}

\newcommand{\eg}{\mbox{\itshape e.g.}\xspace}
\newcommand{\ie}{\mbox{\itshape i.e.}\xspace}
\newcommand{\etal}{\mbox{\itshape et al.}\xspace}

%% file: title-LHCb-PAPER.tex
% $Id: title-LHCb-PAPER.tex 10646 2011-10-12 13:51:38Z uegede $
% ===============================================================================
% Purpose: LHCb-PAPER journal paper title page template
% Author: 
% Created on: 2010-09-25
% ===============================================================================

%%%%%%%%%%%%%%%%%%%%%%%%%
%%%%%  TITLE PAGE  %%%%%%
%%%%%%%%%%%%%%%%%%%%%%%%%
\begin{titlepage}
\pagenumbering{roman}

% Header ---------------------------------------------------
\vspace*{-1.5cm}
\centerline{\large EUROPEAN ORGANIZATION FOR NUCLEAR RESEARCH (CERN)}
\vspace*{0.5cm}
\hspace*{-0.5cm}
\begin{tabular*}{\linewidth}{lc@{\extracolsep{\fill}}r}
\ifthenelse{\boolean{pdflatex}}% Logo format choice
{\vspace*{-3.1cm}\mbox{\!\!\!\includegraphics[width=.14\textwidth]{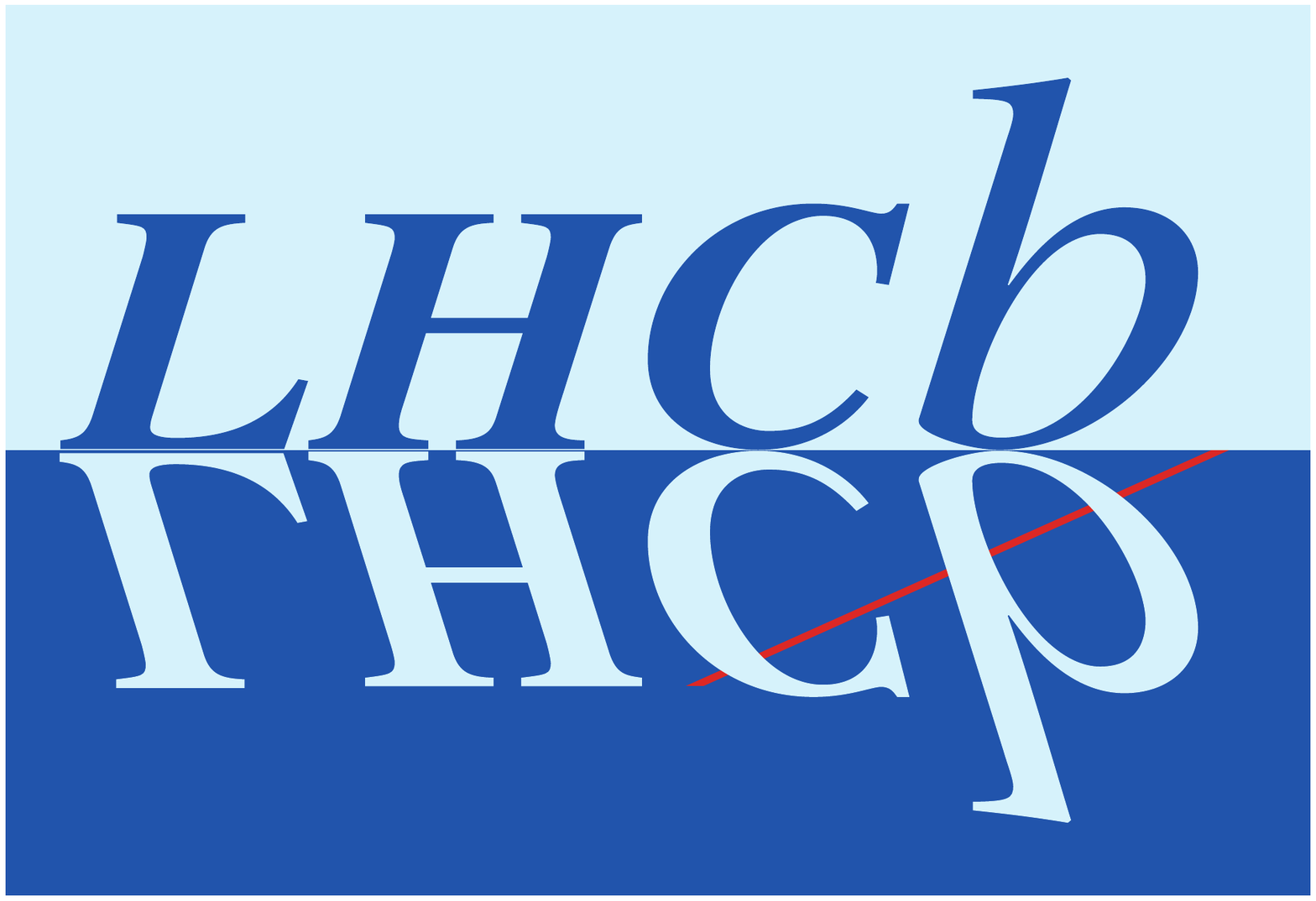}} & &}%
{\vspace*{-1.2cm}\mbox{\!\!\!\includegraphics[width=.12\textwidth]{lhcb-logo.eps}} & &}%
\\
 & & CERN-EP-2016-156 \\  % ID 
 & & LHCb-PAPER-2016-019 \\  % ID 
 & & 29 September 2016 \\ %\today \\ %25 June 2016 \\ % \today \\ %1 May 2014 \\  % Date - Can also hardwire e.g.: 23 March 2010
 & & \\
% not in paper \hline
\end{tabular*}

\vspace*{3.0cm}

% Title --------------------------------------------------
{\bf\boldmath\huge
\begin{center}
Amplitude analysis of $B^+\to J/\psi \phi K^+$ decays
\end{center}
}

\vspace*{1.0cm}

% Authors -------------------------------------------------
\begin{center}
The LHCb collaboration\footnote{Authors are listed at the end of this paper.}
\end{center}

\vspace{0.1cm}

% Abstract -----------------------------------------------
\begin{abstract}
  \noindent
\input{abstract}

\end{abstract}

\vspace*{1.1cm}
\vfill

\begin{center}
  Published in Physical Review D {\bf 95}, 012002 (2017). 
\end{center}

\vspace{0.1cm}

{\footnotesize 
\centerline{\copyright~CERN on behalf of the \lhcb collaboration, license \href{http://creativecommons.org/licenses/by/4.0/
}{CC-BY-4.0}.}}
\vspace*{2mm}

\end{titlepage}

%%%%%%%%%%%%%%%%%%%%%%%%%%%%%%%%
%%%%%  EOD OF TITLE PAGE  %%%%%%
%%%%%%%%%%%%%%%%%%%%%%%%%%%%%%%%

%  empty page follows the title page ----
\newpage
\setcounter{page}{2}
\mbox{~}
\newpage

% Author List ----------------------------
%  You need to get a new author list!
%\input{LHCb_authorlist.tex}

\cleardoublepage

%% file: abstract.tex
The first full amplitude analysis of $B^+\to J/\psi \phi K^+$ with $J/\psi\to\mu^+\mu^-$, $\phi\to K^+K^-$ 
decays is performed with a data sample of 
3 fb$^{-1}$ of $pp$ collision data collected at $\sqrt{s}=7$ and $8$ TeV with the LHCb detector.
The data cannot be described by a model that contains only excited kaon states decaying into
$\phi K^+$, and four $J/\psi\phi$ structures are observed, each with significance over $5$ standard deviations.
The quantum numbers of these structures are determined with significance of at least $4$ standard deviations.
The lightest has mass consistent with, but width much larger than, previous measurements of the claimed $X(4140)$ state.
%and can also be described as a $D_s^{\pm}D_s^{*\mp}$ cusp. 
The model includes significant contributions from a number of expected kaon excitations, including 
the first observation of the $K^{*}(1680)^+\to\phi K^+$ transition.

%% file: paper-body.tex
% --------------------
\def\bujkpp{B^+\to\jpsi K^+\pi^-\pi^+}
\def\bujkkk{B^+\to\jpsi K^+K^-K^+}
\def\bujphik{B^+\to\jpsi \phi K^+}
\def\buxk{B^+\to X(4140) K^+}
\def\BR{{\cal B}}
\def\DLL{{\rm DLL}}
\def\PDF{{\cal P}}
\def\NDOF{\hbox{\rm ndf}}
\def\coskj{\cos(K,\jpsi)}
\def\funone{{\cal F}^{\rm bkg}_1}
\def\funtwo{{\cal F}^{\rm bkg}_2}
% -----------------------------------
\def\kp{K^+}
\def\km{K^-}
\def\cosks{\cos\theta_{\Kstar}}
\def\cospsi{\cos\theta_{\jpsi}}
\def\cosphi{\cos\theta_{\phiz}}
\def\phiksphi{\Delta{\phi}_{\Kstar,\phiz}}
\def\phikspsi{\Delta{\phi}_{\Kstar,\jpsi}}
\def\cosx{\cos\theta_{X}}
\def\phixphi{\Delta\phi_{X,\phiz}}
\def\phixpsi{\Delta\phi_{X,\jpsi}}
\def\cosz{\cos\theta_{Z}}
\def\phizphi{{\Delta\phi}_{Z,\phiz}}
\def\phizpsi{{\Delta\phi}_{Z,\jpsi}}
\def\mphik{m_{\phiz\kaon}}
\def\mjpsiphi{m_{\jpsi\phiz}}
\def\mjpsik{m_{\jpsi\kaon}}
\def\btojpsiphik{B^+\to\jpsi\phiz K^+}
\def\jpsiphik{\jpsi\phiz\kaon}
\def\btojpsikkk{B^+\to\jpsi K^+K^-K^+}
\def\ndf{{\rm ndf}}
\def\Kst{\Kstar}
\def\qbar{\bar{q}}
\def\spc{2pt}
\def\NRKs{{\rm NR}_{\phi K}}
\def\NRX{{\rm NR}_{\jpsi\phi}}
\def\nbin{{\rm N_{bin}}}
\def\chiod{\chi^2_{\rm 1D}/(\nbin-1)}
\def\pod{p_{\rm 1D}}
\def\chitd{\chi^2_{\rm 2D}/(\nbin-1)}
\def\ptd{p_{\rm 2D}}
\def\chisd{\chi^2_{\rm 6D}/(\nbin-1)}
\def\psd{p_{\rm 6D}}
\def\Like{{\cal L}}
\def\dll{\Delta(\!-2\!\ln\Like)}
\def\ndf{{\rm ndf}}
\def\npar{n_{\rm par}}
\def\dnpar{\Delta\npar}
\def\Pars{\vec{\omega}}
\def\PDF{\mathcal{P}}
\def\Mat{\mathcal{M}}
\def\AR{{\cal A}}
\def\Xone{X(4140)}
\def\Xtwo{X(4274)}
\def\Xthree{X(4500)}
\def\Xfour{X(4700)}
\def\9{\phantom{8}}
\def\FiFr{{\rm FF}}
%\def\nslj#1#2#3#4{#1{}^{#2}{\rm #3}_{#4}}
%\def\nlj#1#2#3{#1{\rm #2}_{#3}}
% ----------------- 
%\def\bls#1#2#3{$B_{#1,#2}^{#3}$}
\newcommand{\bls}[3]{$B_{#1,#2}^{#3}$}
\newcommand{\clx}[4]{$(\!#1\!\pm\!#2,#3\!\pm\!#4  )$}
\newcommand{\BLS}[3]{{B_{#1,#2}}^{#3}}
\newcommand{\nslj}[4]{#1{}^{#2}{\rm #3}_{#4}}
\newcommand{\nlj}[3]{#1{\rm #2}_{#3}}
% ------------------------------------------

\section{Introduction}

In 2008 the CDF collaboration presented $3.8\sigma$ evidence for a 
near-threshold $X(4140)\to\jpsi\phi$ mass peak in $\bujphik$ 
decays\footnote{Inclusion of charge-conjugate processes is implied throughout this paper, 
unless stated otherwise.}
also referred to as $Y(4140)$ in the literature, 
with width $\Gamma=11.7 \mev$ \cite{Aaltonen:2009tz}.\footnote{Units with $c=1$ are used.}
Much larger widths are expected for charmonium states at this mass
because of open flavor decay channels \cite{Brambilla:2010cs},
which should also make the kinematically suppressed $X\to\jpsi\phi$ decays undetectable.
Therefore, the observation by CDF triggered wide interest.
It has been suggested that the $X(4140)$ structure could be a molecular state 
\cite{Liu:2009ei,Branz:2009yt,Albuquerque:2009ak,Ding:2009vd,Zhang:2009st,Liu:2009pu,Wang:2009ry,Chen:2015fdn,Karliner:2016ith},
a tetraquark state \cite{Stancu:2009ka,Drenska:2009cd,Wang:2015pea,Anisovich:2015caa,Lebed:2016yvr}, 
a hybrid state \cite{Mahajan:2009pj,Wang:2009ue} 
or a rescattering effect \cite{Liu:2009iw,Swanson:2014tra}. 

The LHCb collaboration did not see evidence for the narrow $X(4140)$ peak 
in the analysis presented in Ref.~\cite{LHCb-PAPER-2011-033}, 
based on a data sample corresponding to 0.37 fb$^{-1}$ of 
integrated luminosity, 
a fraction of that now available.
Searches for the narrow $X(4140)$ 
did not confirm its presence in analyses performed by the 
Belle \cite{Brodzicka:2010zz,ChengPing:2009vu} (unpublished) 
and BaBar \cite{Lees:2014lra} experiments.
The $X(4140)$ structure was observed however
by the CMS ($5\sigma$) \cite{Chatrchyan:2013dma} collaboration.
Evidence for it was also reported 
in $B^+\to\jpsi\phi K^+$ decays
by the D0 ($3\sigma$) \cite{Abazov:2013xda} collaboration.
The D0 collaboration claimed in addition a significant 
signal for prompt $X(4140)$ production in $p\bar p$  
collisions \cite{Abazov:2015sxa}.
The BES-III collaboration did not find evidence for $X(4140)\to\jpsi\phi$ in 
$e^+e^-\to\gamma X(4140)$ and set upper limits on its production 
cross-section at $\sqrt{s}=4.23$, $4.26$ and $4.36$ \gev \cite{Ablikim:2014atq}.
Previous results related to the $X(4140)$ structure 
are summarized in Table~\ref{tab:xprevious}.

\begin{table*}[bhtp]
\caption{\small Previous results related to the $X(4140)\to\jpsi\phi$ mass peak, 
first observed in $B^+\to\jpsi\phi K^+$ decays.
The first (second) significance quoted for Ref.~\cite{Abazov:2015sxa} 
is for the prompt (non-prompt) production components.
The statistical and systematic errors are added in quadrature 
and then used in the weights to calculate the averages, excluding
unpublished results (shown in italics).
The last column gives a fraction of the total $B^+\to\jpsi\phi K^+$ rate 
attributed to the $X(4140)$ structure. 
}
\label{tab:xprevious}
\hbox{
\hbox{\ifthenelse{\boolean{prl}}{}{\quad\hskip-2.5cm}
\hbox{
\ifthenelse{\boolean{prl}}{}{\begin{footnotesize}}
\renewcommand{\arraystretch}{1.2}
\def\1#1{\multicolumn{1}{c}{#1}}
\def\2{}
\def\3{\ifthenelse{\boolean{prl}}{}{\!\!\!}}
\def\pms{\ifthenelse{\boolean{prl}}{\pm}{\!\pm\!}}
\begin{tabular}{ccrllcl}
\hline
Year & Experiment  & \1{$B\to\jpsi\phi K$} & \multicolumn{4}{c}{$X(4140)$ peak} \\
     & luminosity  & \1{yield}        & \1{Mass [\mev]} & \1{Width [\mev]} & 
\1{\ifthenelse{\boolean{prl}}{Significance}{Sign.}} & \1{Fraction \%} \\
\hline\hline
2008 & CDF 2.7 fb$^{-1}$  \cite{Aaltonen:2009tz} & 
$58\pm10$ \2 &  
$4143.0\pms2.9\pms1.2$ \ifthenelse{\boolean{prl}}{\quad}{\3} &
$11.7\,^{+8.3}_{-5.0}\pms3.7$ &
$3.8\sigma$ & \\
%\hline
{\it 2009} & {\it Belle} \cite{Brodzicka:2010zz} &
$\mathit{325\pm21}$ \2 &
$\mathit{4143.0}$ {\it fixed} & 
$\mathit{11.7}$ {\it fixed} &  
$\mathit{1.9\sigma}$ & \\
%\hline
{\it 2011} & {\it CDF 6.0 fb$^{-1}$}  \cite{CDF2011} & 
$\mathit{115\pm12}$ \2 &
$\mathit{4143.4\,^{+2.9}_{-3.0}\pms0.6}$ &
$\mathit{15.3\!^{+10.4}_{-\phantom{0}6.1}\!\pms\!2.5}$ & 
$\mathit{5.0\sigma}$ &
$\mathit{14.9\pms3.9\pms2.4}$ \\
%\hline
2011 & LHCb 0.37 fb$^{-1}$  \cite{LHCb-PAPER-2011-033} &
$346\pm20$ \2 &
$4143.4$ fixed & 
$15.3$ fixed   &
$1.4\sigma$    &
$<7$ @~90\%CL \\
%\hline
2013 &
CMS 5.2 fb$^{-1}$  \cite{Chatrchyan:2013dma} &
$2480\pm160$ &
$4148.0\pms2.4\pms6.3$ &
$28\phantom{.0}\,^{+15\phantom{.0}}_{-11\phantom{.0}}\pm19$ & 
$5.0\sigma$ &
$10\pms3$ (stat.)  \\
%\hline
2013 & D0 10.4 fb$^{-1}$  \cite{Abazov:2013xda} &
$215\pm37$ \2 &
$4159.0\pms4.3\pms6.6$ &
$19.9\pms12.6\,^{+1.0}_{-8.0}$ &
$3.0\sigma$ &
$21\pms8\pms4$ \\
%\hline
2014 & BaBar  \cite{Lees:2014lra} &
$189\pm14$ \2 &
$4143.4$ fixed & 
$15.3$ fixed  &
$1.6\sigma$ &
$<13.3$ @~90\%CL \\
\hline
2015 & D0 10.4 fb$^{-1}$  \cite{Abazov:2015sxa} &
\1{\3$p\bar{p}\to\jpsi\phi...$\3} &
$4152.5\pms1.7\,^{+6.2}_{-5.4}$ &
$16.3\pms5.6\pms11.4$ &
\3$4.7\sigma$ ($5.7\sigma$){\hskip-1cm\quad} &
 \\
\hline
\3 Average\3 &   &  & $4147.1\pms2.4$ & $15.7\pms6.3$ & & \\
\hline
\end{tabular}
\ifthenelse{\boolean{prl}}{}{\end{footnotesize}}
}
}
}
\end{table*} 

In an unpublished update to their $\bujphik$ analysis \cite{CDF2011}, 
the CDF collaboration presented $3.1\sigma$ evidence 
for a second relatively narrow 
$\jpsi\phi$ mass peak near $4274 \mev$.
This observation has also received attention in the literature \cite{He:2011ed,Finazzo:2011he}.
A second $\jpsi\phi$ mass peak was observed by the CMS collaboration at 
a mass which is higher by $3.2$ standard deviations, but the statistical significance of this structure
was not determined \cite{Chatrchyan:2013dma}.
The Belle collaboration saw $3.2\sigma$ evidence for a 
narrow $\jpsi\phi$ peak at $4350.6\,^{+4.6}_{-5.1}\pm0.7 \mev$ 
in two-photon collisions, which implies $J^{PC}=0^{++}$ or $2^{++}$,
and found no evidence for $X(4140)$ in the same analysis \cite{Shen:2009vs}.
The experimental results related to $\jpsi\phi$ mass peaks heavier than $X(4140)$ 
are summarized in Table~\ref{tab:xtwoprevious}.

\begin{table*}[bhtp]
\caption{\small Previous results related to $\jpsi\phi$ mass structures 
heavier than the $X(4140)$ peak.   
The unpublished results are shown in italics.
}
\label{tab:xtwoprevious}
\hbox{
\hbox{\ifthenelse{\boolean{prl}}{}{\quad\hskip-2.2cm}
\hbox{
\ifthenelse{\boolean{prl}}{}{\begin{footnotesize}}
\def\1#1{\multicolumn{1}{c}{#1}}
\def\2{}
\def\3{\ifthenelse{\boolean{prl}}{}{\!\!\!}}
\def\pms{\ifthenelse{\boolean{prl}}{\pm}{\!\pm\!}}
\renewcommand{\arraystretch}{1.2}
\begin{tabular}{ccrllcl}
\hline
Year & Experiment  & \1{$B\to\jpsi\phi K$} & \multicolumn{3}{c}{$X(4274-4351$) peaks(s)} \\
     & luminosity  & \1{yield}        & \1{Mass [\mev]} & \1{Width [\mev]} & 
\1{\ifthenelse{\boolean{prl}}{Significance}{Sign.}}  & \1{Fraction [\%]} \\
\hline\hline
{\it 2011} & {\it CDF 6.0 fb$^{-1}$}  \cite{CDF2011} & 
$\mathit{115\pm12}$ \2 &
$\mathit{4274.4\,^{+8.4}_{-6.7}\pms1.9}$ &
$\mathit{32.3{^{+21.9}_{-15.3}}\!\pms7.6}$ &
$\mathit{3.1\sigma}$ & \\
%\hline
2011 & LHCb 0.37 fb$^{-1}$  \cite{LHCb-PAPER-2011-033} &
$346\pm20$ \2 &
$4274.4$ fixed &
$32.3$ fixed  &
  &
$<\phantom{0}8$ @~90\%CL \\
%\hline
2013 &
CMS 5.2 fb$^{-1}$  \cite{Chatrchyan:2013dma} &
$2480\pm160$ &
$4313.8\pms5.3\pms7.3$ \ifthenelse{\boolean{prl}}{\quad}{\3} &
$38\phantom{.0}\,^{+30\phantom{.0}}_{-15\phantom{.0}}\pms16$ &
  & \\
%\hline
2013 & D0 10.4 fb$^{-1}$  \cite{Abazov:2013xda} &
$215\pm37$ \2 &
$4328.5\pms12.0$ &
$30$ fixed &
  & \\
%\hline
2014 & BaBar \cite{Lees:2014lra} &
$189\pm14$ \2 &
$4274.4$ fixed &
$32.3$ fixed  &
$1.2\sigma$  &
$<18.1$ @~90\%CL \\
%\hline
\hline
2010 & Belle \cite{Shen:2009vs} &
\1{\3$\gamma\gamma\to\jpsi\phi$\3} &
$4350.6\,^{+4.6}_{-5.1}\pms0.7$ &
$13\,^{+18}_{-\phantom{0}9}\pms4$ & 
$3.2\sigma$ & \\
\hline
\end{tabular}
\ifthenelse{\boolean{prl}}{}{\end{footnotesize}}
}
}
}
\end{table*} 

In view of the considerable theoretical interest 
in possible exotic hadronic states decaying to $\jpsi\phi$,
it is important to clarify the rather confusing experimental
situation concerning $\jpsi\phi$ mass structures. 
The data sample used in this work corresponds to an integrated luminosity of $3$~fb$^{-1}$  
collected with the \lhcb detector in $pp$ collisions at 
center-of-mass energies 7 and 8~\tev.
Thanks to the larger signal yield, 
corresponding to $4289\pm151$ reconstructed $B^+\to\jpsi\phi K^+$ decays, 
the roughly uniform efficiency and the relatively low background 
across the entire $\jpsi\phi$ mass range,
this data sample offers the best sensitivity to date, 
not only to probe for the $X(4140)$, $X(4274)$ and other previously 
claimed structures, but also to inspect the high mass region.

All previous analyses were based on naive $\jpsi\phi$ mass ($m_{\jpsi\phi}$)
fits, with Breit--Wigner signal peaks
on top of incoherent background described by ad-hoc functional shapes 
(\eg three-body phase space distribution in $\bujphik$ decays).
While the $m_{\phi K}$ distribution has been observed to be smooth, 
several resonant contributions from 
kaon excitations (hereafter denoted generically as $K^*$) are expected.
It is important to prove that any $m_{\jpsi\phi}$ peaks are not merely
reflections of these conventional resonances.
If genuine $\jpsi\phi$ states are present, it is crucial to
determine their quantum numbers to aid their theoretical interpretation. 
Both of these tasks call for a proper amplitude analysis of $\bujphik$ decays,
in which the observed $m_{\phi K}$ and $m_{\jpsi\phi}$ masses are analyzed
simultaneously with the distributions of decay angles, without which the resolution of different
resonant contributions is difficult, if not impossible. 
The analysis of $\jpsi$ and $\phi$ polarizations via their decays to $\mu^+\mu^-$ 
and $K^+K^-$, respectively, increases greatly the sensitivity of the analysis 
as compared with the Dalitz plot analysis alone. 
In addition to the search for exotic hadrons, which includes $X\to\jpsi\phi$ and
$Z^+\to\jpsi K^+$ contributions, the amplitude analysis of our data 
offers unique insight into the spectroscopy of the poorly experimentally understood 
higher excitations of the kaon system, in their decays to a $\phi K^+$ final state.

In this article, an amplitude analysis of the decay $\bujphik$ is presented for the first time,
with additional results for, and containing more detailed description of, 
the work presented in Ref.~\cite{LHCb-PAPER-2016-018}.

\section{LHCb detector}

The \lhcb detector~\cite{Alves:2008zz,LHCb-DP-2014-002} is a single-arm forward
spectrometer covering the \mbox{pseudorapidity} range $2<\eta <5$,
designed for the study of particles containing \bquark or \cquark
quarks. The detector includes a high-precision tracking system
consisting of a silicon-strip vertex detector surrounding the $pp$
interaction region,
a large-area silicon-strip detector located
upstream of a dipole magnet with bending power of about
$4{\mathrm{\,Tm}}$, and three stations of silicon-strip detectors and straw
drift tubes %~\cite{LHCb-DP-2013-003}\verb!*! 
placed downstream of the magnet.
The tracking system provides a measurement of momentum, \ptot, of charged particles with
relative uncertainty that varies from 0.5\% at low momentum to 1.0\% at $200 \gev$.
The minimum distance of a track to a primary vertex (PV), the impact parameter (IP), 
is measured with a resolution of $(15+29/\pt)\mum$,
where \pt is the component of the momentum transverse to the beam, in\,\gev.
Different types of charged hadrons are distinguished using information
from two ring-imaging Cherenkov detectors. %~\cite{LHCb-DP-2012-003}\verb!*!. 
Photons, electrons and hadrons are identified by a calorimeter system consisting of
scintillating-pad and preshower detectors, an electromagnetic
calorimeter and a hadronic calorimeter. Muons are identified by a
system composed of alternating layers of iron and multiwire
proportional chambers. %~\cite{LHCb-DP-2012-002}\verb!*!.
The online event selection is performed by a trigger, %~\cite{LHCb-DP-2012-004}\verb!*!, 
which consists of a hardware stage, based on information from the calorimeter and muon
systems, followed by a software stage, which applies a full event
reconstruction.

\section{Data selection}

  Candidate events for this analysis are first required to pass the hardware trigger,
  which selects muons with transverse momentum $\pt>1.48\gev$ 
  in the 7\tev data or $\pt>1.76\gev$ in the 8\tev data.
  In the subsequent software trigger, at least 
  one of the final-state particles is required to have $\pt>1.7\gev$ 
  in the 7\tev data or $\pt>1.6\gev$ in the 8\tev data, 
  unless the particle is identified as a muon in which case $\pt>1.0\gev$ is required. 
  The final-state particles that satisfy these transverse momentum criteria 
  are also required to have an impact parameter larger than 100\mum with respect 
  to all of the primary $pp$ interaction vertices~(PVs) in the event. 
  Finally, the tracks of two or more of the final-state particles are required to form 
  a vertex that is significantly displaced from the PVs. 
  In the subsequent offline selection, trigger signals are required to be associated with
  reconstructed particles in the signal decay chain. 

\begin{figure}[tbhp]
  \begin{center}
    \includegraphics*[width=\figsize]{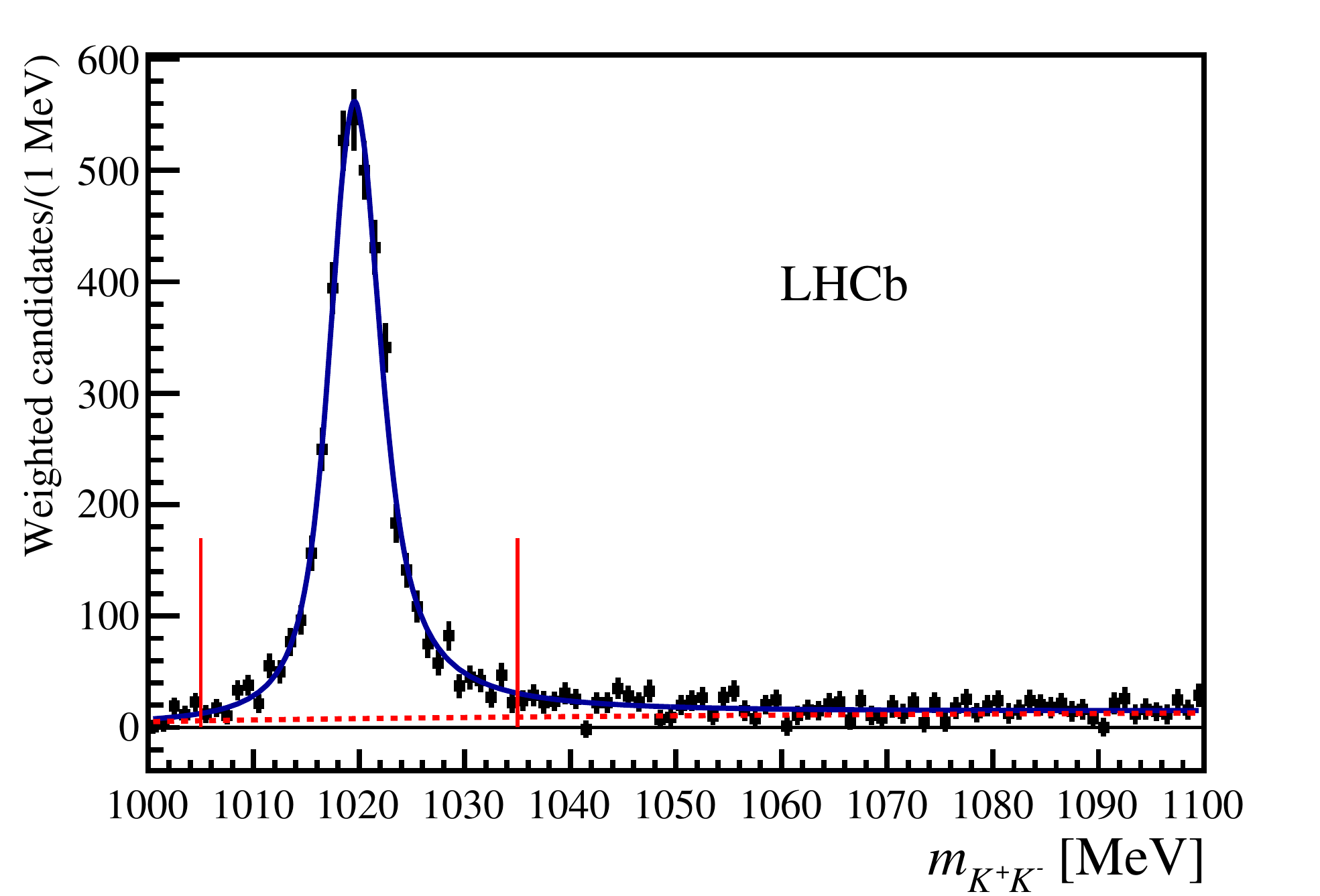}
  \end{center}
  \vskip-0.3cm\caption{\small
  Distribution of $m_{K^+K^-}$ near the $\phi$ peak 
  before the $\phi$ candidate selection.
  Non-$B^+$ backgrounds have been subtracted using 
  \sPlot weights \cite{Pivk:2004ty} obtained from a fit 
  to the $m_{\jpsi K^+K^-K^+}$ distribution.
  The default $\phi$ selection window is indicated with vertical red lines.
  The fit (solid blue line) of a Breit--Wigner $\phi$ signal shape 
  plus two-body phase space function (dashed red line), 
  convolved with a Gaussian resolution function,
  is superimposed. 
  \label{fig:mkk}
  }
\end{figure}

The offline data selection is very similar 
to that described in Ref.~\cite{LHCb-PAPER-2011-033}, with 
$\jpsi\to\mu^+\mu^-$ candidates
required to satisfy the following criteria:
$\pt(\mu)>0.55 \gev$,
$\pt(\jpsi)>1.5 \gev$,   
$\chi^2$ per degree of freedom for the two muons to form a common vertex, $\chi^2_{\rm vtx}(\mu^+\mu^-)/\NDOF<9$,
and mass consistent with the $\jpsi$ meson. 
Every charged track with $\pt>0.25 \gev$,
missing all PVs by at least 3 standard deviations ($\chi^2_{\rm IP}(K)>9$)
and classified as more likely to be a kaon than a pion according to the particle identification system,
is considered a kaon candidate.
The quantity $\chi^2_{\rm IP}(K)$ is defined as the
difference between the \chisq of the PV reconstructed with and
without the considered particle.
Combinations of $K^+K^-K^+$ candidates that are consistent with originating from a common vertex with 
$\chi^2_{\rm vtx}(K^+K^-K^+)/\NDOF<9$ are selected.
We combine $\jpsi$ candidates with $K^+K^-K^+$ candidates to form 
$B^+$ candidates, which must satisfy $\chi^2_{\rm vtx}(\jpsi K^+K^-K^+)/\NDOF<9$, $\pt(B^+)>2 \gev$ 
and have decay time greater than $0.25$~ps. The $\jpsi K^+K^-K^+$ mass is calculated 
using the known $\jpsi$ mass \cite{PDG2014} and the $B^+$ vertex as constraints \cite{Hulsbergen:2005pu}.   

Four discriminating variables ($x_i$) are
used in a likelihood ratio to improve the background suppression:
the minimal $\chi^2_{\rm IP}(K)$,
$\chi^2_{\rm vtx}(\jpsi  K^+K^-K^+)/\NDOF$,
$\chi^2_{\rm IP}(B^+)$,
and the cosine of the largest opening angle between the $\jpsi$ and 
the kaon transverse momenta. 
The latter peaks at positive values for the signal as the $B^+$ meson has high transverse momentum.
Background events in which particles are combined 
from two different $B$ decays peak at negative values,
whilst those due to random combinations of particles are more uniformly distributed.  
The four signal probability density functions (PDFs), $\PDF_{\rm sig}(x_i)$, 
are obtained from simulated $\bujphik$ decays.
The background PDFs, $\PDF_{\rm bkg}(x_i)$, are obtained from 
candidates in data with $\jpsi K^+K^-K^+$ invariant mass between 5.6 and $6.4 \gev$. 
We require $-2 \sum_{i=1}^4 \ln[\PDF_{\rm sig}(x_i)/\PDF_{\rm bkg}(x_i)]<5$,
which retains about 90\%\ of the signal events.

Relative to the data selection described in Ref.~\cite{LHCb-PAPER-2011-033}, the
requirements on transverse momentum for $\mu$ and $B^+$ 
candidates have been lowered 
and the requirement on the multivariate signal-to-background log-likelihood
difference was loosened.
As a result, the $B^+$ signal yield per unit luminosity has increased by about 50\%
at the expense of somewhat higher background. 

\begin{figure}[tbhp]
  \begin{center}
    \includegraphics*[width=\figsize]{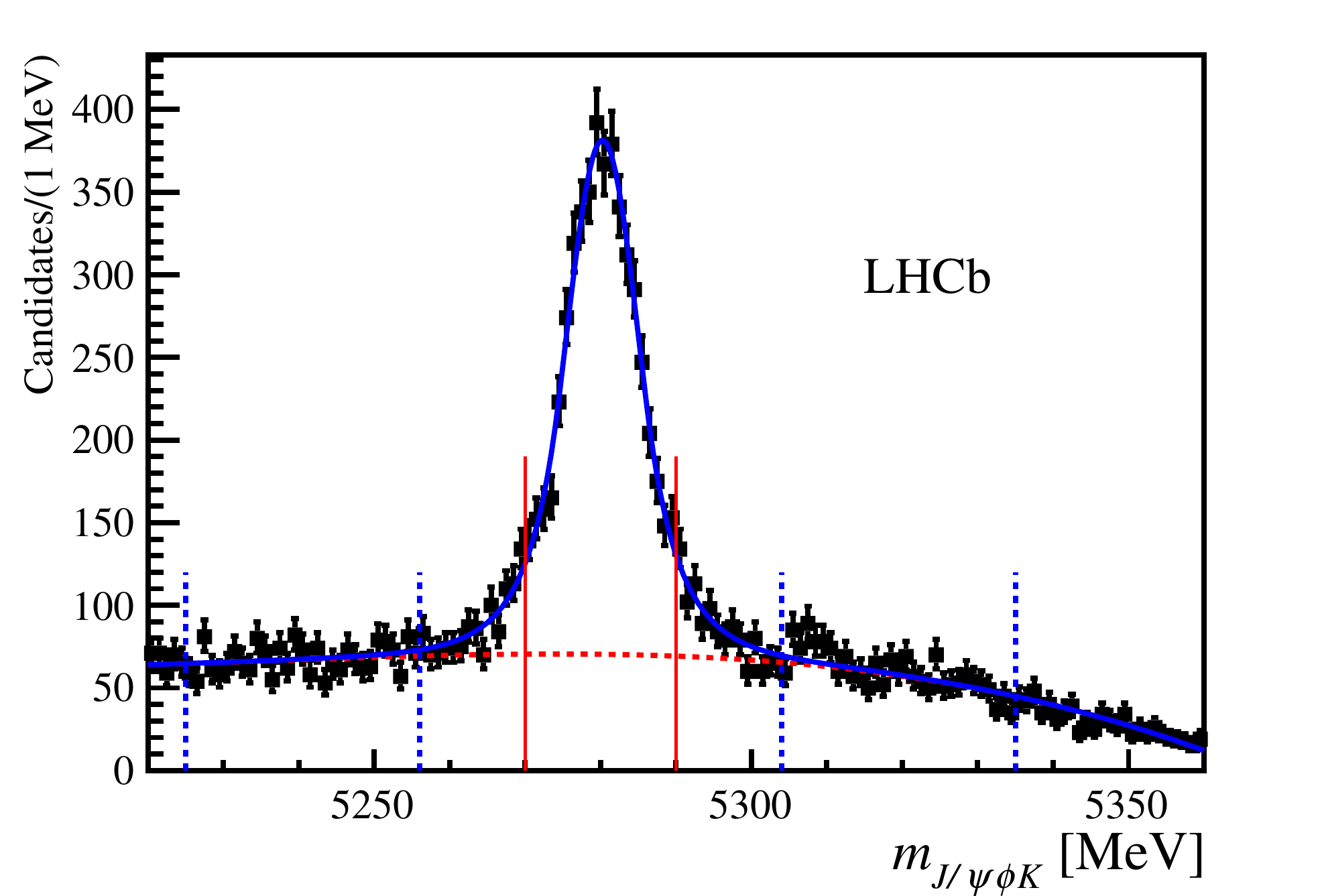}
  \end{center}
  \vskip-0.3cm\caption{\small
    Mass of $B^+\to\jpsi\phi K^+$ candidates in the data (black points with error bars)
    together with the results of the fit (blue line) with a double-sided Crystal Ball shape  
    for the $B^+$ signal on top of a quadratic function for the background (red dashed line). 
    The fit is used to determine the background fraction under the peak in the mass
    range used in the amplitude analysis (indicated with vertical solid red lines). 
    The sidebands used for the background parameterization are indicated with vertical
    dashed blue lines.
  \label{fig:mjpsiphik}
  }
\end{figure}

The distribution of $m_{K^+K^-}$ for the selected 
$B^+\to\jpsi K^+K^-K^+$ candidates
is shown in Fig.~\ref{fig:mkk} (two entries per candidate).
A fit with a P-wave relativistic Breit--Wigner shape on top of
a two-body phase space distribution representing non-$\phi$ background, 
both convolved with 
a Gaussian resolution function with width of $1.2 \mev$
is superimposed. 
Integration of the fit components gives 
$(5.3\pm0.5)\%$ of nonresonant background in the
$|m_{K^+K^-}-1020~\mev|<15 \mev$ region used to define
a $\phi$ candidate. 
To avoid reconstruction ambiguities, we
require that there is exactly one $\phi$ candidate per $\jpsi K^+K^-K^+$ combination, 
which reduces the $B^+$ yield by $3.2\%$. 
The non-$\phi$ $B^+\to\jpsi K^+K^-K^+$ background in 
the remaining sample is small ($2.1\%$)
and neglected in the amplitude model. 
The related systematic uncertainty is estimated by 
tightening the $\phi$ mass selection window to $\pm7 \mev$.  

The mass distribution of the remaining $\jpsi\phi K^+$ 
combinations
is shown in Fig.~\ref{fig:mjpsiphik} together with 
a fit of the $B^+$ signal represented by a symmetric 
double-sided Crystal Ball function \cite{Skwarnicki:1986xj}  
on top of a quadratic function for the background.
The fit yields $4289\pm151$ $B^+\to\jpsi\phi K^+$ 
events. Integration of the fit components in the 
$5270$--$5290 \mev$ region
(twice the $B^+$ mass resolution on each side of its peak)
used in the amplitude fits,
gives a background fraction ($\beta$) of $(23\pm6)\%$.
A Gaussian signal shape and a higher-order polynomial background
function are used to assign systematic uncertainties 
which are included in, and dominate, the uncertainty given above.
The $B^+$ invariant mass sidebands, $5225$--$5256$ and $5304$--$5335 \mev$, are used 
to parameterize the background in the amplitude fit. 

The $B^+$ candidates for the amplitude analysis
are kinematically constrained 
to the known $B^+$ mass \cite{PDG2014} 
and to point to the closest $pp$ interaction vertex.
The measured value of $m_{K^+K^-}$ is used for the $\phi$ candidate mass, 
since the natural width of the $\phi$ resonance is larger than
the detector resolution.

\FloatBarrier

\section{Matrix element model}

We consider the three interfering processes 
corresponding to the following decay sequences:
$B^+\to K^{*+}\jpsi$, $K^{*+}\to\phi K^+$
(referred to as the $\Kstar$ decay chain),
$B^+\to X K^+$, $X\to\jpsi\phi$ 
($X$ decay chain) and
$B^+\to Z^+\phi$, $Z^+\to\jpsi K^+$
($Z$ decay chain), 
all followed by $\jpsi\to\mu^+\mu^-$ and $\phi\to K^+K^-$ decays.
Here, $K^{*+}$, $X$ and $Z^+$ should be understood as 
any $\phi K^+$, $\jpsi\phi$ and $\jpsi K^+$ contribution, respectively.

\begin{figure}[tbhp]
  \begin{center}
    \includegraphics*[width=\figsize]{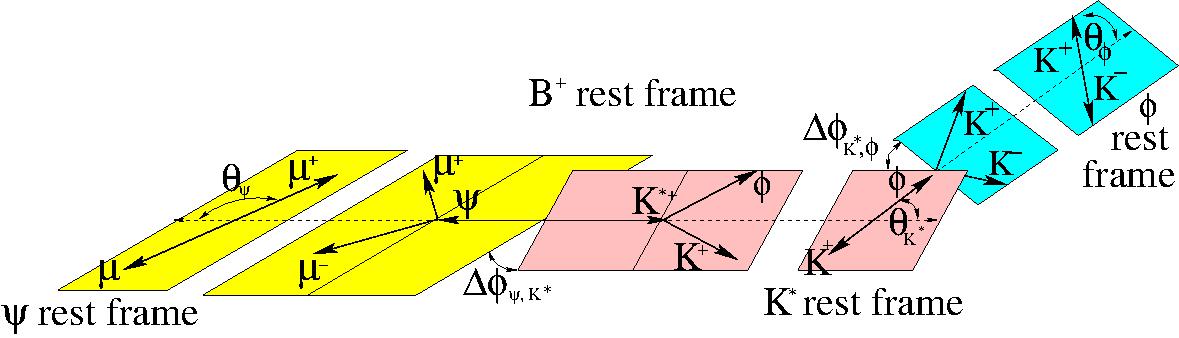}
  \end{center}
  \vskip-0.3cm\caption{\small
      Definition of the $\theta_{\Kstar}$, $\theta_{\jpsi}$, $\theta_{\phi}$, $\phikspsi$ and $\phiksphi$
      angles describing angular correlations in $B^+\to\jpsi K^{*+}$, $\jpsi\to\mu^+\mu^-$, 
      $K^{*+}\to\phi K^+$, $\phi\to K^+K^-$ decays ($\jpsi$ is denoted as $\psi$ in the figure).      
  \label{fig:KstarAngles}
  }
\end{figure}

We construct a model of the matrix element ($\Mat$) 
using the helicity formalism \cite{Jacob:1959at,Richman:1984gh,PhysRevD.57.431} 
in which the 
six independent variables fully describing the $K^{*+}$ decay chain are
$m_{\phi K}$, 
$\theta_{\Kstar}$, $\theta_{\jpsi}$, $\theta_{\phi}$, 
$\phikspsi$ and $\phiksphi$, 
where the helicity angle $\theta_P$ is defined
as the angle in the rest frame of $P$ 
between the momentum of its decay product
and the boost direction from the rest frame of the particle
which decays to $P$,
and $\Delta\phi$ is the angle between the decay planes of
the two particles (see Fig.~\ref{fig:KstarAngles}).
The set of angles is denoted by $\Omega$. 
The explicit formulae for calculation of the angles in $\Omega$
are given in Appendix~\ref{sec:angles}.

The full six-dimensional (6D) matrix element for the $\Kstar$ decay chain is given by
\def\jkstar{j}
\ifthenelse{\boolean{prl}}{\begin{widetext}}{}
\begin{equation}
\begin{split}
\Mat^{\Kstar}_{\Delta\lambda_\mu} \equiv & 
{\sum_{\jkstar}}\,
{\,}
R_j(\mphik)
%|M_{0\,\Kstar\,\jkstar},\Gamma_{0\,\Kstar\,\jkstar})
{\sum_{\lambda_{\jpsi}=-1,0,1}}\,
{\sum_{\lambda_{\phiz}=-1,0,1}}\,\,
{A_{\lambda_{\jpsi}}^{B\to\jpsi\Kstar\,\jkstar}} 
{\,}
{A_{\lambda_{\phiz}}^{\Kstar\to\phiz K\,\jkstar}}
{\,}\times \\%\notag\\ 
& \quad\quad {d^{J_{\Kstar}\,\jkstar}_{\rm \lambda_{\jpsi},\lambda_{\phiz}}(\theta_{\Kstar})}
{\,}
{d^{1}_{\rm \lambda_{\phiz},0}(\theta_{\phiz})}
{\,}
e^{i\lambda_\phiz \phiksphi}
{\,}
{d^{1}_{\rm \lambda_{\jpsi},\Delta\lambda_{\mu}}(\theta_{\jpsi})}
{\,}
e^{i\lambda_\jpsi \phikspsi},
\\%\notag\\
|\Mat^{\Kstar}|^{2}= & {\sum_{\Delta\lambda_{\mu}=\pm1}}\,
\left| 
\Mat^{\Kstar}_{\Delta\lambda_\mu} 
\right|^2,
\end{split}
\label{eq:KstMatrix}
\end{equation}
\ifthenelse{\boolean{prl}}{\end{widetext}}{}%
where the index $\jkstar$ enumerates the different $\Kstarp$ resonances. 
The symbol $J_\Kstar$ denotes the spin of the $\Kstar$ resonance, 
$\lambda$ is the helicity (projection of the particle spin onto its momentum in the rest
frame of its parent) and $\Delta\lambda_\mu\equiv\lambda_{\mup}-\lambda_{\mu^-}$.
The terms $d^{J}_{\rm \lambda_1,\lambda_2}(\theta)$ are the Wigner d-functions, 
$R_j(\mphik)$ is the mass dependence of the contribution 
and will be discussed in more detail later
(usually a complex Breit--Wigner amplitude 
depending on resonance pole mass $M_{0\,\Kstar\,\jkstar}$
and width $\Gamma_{0\,\Kstar\,\jkstar}$).
The coefficients 
${A_{\lambda_{\jpsi}}^{B\to\jpsi\Kstar}}$ and  
${A_{\lambda_{\phiz}}^{\Kstar\to\phiz K}}$ are complex helicity couplings
describing the (weak) $B^+$ and (strong) $K^{*+}$ decay dynamics, respectively.
There are three independent complex ${A_{\lambda_{\jpsi}}^{B\to\jpsi\Kstar}}$ couplings to be fitted ($\lambda_{\jpsi}=-1,0,1$)
per $\Kstar$ resonance, unless $J_\Kstar=0$ in which case there is only one since $\lambda_\jpsi=\lambda_\Kstar$ due to $J_B=0$. 
Parity conservation in the $\Kstar$ decay limits the number of 
independent helicity couplings ${A_{\lambda_{\phiz}}^{\Kstar\to\phiz K}}$.
More generally parity conservation requires
\begin{equation}
A_{-\lambda_B,-\lambda_C}^{A\to B\,C} = P_A\,P_B\,P_C\,(-1)^{J_B+J_C-J_A}\, A_{\lambda_B,\,\lambda_C}^{A\to B\,C}, 
\label{eq:genPar}
\end{equation}
which, for the decay $\Kstarp\to\phiz K^+$, leads to
\begin{equation}
A_{\lambda_{\phiz}}={P_\Kstar}(-1)^{J_{\Kstar}+1}\,A_{-\lambda_{\phiz}}.
\label{eq:KsPhiPar}
\end{equation}
This reduces the number of independent couplings in the $\Kstar$ decay to one or two.
Since the overall magnitude and phase of these couplings can be absorbed in 
${A_{\lambda_{\jpsi}}^{B\to\jpsi\Kstar}}$, in practice the $\Kstar$ decay
contributes zero or one complex parameter to be fitted per $\Kstar$ resonance.

The matrix element for the $X$ decay chain 
can be parameterized using $m_{\jpsi\phi}$ and the  
$\theta_X$, $\theta_{\jpsi}^X$, $\theta_\phi^X$, 
$\Delta\phi_{X,\jpsi}$, $\Delta\phi_{X,\phi}$ angles.  
The angles $\theta_{\jpsi}^X$ and $\theta_\phi^X$ are not the same as 
$\theta_{\jpsi}$ and $\theta_\phi$ in the $K^*$ decay chain, since $\jpsi$ and $\phi$ are produced
in decays of different particles. 
For the same reason, the muon helicity states are different between the 
two decay chains, and an azimuthal rotation by angle $\alpha^X$ 
is needed to align them as discussed below. 
The parameters needed to characterize 
the $X$ decay chain, including $\alpha^X$, do not constitute new degrees of freedom
since they can all be derived from $m_{\phi K}$ and $\Omega$. 
The matrix element for the $X$ decay chain also has unique helicity couplings and is given by
\def\jx{j}
\ifthenelse{\boolean{prl}}{\begin{widetext}}{}
\begin{equation}
\begin{split}
\Mat^X_{\Delta\lambda_\mu} \equiv & 
{\sum_{\jx}}\,
{\,}
R_j(\mjpsiphi)
%|M_{0\,{X\,\jx}},\Gamma_{0\,{X\,\jx}})
{\sum_{\lambda_{\jpsi}=-1,0,1}}\,
{\sum_{\lambda_{\phiz}=-1,0,1}}\,\,
A_{\lambda_{\jpsi},\lambda_{\phiz}}^{X\to\jpsi\phiz\,\jx}\,\times\\%\notag\\
& \quad\quad 
{\,}
{d^{J_X\,\jx}_{\rm {0,\lambda_{\jpsi}-\lambda_{\phiz}}}(\theta_{X})}
{\,}
{d^{1}_{\rm \lambda_{\phiz},0}(\theta_{\phiz}^X)}
{\,}
e^{i\lambda_\phiz \phixphi}
{\,}
{d^{1}_{\rm \lambda_{\jpsi},\Delta\lambda_{\mu}}(\theta_{\jpsi}^X)}
{\,}
e^{i\lambda_\jpsi \phixpsi},
\\%\notag\\
|{\Mat^{K^*+X}}|^{2} = & {\sum_{\Delta\lambda_{\mu}=\pm1}}
\left|
\Mat^{\Kstar}_{\Delta\lambda_\mu} 
+
{e^{i\alpha^X\Delta\lambda_{\mu}}}
\Mat^X_{\Delta\lambda_\mu}
\right|^2,
\end{split}
\label{eq:XMatrix}
\end{equation}
\ifthenelse{\boolean{prl}}{\end{widetext}}{}%
where the index $\jx$ enumerates all $X$ resonances.
To add $\Mat^{\Kstar}_{\Delta\lambda_\mu}$ and $\Mat^X_{\Delta\lambda_\mu}$ coherently
it is necessary to introduce the ${e^{i\alpha^X\Delta\lambda_{\mu}}}$ term,
which corresponds to a rotation about the $\mup$ momentum axis by the angle $\alpha^X$ 
in the rest frame of $\jpsi$ after arriving to it by a boost from the $X$ rest frame.
This realigns the coordinate axes for the muon helicity frame in the 
$X$ and $\Kstar$ decay chains. This issue is discussed in 
Ref.~\cite{Chilikin:2013tch} and at more length in 
Ref.~\cite{LHCb-PAPER-2015-029}.

The structure of helicity couplings in the $X$ decay chain is different  
from the $\Kstar$ decay chain. The decay $B^+\to X K^+$ does not contribute 
any helicity couplings to the fit\footnote{There is one additional coupling, but that can be absorbed
by a redefinition of $X$ decay couplings, which are free parameters.}, 
since $X$ is produced fully polarized $\lambda_X=0$.
The $X$ decay contributes a resonance-dependent 
matrix of helicity couplings
$A_{\lambda_{\jpsi},\lambda_{\phiz}}^{X\to\jpsi\phiz}$.
Fortunately, parity conservation reduces
the number of independent complex couplings 
to one for $J^P_X=0^-$, two for $0^+$, three for $1^+$, four for $1^-$ and $2^-$, and at most five independent couplings for $2^+$.

The matrix element for the $Z^+$ decay chain 
can be parameterized using $m_{\jpsi K}$ and the  
$\theta_Z$, $\theta_{\jpsi}^Z$, $\theta_\phi^Z$, 
$\Delta\phi_{Z,\jpsi}$, $\Delta\phi_{Z,\phi}$ angles.  
The $Z^+$ decay chain also requires a rotation 
to align the muon frames to those used in the $\Kstar$ decay chain and to allow  
for the proper description of interference between the three decay chains. 
The full 6D matrix element is given by
\def\jz{j}
\ifthenelse{\boolean{prl}}{\begin{widetext}}{}
\begin{equation}
\begin{split}
\Mat_{\Delta\lambda_\mu}^Z\equiv &
{\sum_{\jz}}\,
{\,}
R_j(\mjpsik)
%|m_{0\,Z\,\jz},\Gamma_{0\,Z\,jz})
{\sum_{\lambda_{\jpsi}=-1,0,1}}\,
{\sum_{\lambda_{\phiz}=-1,0,1}}\,
A_{\lambda_{\phiz}}^{B\to Z\phi\,\jz}
{\,}
A_{\lambda_{\jpsi}}^{Z\to \jpsi K\,\jz}\,\times \\%\notag\\
& \quad\quad
{\,} 
{d^{J_{Z\,\jz}}_{\rm \lambda_{\jpsi},\lambda_{\jpsi}}(\theta_{Z})}
{\,}
{d^{1}_{\rm \lambda_{\phiz},0}(\theta_{\phiz}^Z)}
{\,}
e^{i\lambda_\phiz \phizphi}
{\,}
{d^{1}_{\rm \lambda_{\jpsi},\Delta\lambda_{\mu}}(\theta_{\jpsi}^Z)}
{\,}
e^{i\lambda_\jpsi \phikspsi},
\\%\notag\\
|\Mat^{\Kstar+X+Z}|^{2} = & {\sum_{\Delta\lambda_{\mu}=\pm1}}
\left|
\Mat^{\Kstar}_{\Delta\lambda_\mu} 
+
{e^{i\alpha^X\Delta\lambda_{\mu}}}
\Mat^X_{\Delta\lambda_\mu}
+
{e^{i\alpha^Z\Delta\lambda_{\mu}}}
\Mat^Z_{\Delta\lambda_\mu}
\right|^2.
\end{split}
\label{eq:ZMatrix}
\end{equation}
\ifthenelse{\boolean{prl}}{\end{widetext}}{}%
Parity conservation in the $Z^+$ decay requires
\begin{equation}
A_{\lambda_{\jpsi}}^{B\to Z\phi}={P_Z}(-1)^{J_{Z}+1}\,A_{-\lambda_{\jpsi}}^{B\to Z\phi}
\label{eq:ZPhiPar}
\end{equation}
and provides a similar reduction of the couplings as discussed for the $\Kstar$ decay chain.

Instead of fitting the helicity couplings $A_{\lambda_B,\lambda_C}^{A\to B\,C}$ as free parameters, after imposing parity 
conservation for the strong decays, it is convenient to express them by an equivalent number of independent
$LS$ couplings ($B_{LS}$), where $L$ is the orbital angular momentum in the decay and $S$ is the total spin of $B$ and $C$,
$\vec{S}=\vec{J}_B+\vec{J}_C$ ($|J_B-J_C|\le S \le J_B+J_C$).
Possible combinations of $L$ and $S$ values are constrained via $\vec{J}_A=\vec{L}+\vec{S}$.
The relation involves the Clebsch-Gordan coefficients
\ifthenelse{\boolean{prl}}{\begin{widetext}}{}
\begin{equation}
{A_{\lambda_B,\lambda_C}^{A\to B\,C}}=\sum_{L} \sum_{S} 
\sqrt{\frac{2L+1}{2J_A+1}}\,
B_{L,S}\,
\left( 
\begin{array}{cc|c}
 J_{B} & J_{C} & S \\
 \lambda_{B} & -\lambda_{C} & \lambda_{B}-\lambda_{C} 
\end{array}
\right)
\,
\left( 
\begin{array}{cc|c}
 L  & S & J_A \\
 0 & \lambda_{B}-\lambda_{C} & \lambda_{B}-\lambda_{C}   
\end{array}
\right).
\label{eq:LS}
\end{equation}
\ifthenelse{\boolean{prl}}{\end{widetext}}{}%
Parity conservation in the strong decays is imposed by
\begin{equation}
P_A=P_B\,P_C\,(-1)^L.
\end{equation}

Since the helicity or $LS$ couplings not only shape the angular distributions but also describe the overall strength and phase
of the given contribution relative to all other contributions in the matrix element, we separate these roles by always
setting the coupling for 
the lowest $L$ and $S$, $B_{L_{\rm min}S_{\rm min}}$, for a given contribution to $(1,0)$ and multiplying    
the sum in Eq.~(\ref{eq:LS}) by a complex fit parameter $\AR$ (this is equivalent to factoring out $B_{L_{\rm min}S_{\rm min}}$).
This has an advantage when interpreting the numerical values of these parameters. 
The value of $\AR_j$ describes the relative magnitude and phase of
the $B_{L_{\rm min}S_{\rm min}\,j}$ to the other contributions, and the fitted $B_{LS\,j}$ values correspond to
the ratios, $B_{LS\,j}/B_{L_{\rm min}S_{\rm min}\,j}$, and determine the angular distributions. 

Each contribution to the matrix element comes with its own $R(m_A)$ function, which gives its dependence on the 
invariant mass of the intermediate resonance $A$ in the decay chain ($A=K^{*+}$, $X$ or $Z^+$).
Usually it is given by the Breit--Wigner amplitude, but there are special cases which we discuss below.
An alternative parameterization of $R(m_A)$ to represent 
coupled-channel cusps is discussed in Appendix~\ref{sec:cusp}.

In principle, the width of the $\phiz$ resonance should also be taken into account.
However, since the $\phiz$ resonance is very narrow ($\Gamma_0=4.3 \mev$, with mass resolution of $1.2\mev$)
we omit the amplitude dependence on the invariant $m_{K^+K^-}$ mass from the $\phi$ decay.

A single resonant contribution in the decay chain $B^+\to A...$, $A\to...$ 
is parameterized by the relativistic Breit--Wigner amplitude
together with Blatt--Weisskopf functions,
\ifthenelse{\boolean{prl}}{\begin{widetext}}{}
\begin{equation}
R(m|M_{0},\Gamma_{0}) =
B'_{L_B}(p,p_0,d) \left(\frac{p}{p_0}\right)^{L_B} \,
BW(m | M_{0}, \Gamma_{0} ) \,
B'_{L_{A}}(q,q_0,d)
\left(\frac{q}{q_0}\right)^{L_{A}},
\label{eq:resshapeBW}
\end{equation}
\ifthenelse{\boolean{prl}}{\end{widetext}}{}%
where
\begin{equation}
BW(m | M_0, \Gamma_0) = \frac{1}{M_0^2-m^2 - i M_0 \Gamma(m)} \,,
\label{eq:breitwigner}
\end{equation}
is the Breit--Wigner amplitude including the mass-dependent width, 
\begin{equation}
\Gamma(m)=\Gamma_0 \left(\frac{q}{q_0}\right)^{2\,L_A+1} \frac{M_0}{m} B'_{L_{A}}(q,q_0,d)^2 \,.
\label{eq:mwidth}
\end{equation}
Here, $p$ is the momentum of the resonance $A$ ($K^{*+}$, $X$ or $Z^+$) 
in the $B^+$ rest frame, and  
$q$ is the momentum of one of the decay products of $A$ in the 
rest frame of the $A$ resonance. 
The symbols $p_0$ and $q_0$ are used to indicate values of these quantities at the resonance peak mass ($m=M_0$). 
The orbital angular momentum in $B^+$ decay is denoted as $L_B$, and that in the decay of the resonance $A$ as $L_A$.
The orbital angular momentum barrier factors, $p^L\,B'_{L}(p,p_0,d)$, 
involve the Blatt--Weisskopf functions \cite{Blatt-Weisskopf-1979,VonHippel:1972fg}: 
\ifthenelse{\boolean{prl}}{\begin{widetext}}{}
{%
\def\1{p\phantom{_0}}
\def\2{p_0}
\small
\begin{eqnarray}
\label{eq:blattw}
B'_{0}(p,p_0,d)&=&1 \,,\\
B'_{1}(p,p_0,d)&=&\sqrt{ \frac{1+(\2\,d)^2}{1+(\1\,d)^2} } \,,\\
B'_{2}(p,p_0,d)&=&\sqrt{ \frac{9+3(\2\,d)^2+(\2\,d)^4}{9+3(\1\,d)^2+(\1\,d)^4} } \,,\\
B'_{3}(p,p_0,d)&=&\sqrt{ \frac{225+45(\2\,d)^2+6(\2\,d)^4+(\2\,d)^6}{225+45(\1\,d)^2+6(\1\,d)^4+(\1\,d)^6} } \,,\\
B'_{4}(p,p_0,d)&=&\sqrt{ \frac{11025+1575(\2\,d)^2+135(\2\,d)^4+10(\2\,d)^6+(\2\,d)^8}{11025+1575(\1\,d)^2+135(\1\,d)^4+10(\1\,d)^6+(\1\,d)^8} } \,,
%\\
%B'_{5}(p,p_0,d)&=&\sqrt{ \frac{893025+99225(\2\,d)^2+6300(\2\,d)^4+315(\2\,d)^6+15(\2\,d)^8+(\2\,d)^{10}}{
%                             893025+99225(\1  \,d)^2+6300(\1  \,d)^4+315(\1  \,d)^6+15(\1  \,d)^8+(\1  \,d)^{10}} } \,,
\end{eqnarray}
}%
\ifthenelse{\boolean{prl}}{\end{widetext}}{}%
which account for the centrifugal barrier in the decay 
and depend on the momentum of the decay products in the rest frame of the decaying particle ($p$) 
as well as the size of the decaying particle ($d$).  
In this analysis we set this parameter to a nominal value of 
$d=3.0 \gev^{-1}$, and vary it in between $1.5$ and $5.0 \gev^{-1}$ in the evaluation of the
systematic uncertainty. 

In the helicity approach, each helicity state is a mixture of many different $L$ values.
We follow the usual approach of using in Eq.~(\ref{eq:resshapeBW}) the minimal 
$L_B$ and $L_A$ values allowed by the quantum numbers of the given resonance $A$, while higher values
are used to estimate the systematic uncertainty.

We set $BW(m) = 1.0$ for the nonresonant (NR) contributions,
which means assuming that both magnitude and phase have negligible $m$ dependence.
As the available phase space in the $B^+\to\jpsi\phi K^+$ decays is small 
(the energy release is only 12\%\ of the $B^+$ mass)
this is a well-justified assumption. We consider possible 
mass dependence of NR amplitudes as a source of
systematic uncertainties.  

\section{Maximum likelihood fit of amplitude models}
\label{sec:maxlike}

The signal PDF, $\PDF_{\rm sig}$,
is proportional to the matrix element squared, which 
is a function of six independent variables: 
$m_{\phi K}$ 
and the independent angular variables in the $\Kstar$ decay chain $\Omega$. 
The PDF also depends on the fit parameters, $\Pars$, which include the 
helicity couplings, and masses and widths of resonances. 
The two other invariant masses, $m_{\phi K}$ and $m_{\jpsi K}$, and 
the angular variables describing the $X$ and $Z^+$ decay chains 
depend on $m_{\phi K}$ and $\Omega$, therefore they do not represent independent dimensions. 
The signal PDF is given by:
\ifthenelse{\boolean{prl}}{\begin{widetext}}{}
\begin{equation}
\label{eq:sigpdfsplot}
\frac{d\PDF}{d m_{\phi K}\,d\Omega} \equiv \PDF_{\rm sig}(m_{\phi K},\Omega|\Pars)
=\frac{1}{I(\overrightarrow{w})}\big|\Mat(m_{\phi K},\Omega| \Pars)\big|^2
\Phi(m_{\phi K})
\epsilon(m_{\phi K},\Omega),
\end{equation}
\ifthenelse{\boolean{prl}}{\end{widetext}}{}%
where $\Mat(m_{\phi K},\Omega|\Pars)$ is the matrix element given by Eq.~(\ref{eq:ZMatrix}). 
$\Phi(m_{\phi K})=p\,q$ is the phase space function, where $p$ is the momentum of
the $\phi K^+$ (\ie $\Kstar$) system in the $B^+$ rest frame, and $q$ is the $K^+$ momentum 
in the $K^{*+}$ rest frame.
The function $\epsilon(m_{\phi K},\Omega)$ is the signal efficiency, and
$I(\Pars)$ is the normalization integral,
\ifthenelse{\boolean{prl}}{\begin{widetext}}{}
\begin{equation}
I(\Pars)\equiv \int \PDF_{\rm sig}(m_{\phi K},\Omega)\,d m_{\phi K}\,d\Omega \propto
\frac{\sum_j w_j^{\rm MC}\big|\Mat(m_{Kp~j},\Omega_j|\Pars)\big|^2}{\sum_j w_j^{\rm MC}},
\label{eq:signor}
\end{equation}
\ifthenelse{\boolean{prl}}{\end{widetext}}{}%
where the sum is over simulated events, which are
generated uniformly in $B^+$ decay phase space 
and passed through the detector simulation \cite{LHCb-PROC-2011-006}
and data selection.
In the simulation, $pp$ collisions producing $B^+$ mesons are generated using
\pythia~\cite{Sjostrand:2006za}  with a specific \lhcb
configuration~\cite{LHCb-PROC-2010-056}. 
The weights $w_j^{\rm MC}$ introduced in Eq.~(\ref{eq:signor}) contain 
corrections to the $B^+$ production kinematics in the generation 
and to the detector response 
to bring the simulations into better agreement with the data.
Setting $w_j^{\rm MC}=1$ is one of the 
variations considered when evaluating systematic uncertainties.
The simulation sample contains 132\,000 events,
approximately 30 times the signal size in data.
This procedure folds the detector response into the model 
and allows a direct determination of the parameters of interest from the uncorrected data. 
The resulting log-likelihood 
sums over the data events (here for illustration, $\PDF=\PDF_{\rm sig}$),
\ifthenelse{\boolean{prl}}{\begin{widetext}}{}
{%
\def\1{\ifthenelse{\boolean{prl}}{}{\!\!}}
\begin{equation}
\begin{split}
\ln L(\Pars) & = \sum_i \ln \PDF_{\rm sig}(m_{Kp~i},\Omega_i|\Pars)  \\%\notag\\
 & = \sum_i \ln \big|\Mat(m_{Kp~i},\Omega_i|\Pars)\big|^2  - N\ln I(\Pars) + \sum_i \ln[ \Phi(m_{Kp~i})\epsilon(m_{Kp~i},\Omega_i) ],
\end{split}
\end{equation}
}%
\ifthenelse{\boolean{prl}}{\end{widetext}}{}%
where the last term does not depend on $\Pars$ and can be dropped ($N$ is the total number of the events in the fit).   

In addition to the signal PDF, $\PDF_{\rm sig}(m_{\phi K},\Omega|\Pars)$,
the background PDF, $\PDF_{\rm bkg}(m_{\phi K},\Omega)$ 
determined from the $B^+$ mass peak sidebands, is included.
We minimize the negative log-likelihood defined as
\ifthenelse{\boolean{prl}}{\begin{widetext}}{}
\begin{equation}
\begin{split}
& -\ln L(\Pars)=
-\sum_i\ln \left[ (1-\beta)\, \PDF_{\rm sig}(m_{\phi K~i},\Omega_i|\Pars)
    + \beta\, \PDF_{\rm bkg}(m_{\phi K~i},\Omega_i) \right] \\%\nonumber \\
&=-\sum_i\ln \left[ (1-\beta)\,\frac{\big|\Mat(m_{\phi K~i},\Omega_i|\Pars)\big|^2 \Phi(m_{\phi K~i})\epsilon(m_{\phi K~i},\Omega_i)}{I(\Pars)}
+\beta\, \frac{\PDF_{\rm bkg}^{u}(m_{\phi K~i},\Omega_i)}{I_{\rm bkg}} \right] \\%\nonumber \\
%&=-\sum_i\ln \left\{ \frac{(1-\beta)\,\Phi(m_{\phi K~i})\epsilon(m_{\phi K~i},\Omega_i)}{I(\Pars)}\,\left[\big|\Mat(m_{\phi K~i},\Omega_i|\Pars)\big|^2 
%\right.\right. 
%\nonumber\\
%& \quad\quad\quad\quad\quad\quad\quad\quad\quad\quad\quad\quad\quad\quad\quad\quad\quad\quad\quad\quad \left.\left.
%+ \frac{\beta\,I(\Pars)}{(1-\beta)\,I_{\rm bkg}}\, \frac{\PDF_{\rm bkg}^{u}(m_{\phi K~i},\Omega_i)}{\Phi(m_{\phi K~i})\epsilon(m_{\phi K~i},\Omega_i)} 
%\right] \right\} \\%\nonumber \\
&=-\sum_i\ln \left[\big|\Mat(m_{\phi K~i},\Omega_i|\Pars)\big|^2  
+ \frac{\beta\,I(\Pars)}{(1-\beta) I_{\rm bkg}}\,\frac{\PDF_{\rm bkg}^{u}(m_{\phi K~i},\Omega_i)}{\Phi(m_{\phi K~i})\epsilon(m_{\phi K~i},\Omega_i)}\right]
+N\ln I(\Pars)+{\rm const.},
\end{split}
\end{equation}
\ifthenelse{\boolean{prl}}{\end{widetext}}{}%
where $\beta$ is the background fraction in the peak region determined from the fit to
the $m_{\jpsi \phi K}$ distribution (Fig.~\ref{fig:mjpsiphik}), 
$\PDF_{\rm bkg}^{u}(m_{\phi K},\Omega)$ is the unnormalized background density proportional to the density of sideband events,
with its normalization determined by\footnote{Notice that the distribution of MC events 
includes both the $\Phi(m_{\phi K})$ and $\epsilon(m_{\phi K},\Omega)$ factors, which cancel their product in the numerator.}
\ifthenelse{\boolean{prl}}{\begin{widetext}}{}
\begin{equation}
I_{\rm bkg}\equiv \int \PDF_{\rm bkg}^{u}(m_{\phi K})\,d m_{\phi K}\,d\Omega \, \propto \,
\frac{
\sum_j w_j^{\rm MC} \frac{\PDF_{\rm bkg}^{u}(m_{\phi K~j},\Omega_j)}{\Phi(m_{\phi K~i})\epsilon(m_{\phi K~j},\Omega_j)}}{
\sum_j w_j^{\rm MC}}.
\label{eq:bkgnor}
\end{equation}
\ifthenelse{\boolean{prl}}{\end{widetext}}{}%
The equation above implies that the background term is efficiency corrected,
so it can be added
to the efficiency-independent signal probability expressed by $\left|\Mat\right|^2$.
This way the efficiency parameterization, $\epsilon(m_{\phi K},\Omega)$, 
becomes a part of the background description which affects only a small part 
of the total PDF.

\begin{figure}[tbhp]
  \begin{center}
    \includegraphics*[width=\figsize]{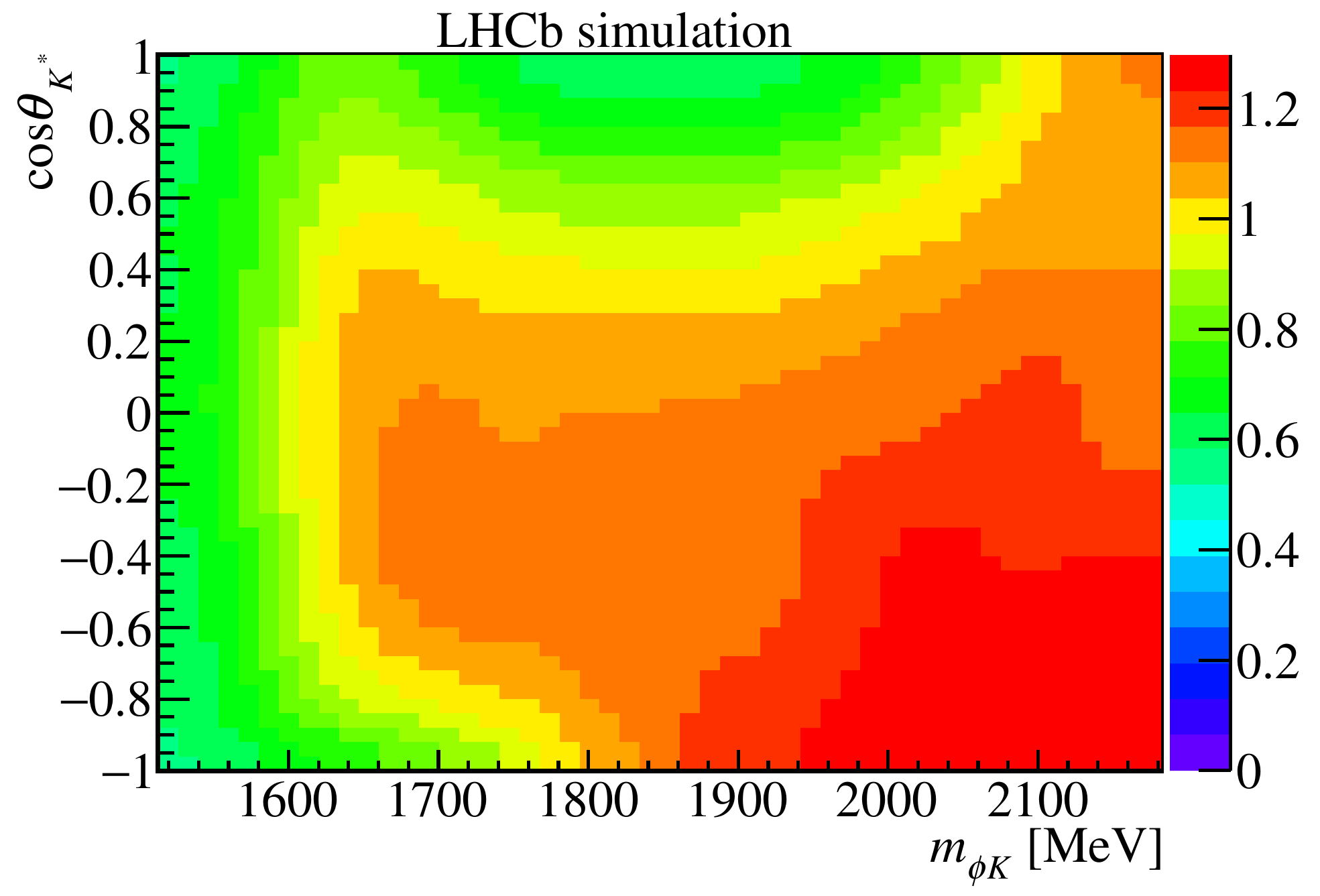} \\
    \includegraphics*[width=\figsize]{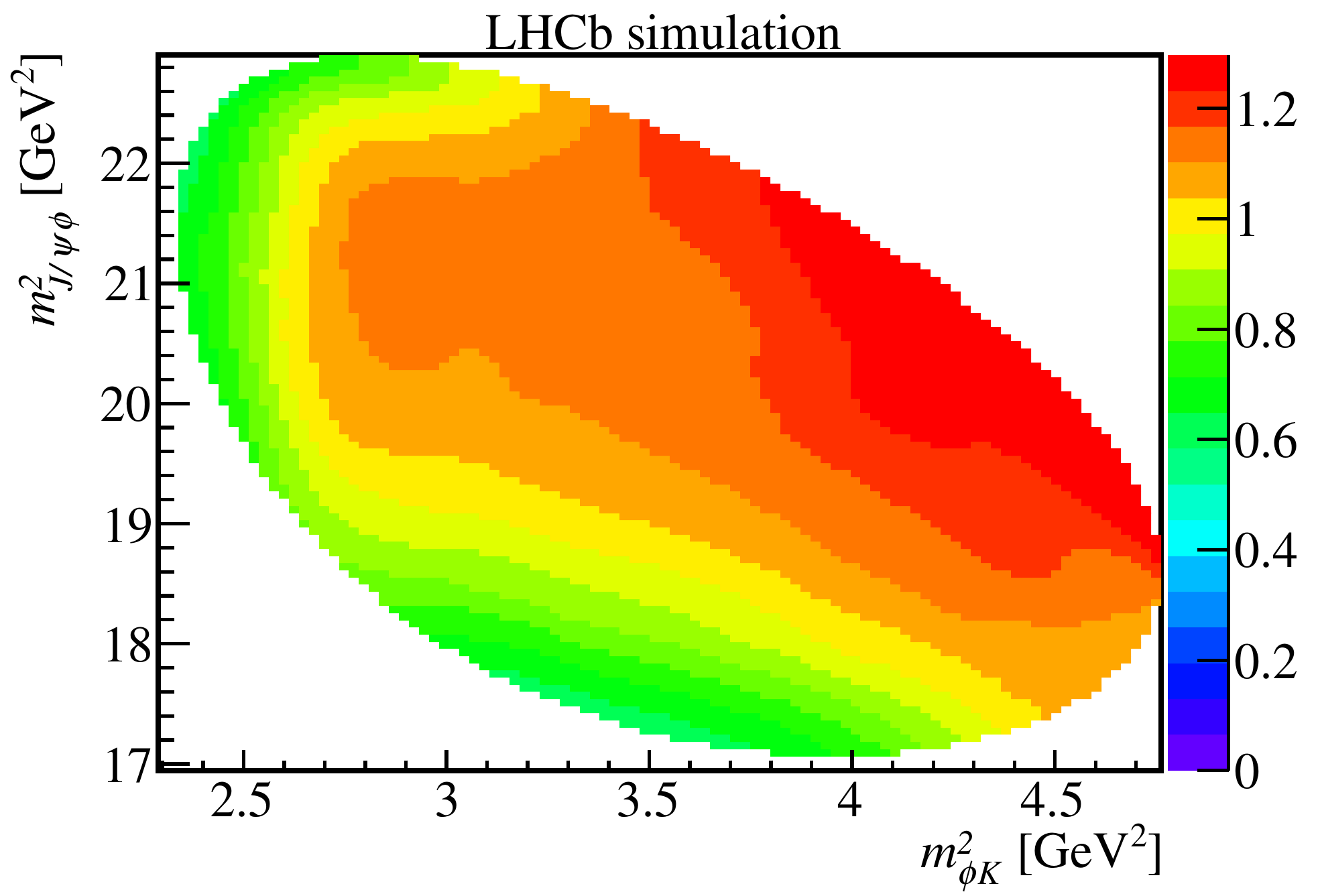}
  \end{center}
  \vskip-0.3cm\caption{\small
     Parameterized efficiency $\epsilon_1(m_{\phi K},\cosks)$ function (top) 
     and its representation in the Dalitz plane $({\mphik^2},{\mjpsiphi^2})$ (bottom).
     Function values corresponding to the color encoding are given on the right.
     The normalization arbitrarily corresponds to unity when averaged over the phase space.     
  \label{fig:effm}
  }
\end{figure}

\begin{figure}[bthp]
  \begin{center}
    \includegraphics*[width=\figsize]{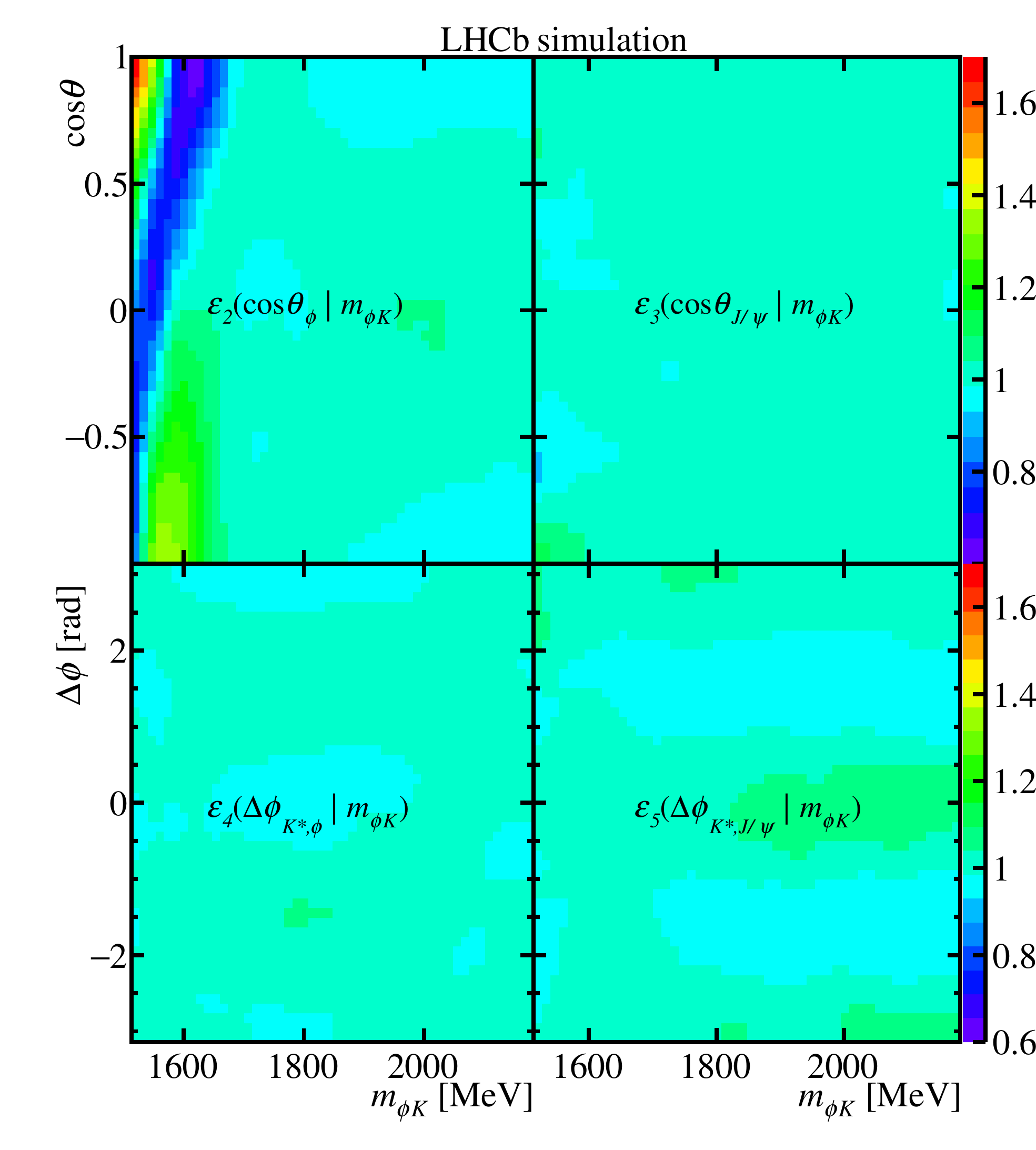} 
  \end{center}
  \vskip-0.3cm\caption{\small
     Parameterized efficiency 
     $\epsilon_2(\cosphi|m_{\phi K})$, $\epsilon_3(\cospsi|m_{\phi K})$,  
     $\epsilon_4(\phiksphi|m_{\phi K})$, $\epsilon_5(\phikspsi|m_{\phi K})$
     functions.
     Function values corresponding to the color encoding are given on the right.
     By construction each function integrates to unity at each $m_{\phi K}$ value.  
     The structure in $\epsilon_2(\cosphi|m_{\phi K})$ present 
     between 1500 and $1600 \mev$ is an artifact of 
     removing $\btojpsikkk$ events in which both $K^+K^-$ combinations 
     pass the $\phi$ mass selection window. 
  \label{fig:effepsilon2345}
  }
\end{figure}

The efficiency parameterization
in the background term is assumed to factorize as
\ifthenelse{\boolean{prl}}{\begin{widetext}}{}
\begin{equation}
\begin{split}
\lefteqn{\epsilon(m_{\phi K},\Omega)=\epsilon_1(m_{\phi K},\cosks)\,\,\epsilon_2(\cosphi|m_{\phi K})\times} \\%\notag\\
&\quad\quad\quad\quad \epsilon_3(\cospsi|m_{\phi K})\,\, \epsilon_4(\phiksphi|m_{\phi K})\,\,\epsilon_5(\phikspsi|m_{\phi K}).
\end{split}
\label{eq:peff}
\end{equation}
\ifthenelse{\boolean{prl}}{\end{widetext}}{}%
The $\epsilon_1(m_{\phi K},\cosks)$ term is obtained by binning 
a two-dimensional (2D) histogram of the simulated signal events. 
Each event is given a $1/(p\,q)$ weight, 
since at the generator level the phase space is flat in $\cosks$ 
but has a $p\,q$ dependence on $m_{\phi K}$.
A bi-cubic function is used to interpolate between bin centers.
The $\epsilon_1(m_{\phi K},\cosks)$ efficiency and
its visualization across the normal Dalitz plane are shown in Fig.~\ref{fig:effm}.
The other terms are again built from 2D histograms, 
but with each bin divided by the number of simulated events in the corresponding 
$m_{\phi K}$ slice to remove the dependence on this mass (Fig.~\ref{fig:effepsilon2345}).

\begin{figure}[bthp]
  \begin{center}
    \includegraphics*[width=\figsize]{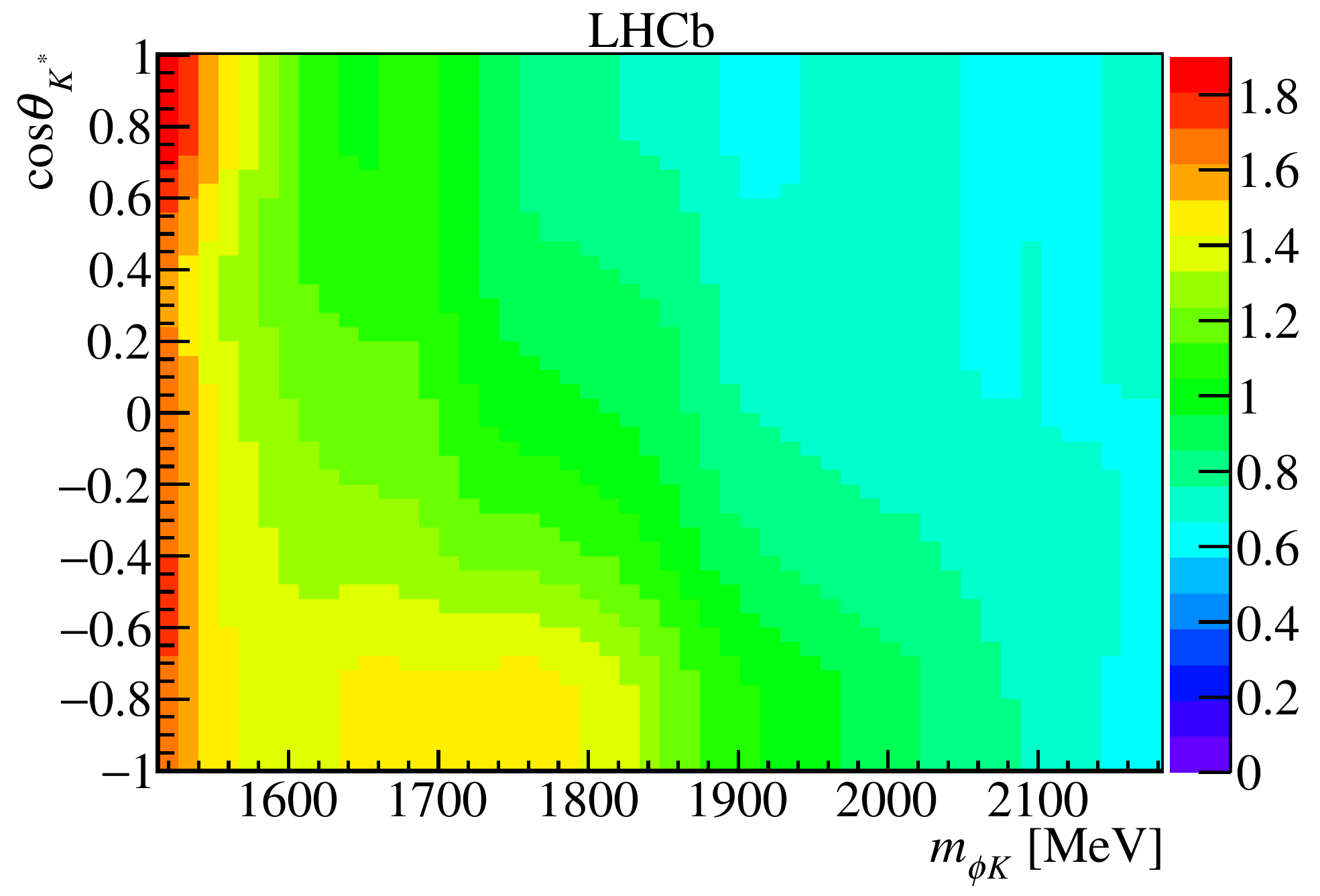} \\
    \includegraphics*[width=\figsize]{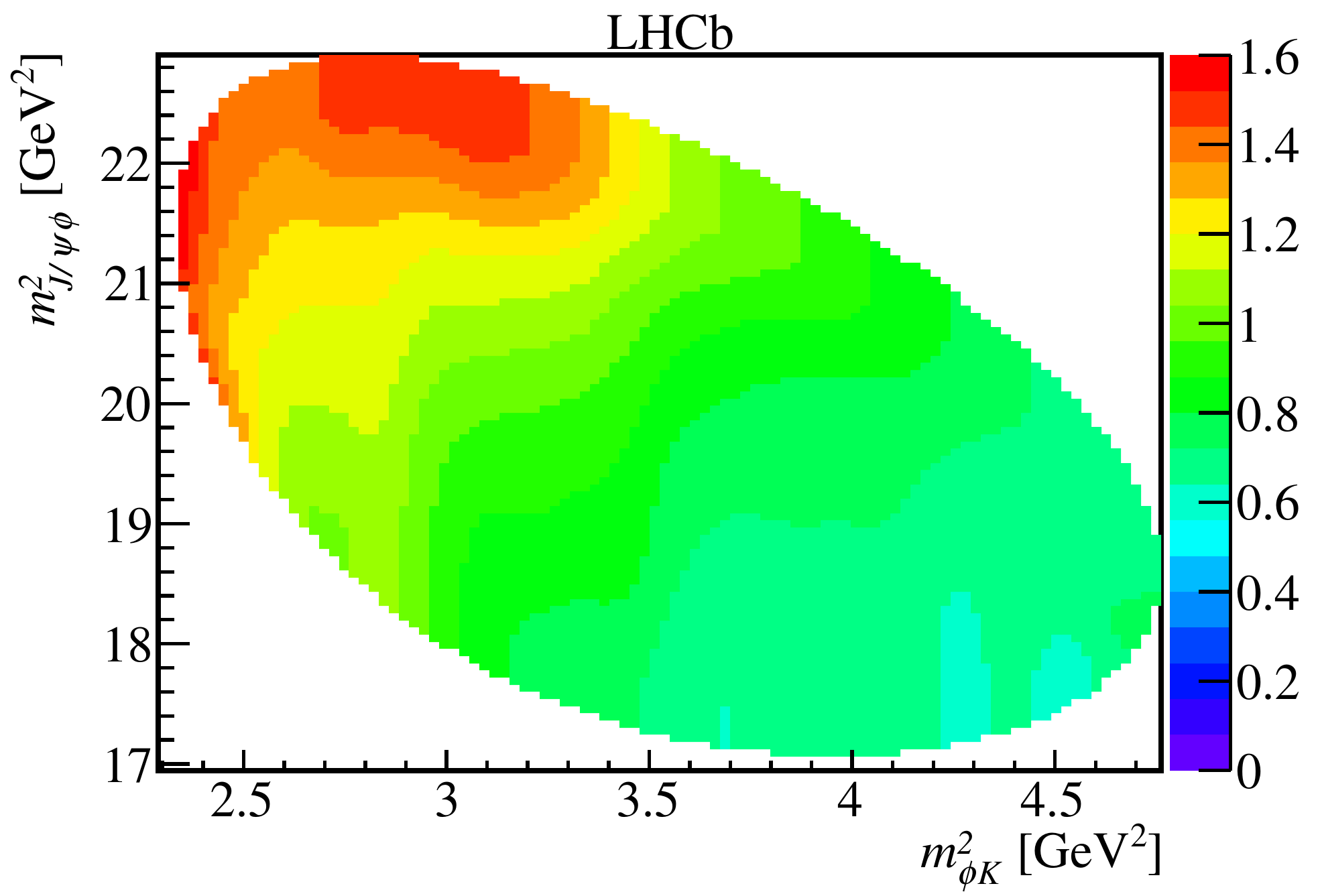}
  \end{center}
  \vskip-0.3cm\caption{\small
     Parameterized background ${P_{\rm bkg}}_1(m_{\phi K},\cosks)$ function (top) 
     and its representation in the Dalitz plane $({\mphik^2},{\mjpsiphi^2})$ (bottom).
     Function values corresponding to the color encoding are given on the right.
     The normalization arbitrarily corresponds to unity when averaged over the phase space.
  \label{fig:bkgdaleffs}
  }
\end{figure}

\begin{figure}[tbhp]
  \begin{center}
    \includegraphics*[width=\figsize]{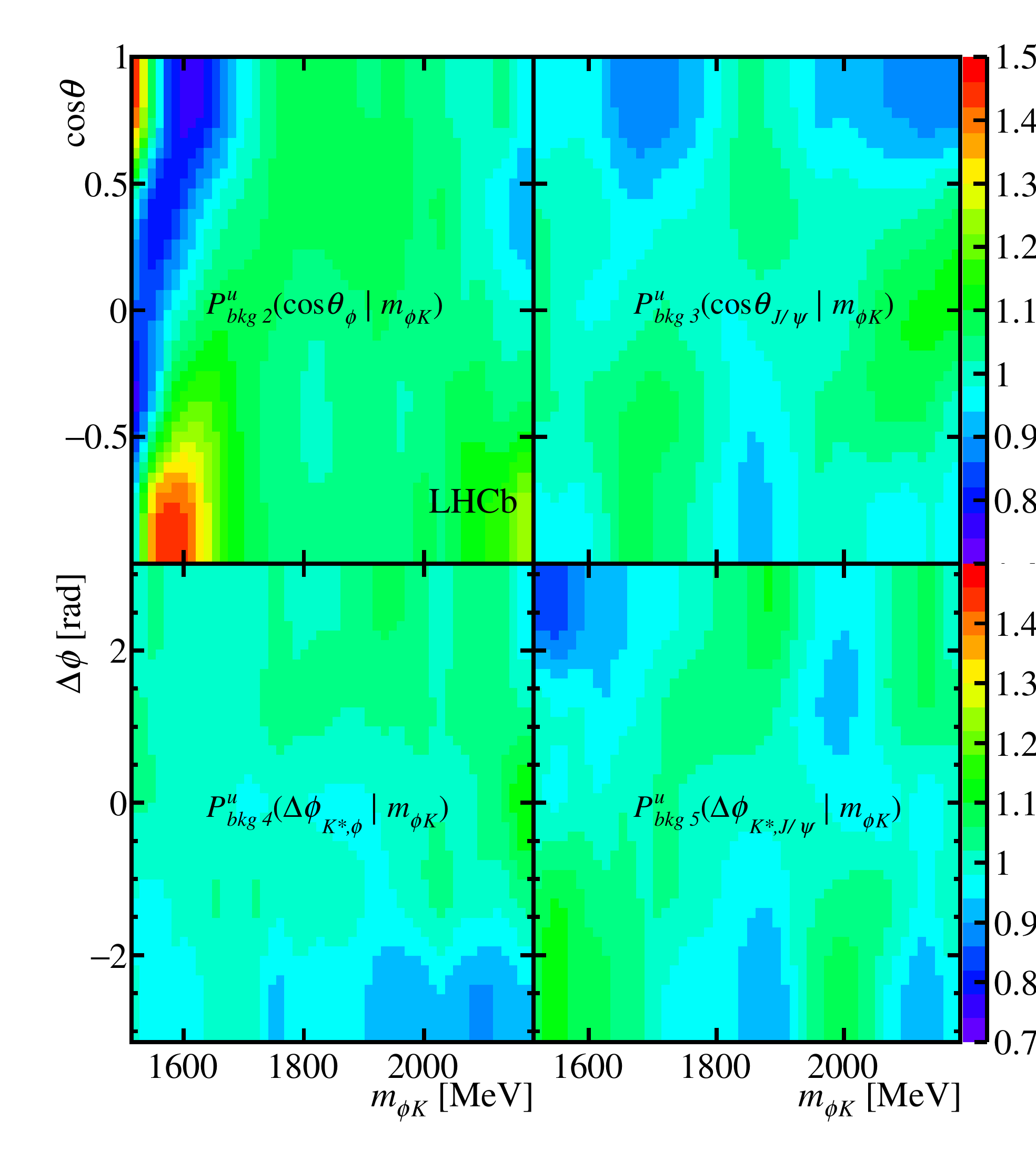} 
  \end{center}
  \vskip-0.3cm\caption{\small
     Parameterized background functions: $P^u_{\rm bkg\,2}(\cosphi|m_{\phiz\kaon})$,
     $P^u_{\rm bkg\,3}(\cospsi|m_{\phiz\kaon})$, $P^u_{\rm bkg\,4}(\phiksphi|m_{\phiz\kaon})$,
     $P^u_{\rm bkg\,5}(\phikspsi|m_{\phiz\kaon})$.
     Function values corresponding to the color encoding are given on the right.
     By construction each function integrates to unity at each $m_{\phi K}$ value.  
  \label{fig:bkgeffs2345}
  }
\end{figure}

The background PDF, $\PDF_{\rm bkg}^{u}(m_{\phi K},\Omega)/\Phi(m_{\phi K})$, 
is built using the same approach,
\ifthenelse{\boolean{prl}}{\begin{widetext}}{}
\begin{equation}
\begin{split}
\lefteqn{
\frac{{\PDF_{\rm bkg}^{u}}(m_{\phi K},\Omega)}{\Phi(m_{\phi K})}=
{P_{\rm bkg}}_1(m_{\phi K},\cosks)\,\,{P_{\rm bkg}}_2(\cosphi|m_{\phi K})\times}
\\%\notag\\
&\quad\quad\quad{P_{\rm bkg}}_3(\cospsi|m_{\phi K})\,\, {P_{\rm bkg}}_4(\phiksphi|m_{\phi K})\,\,{P_{\rm bkg}}_5(\phikspsi|m_{\phi K}).
\end{split}
\label{eq:pbkg}
\end{equation}
\ifthenelse{\boolean{prl}}{\end{widetext}}{}%
The background function ${P_{\rm bkg}}_1(m_{\phi K},\cos\theta_\Kstar)$ is 
shown in Fig.~\ref{fig:bkgdaleffs} and the other terms are shown in 
Fig.~\ref{fig:bkgeffs2345}.

The fit fraction (\FiFr) of any component $R$ is defined as,
\begin{equation}
\FiFr = \frac{
\int \left|\Mat^R(m_{\phi K},\Omega)\right|^2 \Phi(m_{\phi K})\,dm_{\phi K}d\Omega
}{
\int \left|\Mat(m_{\phi K},\Omega)\right|^2 \Phi(m_{\phi K})\,dm_{\phi K}d\Omega
},
\label{eq:ff}
\end{equation}
where in $\Mat^R$ 
all terms except those associated with the $R$ amplitude are set to zero. 

\ifthenelse{\boolean{prl}}{}{\FloatBarrier}%

\section{Background-subtracted and efficiency-corrected distributions}

\begin{figure}[tbhp]
  \begin{center}
    \includegraphics*[width=\figsize]{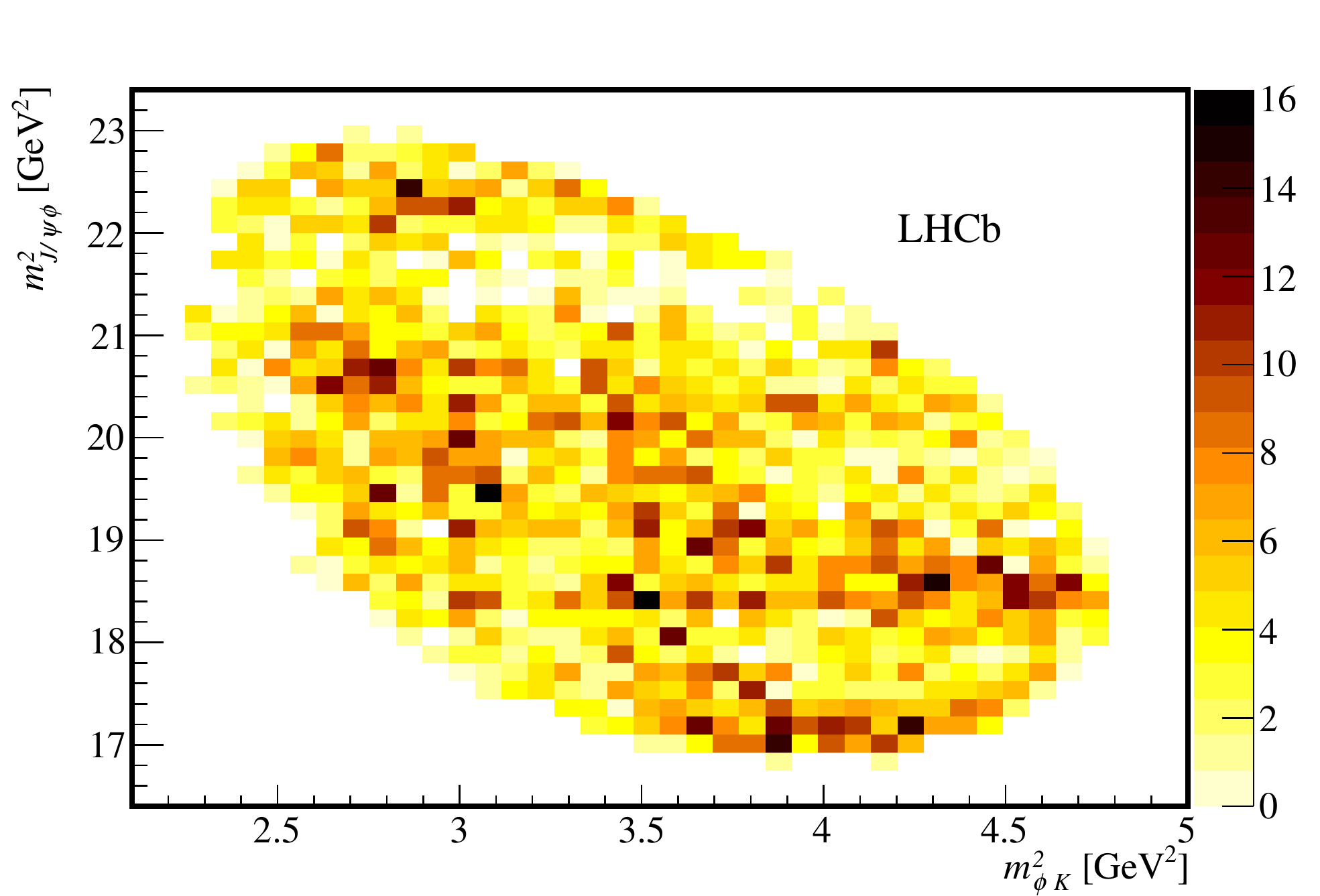}
  \end{center}
  \vskip-0.5cm\caption{\small
   Background-subtracted and efficiency-corrected data yield in 
   the Dalitz plane of $({\mphik^2},{\mjpsiphi^2})$.    
   Yield values corresponding to the color encoding are given on the right.
  \label{fig:phik_vs_psiphi}
  }
\end{figure}

\begin{figure}[tbhp]
  \begin{center}
    \includegraphics*[width=\figsize]{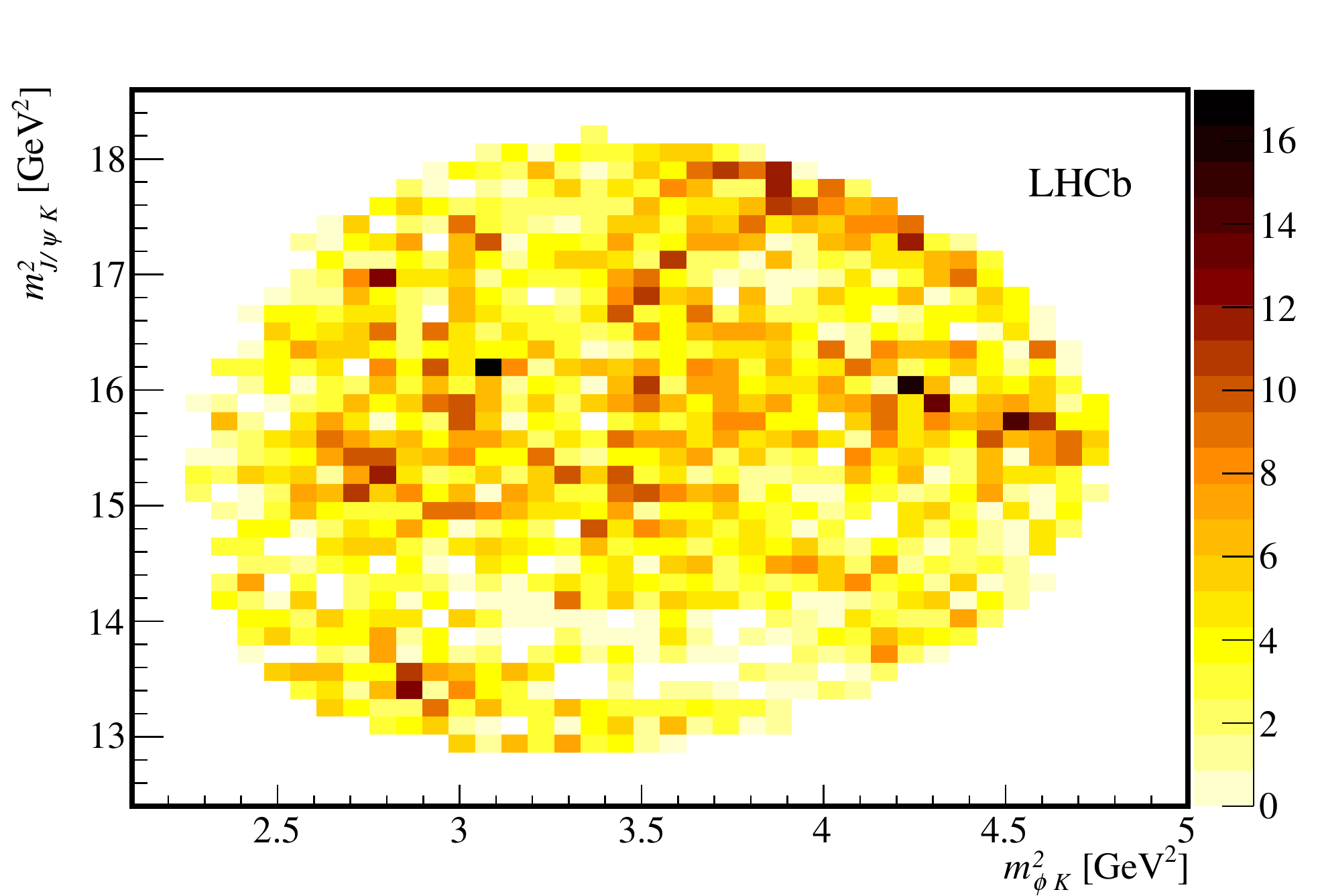}
  \end{center}
  \vskip-0.5cm\caption{\small
   Background-subtracted and efficiency-corrected data yield in 
   the Dalitz plane of $({\mphik^2},{\mjpsik^2})$. 
   Yield values corresponding to the color encoding are given on the right.
  \label{fig:phik_vs_psik}
  }
\end{figure}

\begin{figure}[tbhp]
  \begin{center}
    \includegraphics*[width=\figsize]{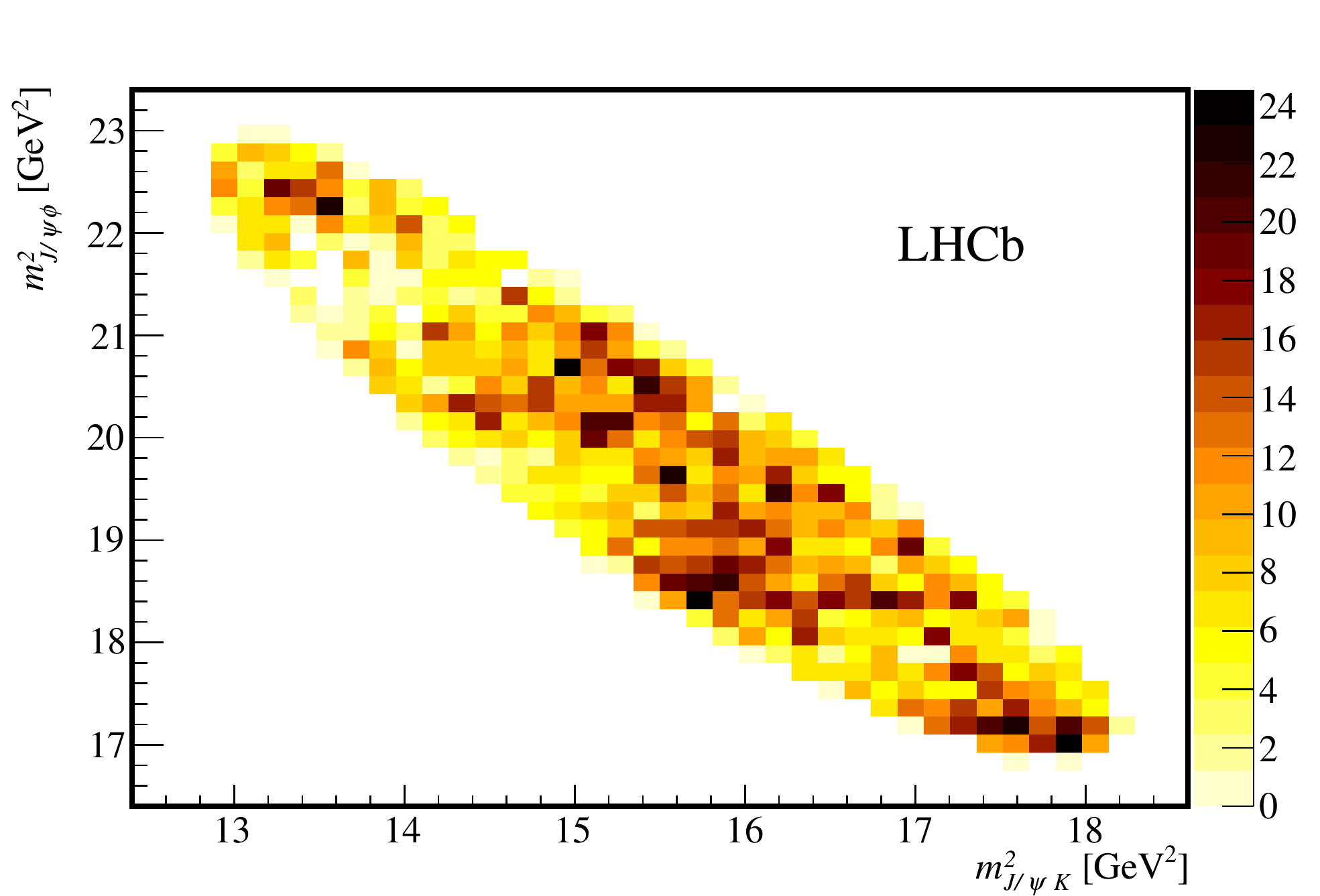}
  \end{center}
  \vskip-0.5cm\caption{\small
   Background-subtracted and efficiency-corrected data yield in 
   the Dalitz plane of $({\mjpsik^2},{\mjpsiphi^2})$. 
   Yield values corresponding to the color encoding are given on the right.
  \label{fig:psik_vs_psiphi}
  }
\end{figure}

\begin{figure}[tbhp]
  \begin{center}
    \includegraphics*[width=\figsize]{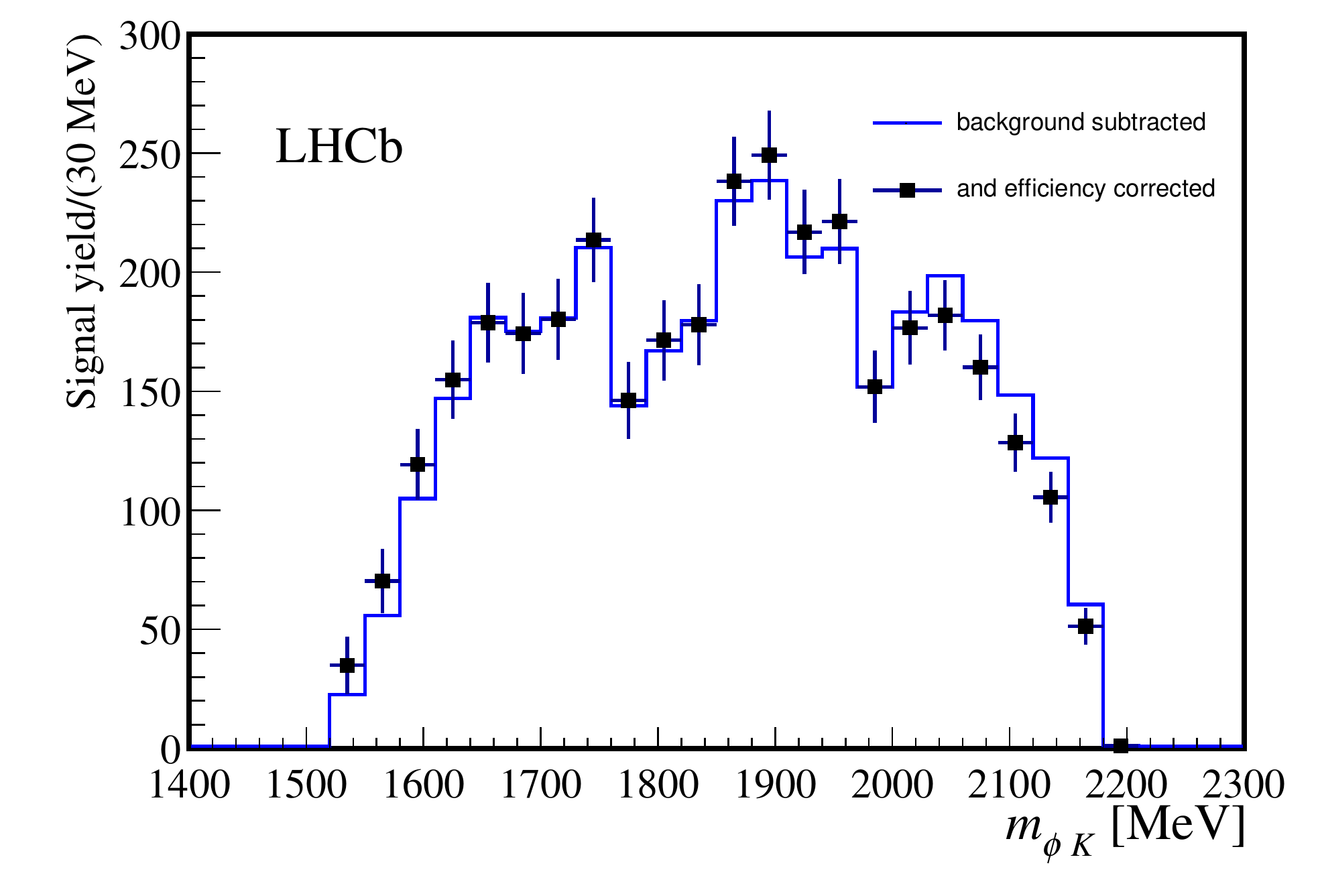} 
  \end{center}
  \vskip-1.0cm\caption{\small
   Background-subtracted (histogram) and efficiency-corrected (points) distribution of $m_{\phi K}$.
   See the text for the explanation of the efficiency normalization.
  \label{fig:phik}
  }
\end{figure}

\begin{figure}[bthp]
  \begin{center}
    \includegraphics*[width=\figsize]{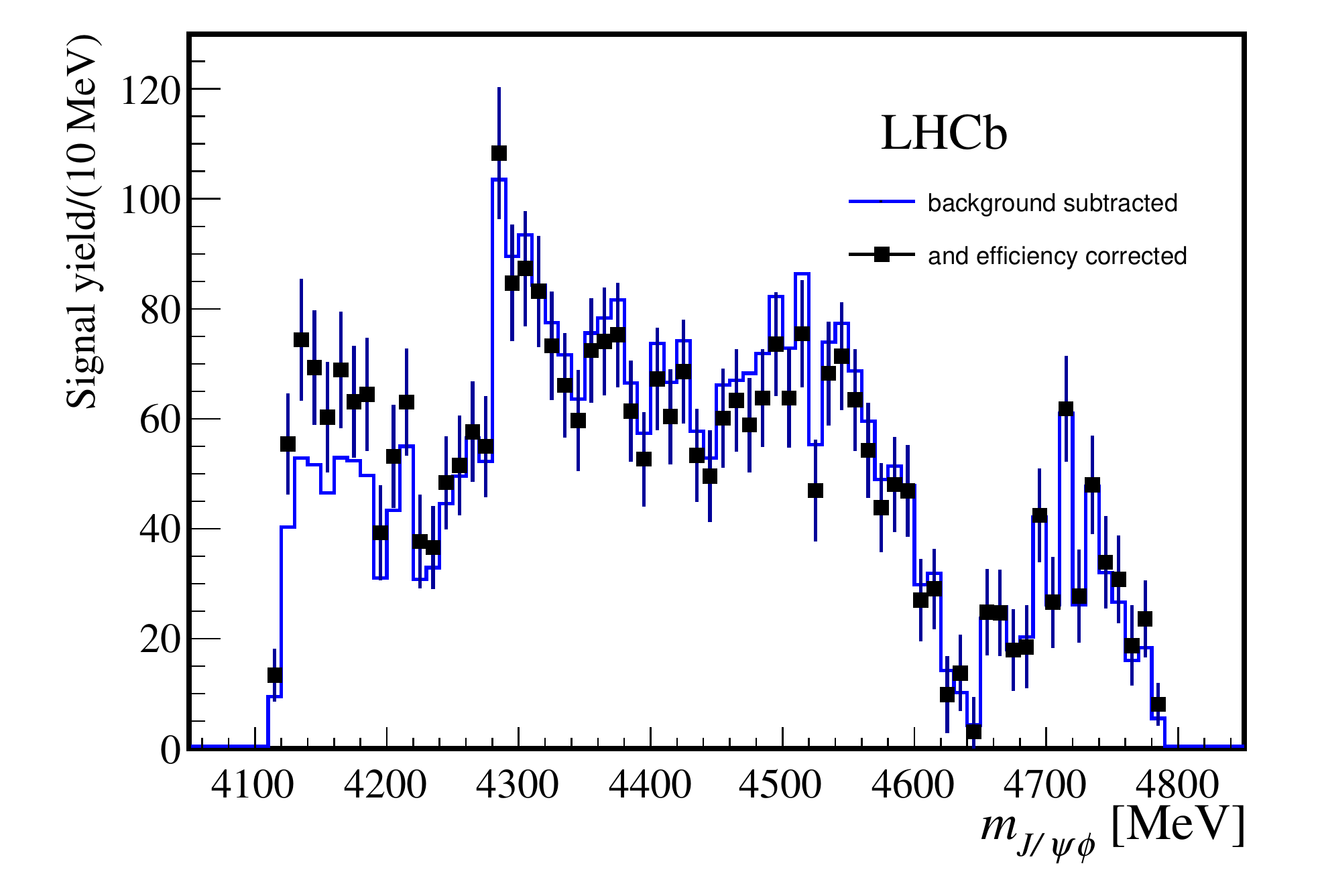} 
  \end{center}
  \vskip-1.0cm\caption{\small
   Background-subtracted (histogram) and efficiency-corrected (points) distribution of $m_{\jpsi\phi}$.
   See the text for the explanation of the efficiency normalization.
  \label{fig:psiphi}
  }
\end{figure}

\begin{figure}[tbhp]
  \begin{center}
    \includegraphics*[width=\figsize]{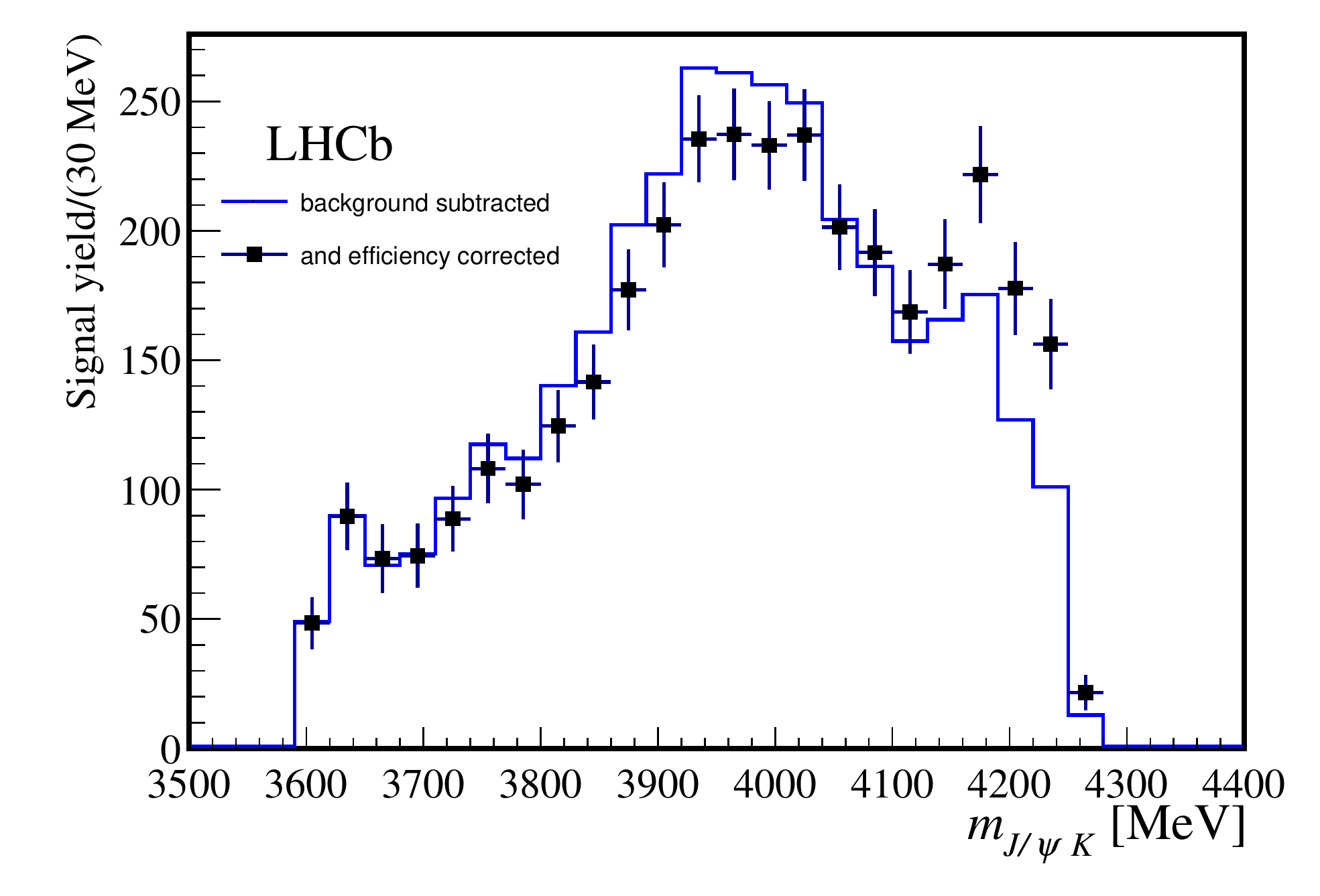} 
  \end{center}
  \vskip-1.0cm\caption{\small
   Background-subtracted (histogram) and efficiency-corrected (points) distribution of $m_{\jpsi K}$.
   See the text for the explanation of the efficiency normalization.
  \label{fig:psik}
  }
\end{figure}

The background-subtracted and efficiency-corrected Dalitz plots
are shown in Figs.~\ref{fig:phik_vs_psiphi}--\ref{fig:psik_vs_psiphi}
and the mass projections are shown in Figs.~\ref{fig:phik}--\ref{fig:psik}.
The latter indicates that the efficiency corrections are rather minor.
The background is eliminated by subtracting the scaled $B^+$ sideband distributions.
The efficiency corrections are achieved by weighting events according to the inverse of 
the parameterized 6D efficiency given by Eq.~(\ref{eq:peff}).
The efficiency-corrected signal yield remains similar to the signal candidate count, 
because we normalize the efficiency to unity when averaged over the phase space. 

While the $m_{\phi K}$ distribution (Fig.~\ref{fig:phik}) 
does not contain any obvious resonance peaks, it would 
be premature to conclude that there are none since all $K^{*+}$ resonances expected in this
mass range belong to higher excitations, and therefore should be broad. In fact the narrowest
known $K^{*+}$ resonance in this mass range has a width of approximately $150 \mev$ \cite{PDG2014}.
Scattering experiments sensitive to $K^{*}\to\phi K$ decays also showed a smooth 
mass distribution, which revealed some resonant activity only after 
partial-wave analysis \cite{Armstrong:1982tw,Frame:1985ka,Kwon:1993xb}.
Therefore, studies of angular distributions in correlation with $m_{\phi K}$ are necessary.
Using full 6D correlations results in the best sensitivity.

The $m_{\jpsi\phi}$ distribution (Fig.~\ref{fig:psiphi}) contains several peaking structures, 
which could be exotic or could be reflections of conventional $K^{*+}$ resonances. 
There is no narrow $X(4140)$ peak just above the kinematic threshold,
consistent with 
the LHCb analysis presented in Ref.~\cite{LHCb-PAPER-2011-033}, %based on 0.37 fb$^{-1}$, 
however we observe a broad enhancement.
A peaking structure is observed at about $4300 \mev$. 
The high mass region is inspected with good sensitivity for the first time,
with the rate having a minimum near $4640 \mev$ with two broad peaks on each side.

The $m_{\jpsi K}$ distribution (Fig.~\ref{fig:psik}) peaks broadly in the middle and
has a high-mass peak, which is strongly correlated with the low-mass $m_{\jpsi\phi}$ 
enhancement (Fig.~\ref{fig:psik_vs_psiphi}). 

As explained in the previous section, the amplitude fits 
are performed by maximizing the unbinned likelihood 
on the selected signal candidates including background
events and without the efficiency weights. 
To properly represent the fit quality, the fit projections in the
following sections show the fitted data sample, 
\ie including the background 
and without the parameterized 
efficiency corrections applied to the signal events.

\ifthenelse{\boolean{prl}}{}{\FloatBarrier}%

\section{Amplitude model with only $\phi K^+$ contributions}

\begin{figure}[tbhp]
  \begin{center}
\ifthenelse{\boolean{prl}}{
    \includegraphics*[width=\figsize]{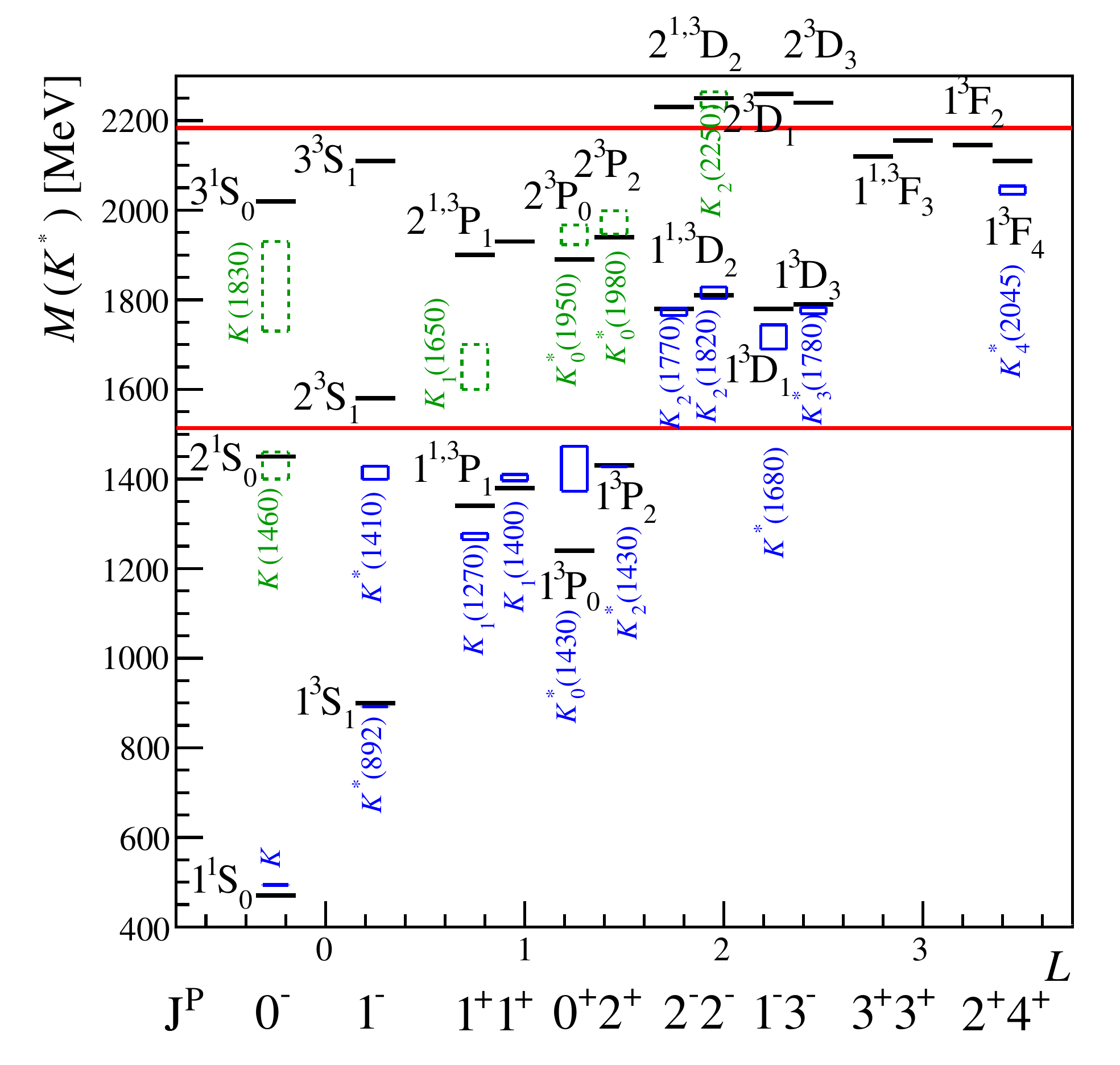} 
}{
    \includegraphics*[width=1.1\figsize]{kstars-nores.pdf} 
}
  \end{center}
  \vskip-0.3cm\caption{\small
   Kaon excitations predicted by Godfrey--Isgur \cite{Godfrey:1985xj} (horizontal black lines)
   labeled with their intrinsic quantum numbers: $n{}^{2S+1}L_J$ (see the text).
   Well established states are shown 
   with narrower solid blue boxes extending
   to $\pm1\sigma$ in mass and labeled with their PDG names \cite{PDG2014}.
   Unconfirmed states are shown with dashed green boxes.
   The long horizontal red lines indicate the $\phi K$ mass range probed in $B^+\to\jpsi\phi K^+$ decays. 
  \label{fig:kaons}
  }
\end{figure}

We first try to describe the data with kaon excitations alone.
Their mass spectrum as predicted in the relativistic potential model 
by Godfrey--Isgur \cite{Godfrey:1985xj} is shown in Fig.~\ref{fig:kaons}
together with the experimentally determined masses 
of both well-established and unconfirmed $K^*$ resonances \cite{PDG2014}.
Past experiments on $K^{*}$ 
states decaying to $\phi K$ \cite{Armstrong:1982tw,Frame:1985ka,Kwon:1993xb} 
had limited precision, especially at high masses, gave somewhat inconsistent results,
and provided evidence for only a few of the states 
expected from the quark model in the $1513$--$2182 \mev$ range probed in our data set.
However, except for the $J^P=0^+$ states which cannot decay to $\phi K$ because of
angular momentum and parity conservation, all other kaon excitations 
above the $\phi K$ threshold are predicted to decay to this final state \cite{Kokoski:1985is}.
In $B^+$ decays,
production of high spin states, like the $K^*_3(1780)$ or $K^*_4(2045)$ resonances, is expected to 
be suppressed by the high orbital angular momentum required to produce them. 

We have used the predictions of the Godfrey--Isgur model as a guide to the quantum numbers of the
$K^{*+}$ states to be included in the model.  
The masses and widths of all states are left free; thus our fits do not depend on detailed
predictions of Ref.~\cite{Godfrey:1985xj}, 
nor on previous measurements.
We also allow a constant nonresonant amplitude with $J^P=1^+$, since such $\phi K^+$ 
contributions can be produced, and can decay, in S-wave.
Allowing the magnitude of the nonresonant amplitude to vary 
with $m_{\phi K}$ does not improve fit qualities.  

\begin{figure}[tbhp]
  \begin{center}
    \includegraphics*[width=\figsize]{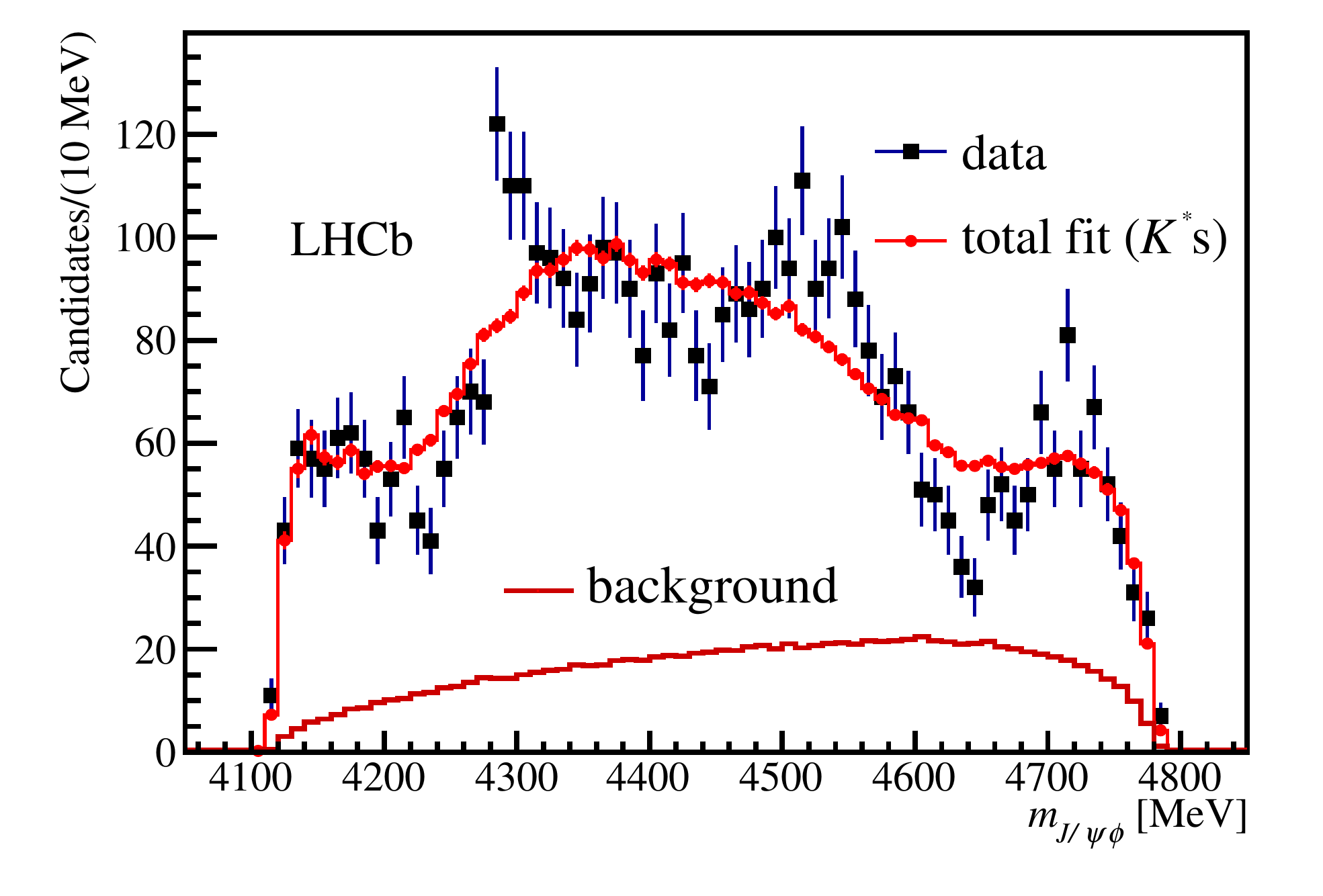} 
  \end{center}
  \vskip-0.3cm\caption{\small
   Distribution of $m_{\jpsi\phi}$ for the data and the fit results  
   with a model containing only $K^{*+}\to\phi K^+$ contributions.
  \label{fig:mjpsiphikstaronly}
  }
\end{figure}

While it is possible to describe the $m_{\phi K}$ and $m_{\jpsi K}$ distributions 
well with $\Kstar$ contributions alone, the fit projections onto $m_{\jpsi\phi}$ 
do not provide an acceptable description of the data.
For illustration we show in Fig.~\ref{fig:mjpsiphikstaronly}
the projection of a fit with the following composition:
a nonresonant term plus candidates for two $\nlj{2}{P}{1}$, 
two $\nlj{1}{D}{2}$, and one of each of $\nslj{1}{3}{F}{3}$, $\nslj{1}{3}{D}{1}$, 
$\nslj{3}{3}{S}{1}$, $\nslj{3}{1}{S}{0}$, 
$\nslj{2}{3}{P}{2}$, $\nslj{1}{3}{F}{2}$, $\nslj{1}{3}{D}{3}$ 
and $\nslj{1}{3}{F}{4}$ states, 
labeled here with their intrinsic quantum numbers: 
$n{}^{2S+1}L_J$ 
($n$ is the radial quantum number,
 $S$ the total spin of the valence quarks, 
 $L$ the orbital angular momentum between quarks, 
 and $J$ the total angular momentum of the bound state). 
The fit contains 104 free parameters. 
The $\chi^2$ value (144.9/68 bins) 
between the fit projection and the observed $m_{\jpsi\phi}$ distribution 
corresponds to a p value below $10^{-7}$. 
Adding more resonances does not change the conclusion that non-$\Kstar$ 
contributions are needed to describe the data.

%\FloatBarrier

\ifthenelse{\boolean{prl}}{}{\FloatBarrier}%

\section{Amplitude model with $\phi K^+$ and $\jpsi\phi$ contributions}

We have explored adding $X$ and $Z^+$ contributions of various quantum numbers 
to the fit models.
Only $X$ contributions lead to significant improvements in the description of the data.
The default resonance model 
is described in detail below and is summarized in Table~\ref{tab:baselineinc},
where the results are also compared with the previous measurements and the theoretical predictions
for $\bar{s} u$ states \cite{Godfrey:1985xj}. 
The model contains seven $K^{*+}$ states, four $X$ states and 
$\phi K^+$ and $\jpsi\phi$ nonresonant components. 
There are 98 free parameters in this fit.
Projections of the fit onto the mass variables are
displayed in Fig.~\ref{fig:defmasses}.
The $\chi^2$ value (71.5/68 bins) 
between the fit projection and the observed $m_{\jpsi\phi}$ distribution 
corresponds to a p value of 22\%. 
Projections onto angular variables are shown 
in Figs.~\ref{fig:defksangles}--\ref{fig:defzangles}.
Projections onto masses in different regions 
of the Dalitz plot
can be found in Fig.~\ref{fig:msliced}.
Using adaptive binning\footnote{The adaptive binning procedure maintains 
uniform and adequate bin contents.} 
on the Dalitz plane $m_{\phi K}^2$ vs.\ $m_{\jpsi\phi}^2$
(or extending the binning to all six fitted dimensions)
the $\chi^2$ value of 438.7/496 bins (462.9/501 bins)
gives a p value of 17\% (2.3\%). 
The $\chi^2$ PDFs used to obtain the p values 
have been obtained with simulations of pseudoexperiments generated from the
default amplitude model. % by finding effective $\ndf$ values. 

The systematic
uncertainties 
are obtained from the sum in quadrature of the changes observed
in the fit results
when: the $K^{*+}$ and $X(4140)$ models are varied;
the Breit--Wigner amplitude parameterization is modified;
only the left or right $B^+$ mass peak sidebands are used for the background
parameterization;
the $\phi$ mass selection window is made narrower 
by a factor of two (to reduce the non-$\phi$ background fraction);
the signal and background shapes are varied in the fit to $m_{\jpsi\phi K}$ which
determines the background fraction $\beta$;
the weights assigned to simulated events, in order to improve
agreement with the data on $B^+$ production characteristics     
and detector efficiency, are removed.
More detailed discussion of the systematic uncertainties can be found 
in Appendix~\ref{sec:sys}. 
   
The significance of each (non)resonant contribution is calculated assuming 
that $\dll$, after the contribution is included in the fit,
follows a $\chi^2$ distribution  
with the number of degrees of freedom ($\ndf$) equal to the 
number of free parameters in its parameterization.
The value of $\ndf$ is doubled when $M_0$ and  $\Gamma_0$ are free parameters in the fit.
The validity of this assumption has been verified using simulated pseudoexperiments.
The significances of the $X$ contributions are given after accounting for systematic variations. 
Combined significances of exotic contributions, determined by removing more than one exotic contribution at a time, 
are much larger than their individual significances given in Table~\ref{tab:baselineinc}.
The significance of the spin-parity determination for each $X$ state is determined as 
described in Appendix~\ref{sec:spinana}.

The longitudinal ($f_L$) and transverse ($f_\perp$) polarizations are
calculated for $K^{*+}$ contributions according to
\def\bafl#1{A_{{#1}}^{B\to\jpsi\Kstar}}
\def\afl#1{A_{\lambda={#1}}^{B\to\jpsi\Kstar}}
\def\aafl#1{\left|A_{\lambda={#1}}^{B\to\jpsi\Kstar}\right|^2}
\def\baafl#1{\left|A_{{#1}}^{B\to\jpsi\Kstar}\right|^2}
\begin{equation}
%\begin{split}
f_L=%&
\frac{ \aafl{0} }{ \aafl{-1} + \aafl{0} + \aafl{+1} }, \\%\notag\\
%f_\perp=&\frac{ \baafl{\perp} }{ \aafl{-1} + \aafl{0} + \aafl{+1} },
%\end{split}
\end{equation}
\begin{equation}
%\begin{split}
%f_L=&\frac{ \aafl{0} }{ \aafl{-1} + \aafl{0} + \aafl{+1} }, \\%\notag\\
f_\perp=%&
\frac{ \baafl{\perp} }{ \aafl{-1} + \aafl{0} + \aafl{+1} },
%\end{split}
\end{equation}
where
\begin{equation}
\bafl{\perp} = \frac{\afl{+1} - \afl{-1}}{\sqrt{2}}. 
\end{equation}
  
\begin{table*}[tbhp]
\def\9{\phantom{2}}
\caption{\small 
         Results for significances, masses, widths and fit fractions of
         the components included in the default amplitude model.
         The first (second) errors are statistical (systematic).
         Errors on $f_L$ and $f_\perp$ are statistical only.
         Possible interpretations in terms of kaon excitation levels are given, 
         with notation $n{}^{2S+1}L_J$, together with 
         the masses predicted in the Godfrey-Isgur model \cite{Godfrey:1985xj}.
         Comparisons with the previously experimentally 
         observed kaon excitations \cite{PDG2014} and $X\to\jpsi\phi$ structures 
         are also given.
}
\label{tab:baselineinc}
\hbox{
\hbox{\ifthenelse{\boolean{prl}}{}{\quad\hskip-3.2cm}
\hbox{
\ifthenelse{\boolean{prl}}{
\renewcommand{\arraystretch}{1.2}
}{
\renewcommand{\arraystretch}{1.05}
}
\def\1#1{\multicolumn{1}{l}{#1}}
\def\2#1{\multicolumn{1}{r}{#1}}
\def\6#1{\multicolumn{1}{c}{#1}}
\def\3#1{{\small #1}}
\def\4{\phantom{\pm888}}
\def\5{\phantom{8}}
\ifthenelse{\boolean{prl}}{
\def\7#1{\quad #1 \quad}
}{
\def\7#1{#1}
}
\def\GI#1{\1{#1}}   
%\begin{center}
%\resizebox{\textwidth}{!}{
\begin{tabular}{ccllrcc}
\hline
Contri- &  sign.   & \multicolumn{5}{c}{Fit results}         \\ 
bution  & \3{or Ref.}  &  \6{$M_0$ [\mev]} & \6{$\Gamma_0$ [\mev]} & \6{\FiFr~\%} &  $f_L$ & $f_\perp$ \\
\hline\hline
\1{All $K(1^+)$}   &  $8.0\sigma$ &    &    & $42\!\pm\!\5 8\,^{+\5 5}_{-\5 9}$  &      &   \\      
\2{$\NRKs$}        &              &    &    & $16\!\pm\!13\,^{+35}_{-\5 6}$      & \7{$0.52\pm0.29$}     &  \7{$0.21\pm0.16$}  \\
\2{$K(1^+)$}       &  $7.6\sigma$ & $1793\!\pm\!59\,^{+153}_{-101}$ & $365\!\pm\!157\,^{+138}_{-215}$ & 
                                              $12\!\pm\!10\,^{+17}_{-\5 6}$  &  \7{$0.24\pm0.21$} & \7{$0.37\pm0.17$}  \\
\2{$\nslj{2}{1}{P}{1}$}     &  \cite{Godfrey:1985xj}            & \GI{$1900$}    &               &    & & \\  
\2{$K_1(1650)$}      &  \cite{PDG2014}    &  $1650\!\pm\!50$ &   $150\!\pm\!\9 50$  &  & & \\
\2{$K^{'}(1^+)$}   &  $1.9\sigma$ & $1968\!\pm\!65\,^{+\5 70}_{-172}$ & $396\!\pm\!170\,^{+174}_{-178}$ & $23\!\pm\!20\,^{+31}_{-29}$  & 
           \7{$0.04\pm0.08$} & \7{$0.49\pm0.10$} \\
\2{$\nslj{2}{3}{P}{1}$}     &   \cite{Godfrey:1985xj}           & \GI{$1930$}    &               &    & & \\  
\hline
\1{All $K(2^-)$}   & $5.6\sigma$ &             &      & $11\!\pm\!\5 3\,^{+\5 2}_{-\5 5}$  & & \\
\2{$K(2^-)$}       & $5.0\sigma$ & $1777\!\pm\!35\,^{+122}_{-\5 77}$ & $217\!\pm\!116\,^{+221}_{-154}$ &  & \7{$0.64\pm0.11$} & \7{$0.13\pm0.13$}   \\
\2{$\nslj{1}{1}{D}{2}$}     &    \cite{Godfrey:1985xj}          & \GI{$1780$}     &               &    & & \\  
\2{$\3{K_2(1770)}$} &    \cite{PDG2014}         & $1773\!\pm\!\9 8$ & $188\!\pm\!\9 14$ &    & & \\ 
\2{$K^{'}(2^-)$}  &  $3.0\sigma$ & $1853\!\pm\!27\,^{+\9 18}_{-\9 35}$ & $167\!\pm\!\9 58\,^{+\9 83}_{-\9 72}$ &  & \7{$0.53\pm0.14$} & \7{$0.04\pm0.08$} \\
\2{$\nslj{1}{3}{D}{2}$}    &  \cite{Godfrey:1985xj}             & \GI{$1810$}    &               &    & & \\  
\2{$\3{K_2(1820)}$}   &    \cite{PDG2014}       & $1816\!\pm\!13$ & $276\!\pm\!\9 35$ &    & & \\ 
\hline 
\1{$K^*(1^-)$}    &  $8.5\sigma$ & $1722\!\pm\!20\,^{+\5 33}_{-109}$ & $354\!\pm\!\9 75\,^{+140}_{-181}$ & $6.7\!\pm\!1.9\,^{+3.2}_{-3.9}$ & \7{$0.82\pm0.04$} & \7{$0.03\pm0.03$} \\
\2{$\nslj{1}{3}{D}{1}$}    &   \cite{Godfrey:1985xj}           & \GI{$1780$}      &               &    & & \\  
\2{$\3{K^*(1680)}$}  &   \cite{PDG2014}        & $1717\!\pm\!27$ & $322\!\pm\!110$   &    & & \\ 
\hline
\1{$K^*(2^+)$}   &  $5.4\sigma$ & $2073\!\pm\!94\,^{+245}_{-240}$ & $678\!\pm\!311\,^{+1153}_{-\9 559}$ & $2.9\!\pm\!0.8\,^{+1.7}_{-0.7}$ & \7{$0.15\pm0.06$} & \7{$0.79\pm0.08$} \\
\2{$\nslj{2}{3}{P}{2}$}   &  \cite{Godfrey:1985xj}            & \GI{$1940$}      &               &    & & \\  
\2{${K^*_2(1980)}$}  &  \cite{PDG2014}        & $1973\!\pm\!26$ & $373\!\pm\!\9 69$ &    & & \\ 
\hline
\1{$K(0^-)$}   &   $3.5\sigma$ & $1874\!\pm\!43\,^{+\9 59}_{-115}$ & $168\!\pm\!\9 90\,^{+280}_{-104}$ & $2.6\!\pm\!1.1\,^{+2.3}_{-1.8}$ & 
               \7{$1.0\phantom{5\pm0.55}$} &  \\
\2{$\nslj{3}{1}{S}{0}$} &  \cite{Godfrey:1985xj}             & \GI{$2020$}    &               &    & & \\
\2{${K(1830)}$}  &  \cite{PDG2014}         & $\sim1830$& $\sim 250\4$ &   & &  \\ 
\hline
\hline
\1{All $X(1^+)$}             &                            &                 &                & $16\!\pm\!3\9\9\,^{+\9 6}_{-\9 2}$ & & \\                                                   
$\Xone$  & $8.4\sigma$ & $4146.5\!\pm\!4.5\,^{+4.6}_{-2.8}$ & $83\!\pm\!21\,^{+21}_{-14}$ & $13.0\!\pm\!3.2\,^{+4.8}_{-2.0}$ & & \\                                                
ave.          &  {\!\!\!\!\!\!Table~\ref{tab:xprevious}}     & $4147.1\!\pm\!2.4$ & $15.7\!\pm\!6.3$ &               & & \\
$\Xtwo$  & $6.0\sigma$ & $4273.3\!\pm\!8.3\,^{+17.2}_{-\9 3.6}$ & $56\!\pm\!11\,^{+\9 8}_{-11}$  
     &  $7.1\!\pm\!2.5\,^{+3.5}_{-2.4}$ & & \\
CDF          & \cite{CDF2011}    &  $4274.4\,^{+8.4}_{-6.7}\pm1.9$  & $32\,^{+22}_{-15}\pm8$ &      & &          \\
CMS          & \cite{Chatrchyan:2013dma} & $4313.8\!\pm\!5.3\!\pm\!7.3$ & $38\,^{+30}_{-15}\pm16$        & & &               \\
\hline
\1{All $X(0^+)$}             &                            &                 &                & $28\!\pm\!\9 5\!\pm\!\9 7$   & & \\                                                   
$\NRX$      &  $6.4\sigma$   &   &   & $46\!\pm\!11\9\,^{+11}_{-21}$ & & \\
$\Xthree$   &  $6.1\sigma$  & $4506\!\pm\!11\,^{+12}_{-15}$ & $\5 92\!\pm\!21\,^{+21}_{-20}$  & $6.6\!\pm\!2.4\,^{+3.5}_{-2.3}$ & & \\
$\Xfour$    &  $5.6\sigma$  & $4704\!\pm\!10\,^{+14}_{-24}$ & $120\!\pm\!31\,^{+42}_{-33}$ & $12\!\pm\!\9 5\9\,^{+\9 9}_{-\9 5}$ & & \\           
\hline
\end{tabular}
%}
%\end{center}
}
}
}
\end{table*}

Among the $K^{*+}$ states, the $J^P=1^+$ partial wave has the largest
total fit fraction (given by Eq.~(\ref{eq:ff})). We describe it with three 
heavily interfering contributions: a nonresonant term and two
resonances. The significance of the nonresonant amplitude cannot be quantified,
since when it is removed one of the resonances becomes very broad, taking over its role.
Evidence for the first $1^+$ resonance is significant ($7.6\sigma$).
We include a second resonance in the model, even though it is not
significant ($1.9\sigma$), because two states are expected in the quark model.
We remove it as a systematic variation.       
The $1^+$ states included in our model
appear in the  mass range where two $\nlj{2}{P}{1}$ states are predicted 
(see Table~\ref{tab:baselineinc}), 
and where the $K^-p\to\phi K^-p$ scattering experiment found evidence 
for a $1^+$ state 
with $M_0\sim1840 \mev$, 
$\Gamma_0\sim250 \mev$ \cite{Armstrong:1982tw},
also seen in the $K^-p\to K^-\pi^+\pi^-p$ scattering data \cite{Daum:1981hb}. 
Within the large uncertainties the lower mass state is 
also consistent with the unconfirmed $K_1(1650)$ state \cite{PDG2014}, 
based on evidence from the $K^-p\to\phi K^-p$ scattering experiment \cite{Frame:1985ka}.

\ifthenelse{\boolean{prl}}{ 
\begin{figure}[bthp] 
  \begin{center}
    \includegraphics*[width=\figsize]{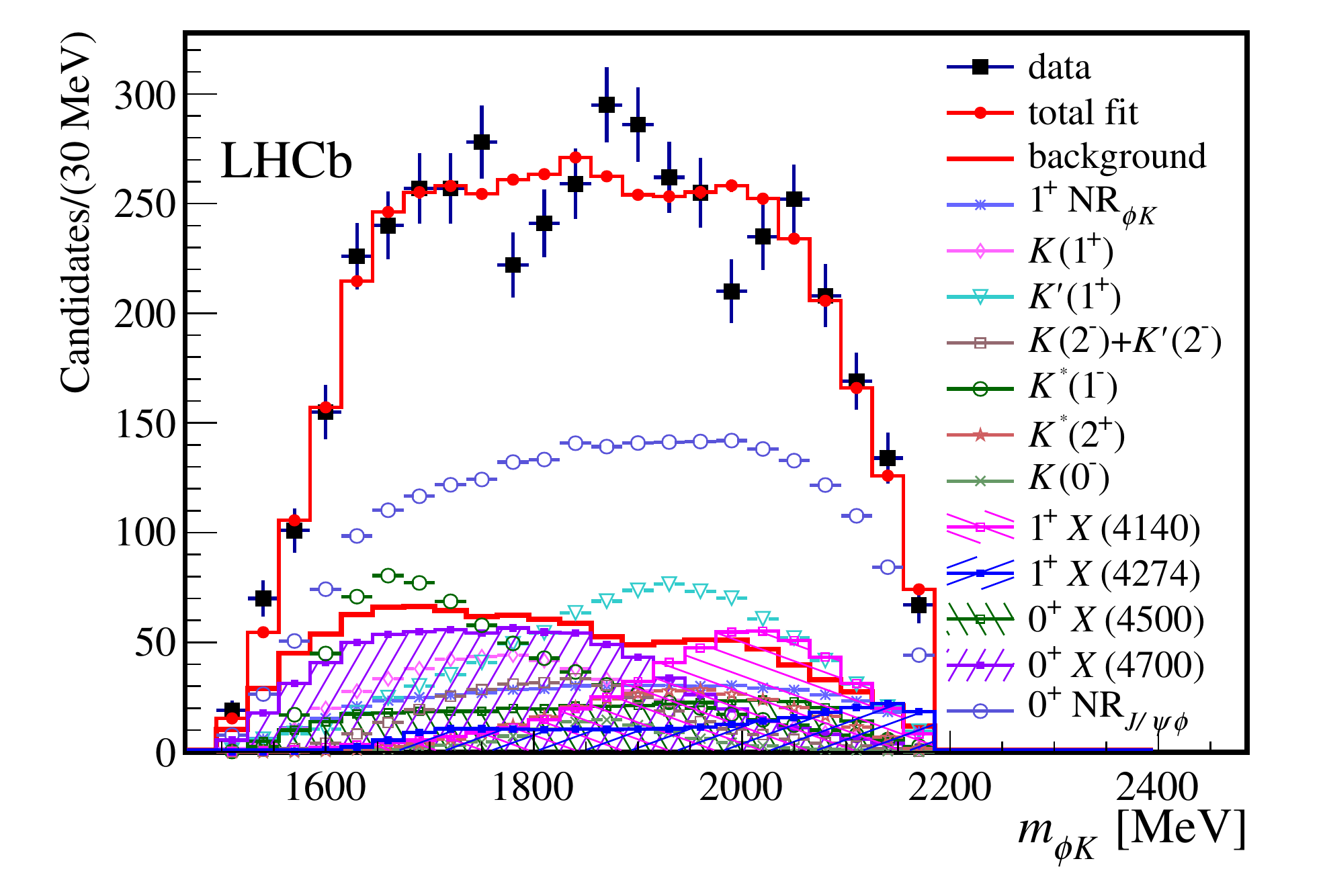} \\[-0.2cm]
    \includegraphics*[width=\figsize]{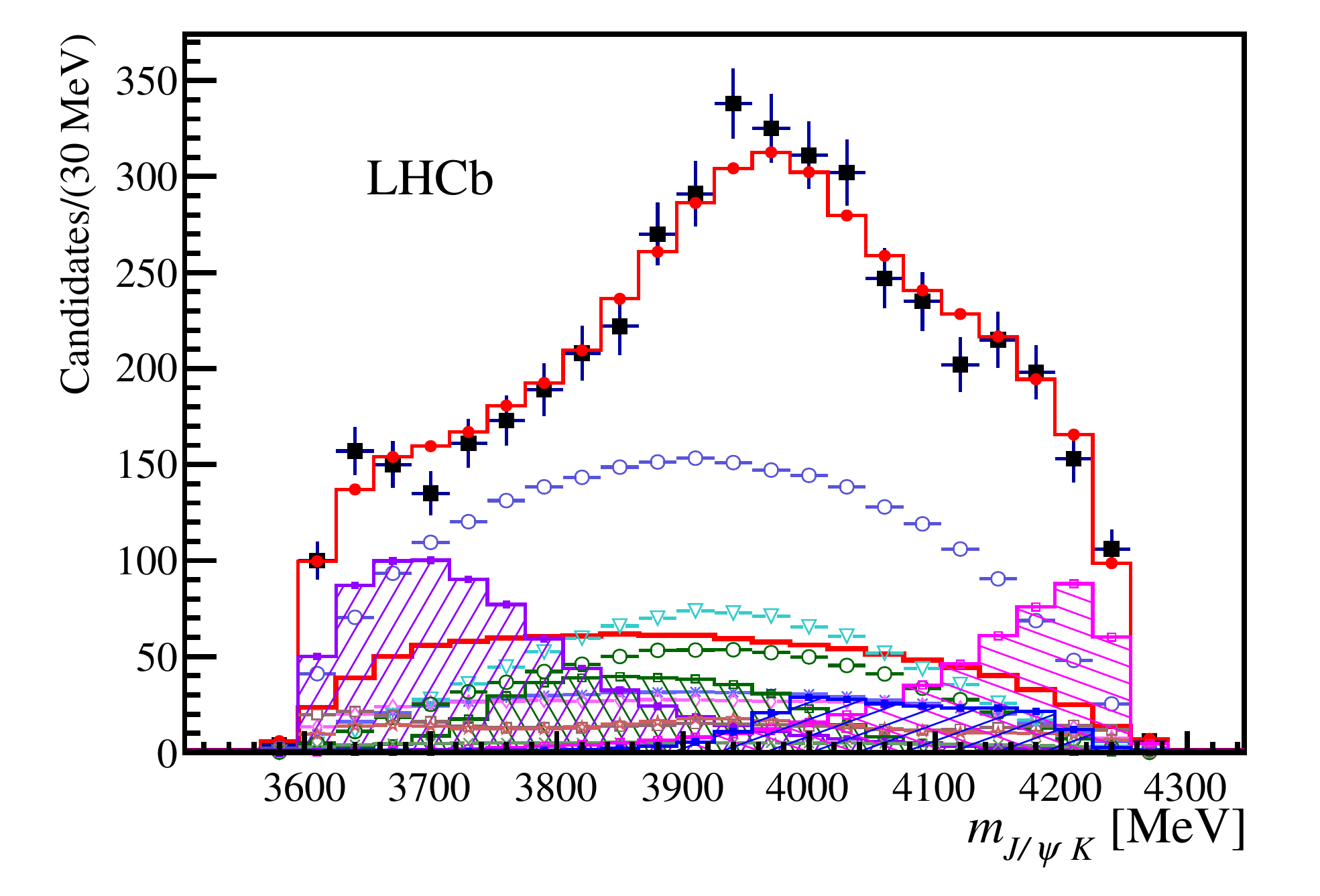} \\[-0.2cm]
    \includegraphics*[width=\figsize]{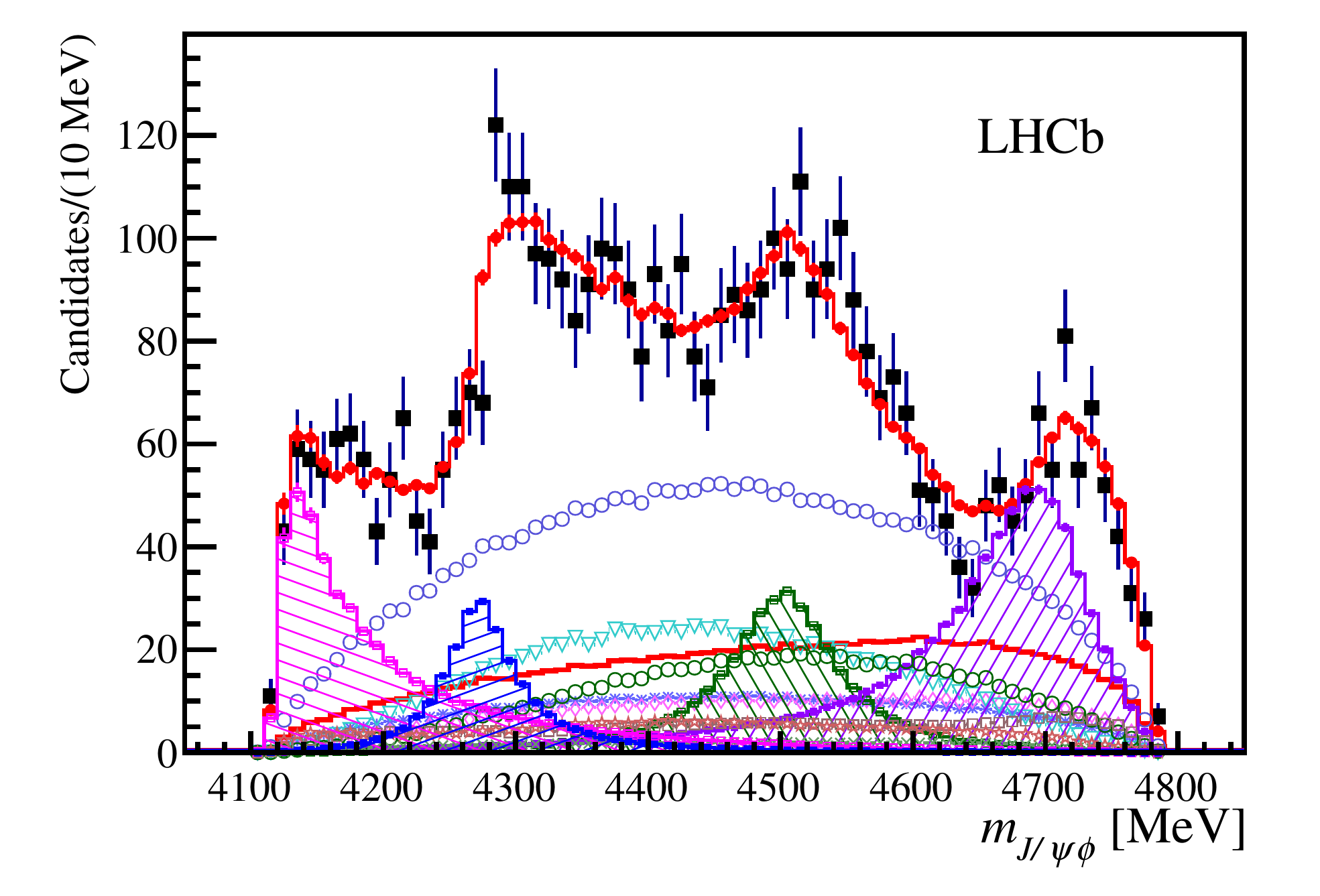} 
  \end{center}
  \vskip-0.3cm\caption{\small
    Distributions of (top) $\phiz K^+$, (middle) $\jpsi K^+$ and (bottom) $\jpsi\phi$ 
    invariant masses for the $\bujphik$ data (black data points) 
    compared with the results of the default amplitude fit 
    containing $K^{*+}\to\phi K^+$ and $X\to\jpsi\phi$ contributions.
    The total fit is given by the red points with error bars. Individual fit components
    are also shown.
    Displays of $m_{\jpsi\phi}$ and of $m_{\jpsi K}$ masses in slices of $m_{\phi K}$ are 
    shown in Fig.~\ref{fig:msliced}.
  \label{fig:defmasses}
  }
\end{figure} 
}{
\begin{figure}[bthp]  
  \begin{center}
    \hbox{\hskip-2.9cm \includegraphics*[width=0.75\figsize]{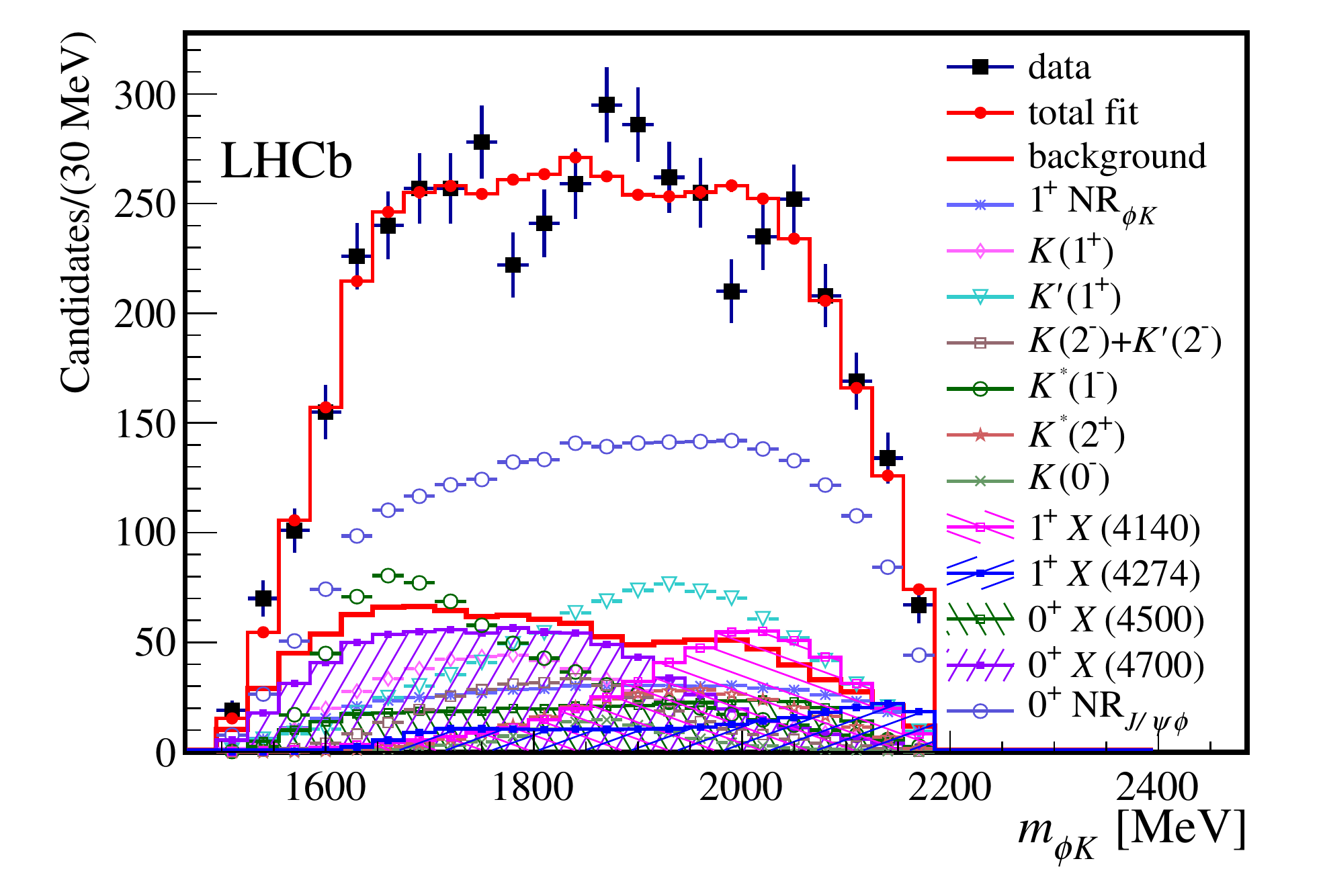} \hskip-0.7cm 
    \includegraphics*[width=0.75\figsize]{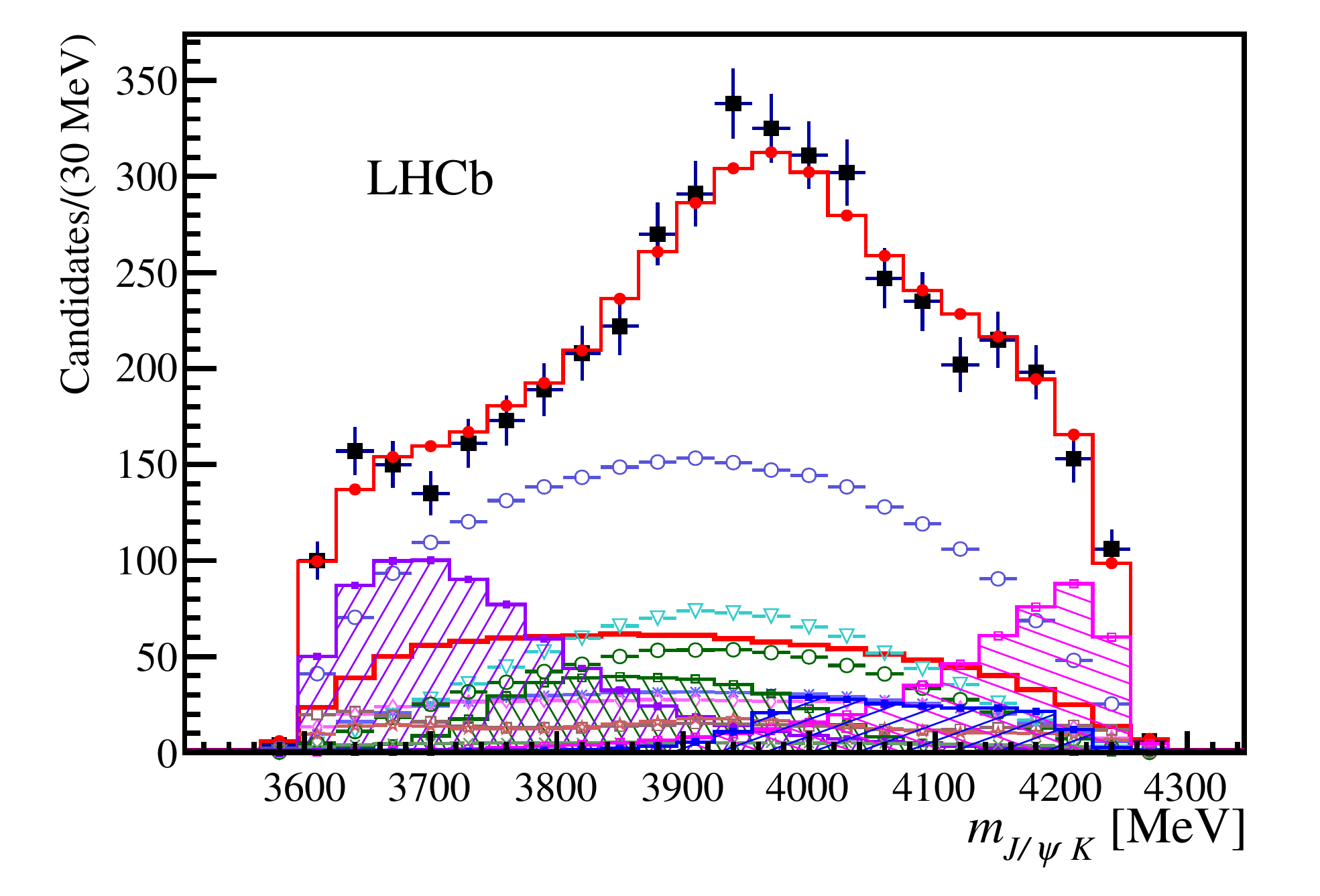} \hskip-2.1cm}\quad \\
    \includegraphics*[width=\figsize]{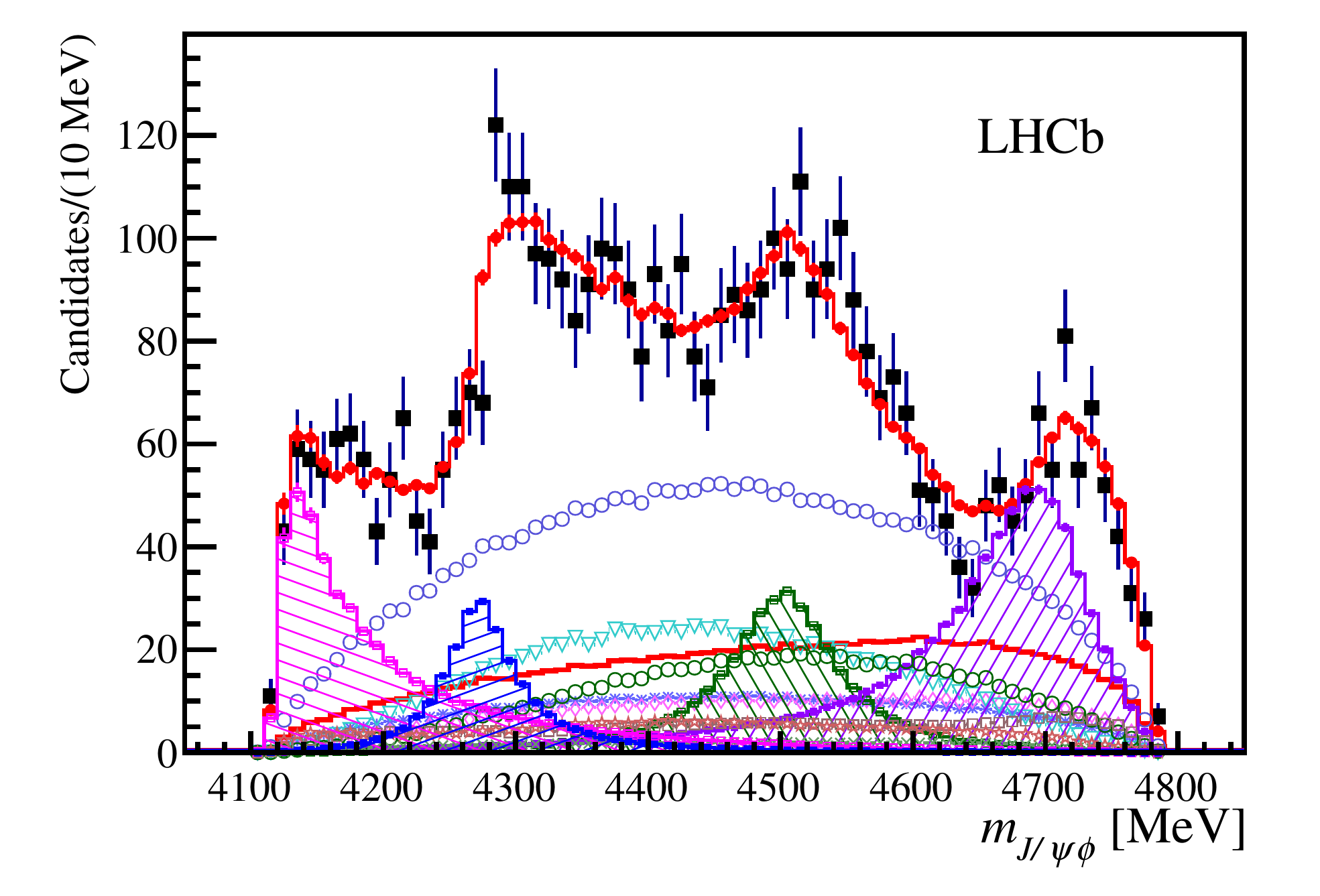} 
  \end{center}
  \vskip-0.3cm\caption{\small
    Distributions of (top left) $\phiz K^+$, (top right) $\jpsi K^+$ and (bottom) $\jpsi\phi$ 
    invariant masses for the $\bujphik$ data (black data points) 
    compared with the results of the default amplitude fit 
    containing $K^{*+}\to\phi K^+$ and $X\to\jpsi\phi$ contributions.
    The total fit is given by the red points with error bars. Individual fit components
    are also shown.
    Displays of $m_{\jpsi\phi}$ and of $m_{\jpsi K}$ masses in slices of $m_{\phi K}$ are 
    shown in Fig.~\ref{fig:msliced}.
  \label{fig:defmasses}
  }
\end{figure} 
}

There is also a substantial $2^-$ contribution to the amplitude model.
When modeled as a single resonance ($5.0\sigma$ significant), 
$M_0=1889\pm27 \mev$ and $\Gamma_0=376\pm94 \mev$ are obtained in agreement
with the evidence from the  $K^-p\to\phi K^-p$ scattering data which yielded 
a mass of around $1840 \mev$ and a width of order $250 \mev$ \cite{Armstrong:1982tw}.
The $K^+p\to\phi K^+p$ scattering data also supported such a state at $1810\pm20 \mev$, but with a 
narrower width, $140\pm40 \mev$ \cite{Frame:1985ka}.
Since two closely spaced $2^-$ states are established 
from other decay modes \cite{PDG2014}, 
and since two $\nlj{1}{D}{2}$ states are predicted, we allow two resonances in the default fit.
The statistical significance of the second state is $3\sigma$.
The masses and widths obtained by the fit to our data are in good agreement 
with the parameters of the $K_2(1770)$ and $K_2(1820)$ states 
and in agreement with the predicted masses of the $\nlj{1}{D}{2}$ states 
(Table~\ref{tab:baselineinc}). 
The individual fit fractions are poorly defined, and not quoted, 
because of large destructive interferences. 
There is no evidence for an additional $2^-$ state in our data 
(which could be the expected $\nlj{2}{D}{2}$ state \cite{Godfrey:1985xj}), 
but we consider the inclusion of such a state among the systematic variations.  

The most significant $K^{*+}$ resonance in our data is a vector state ($8.5\sigma$).  
Its mass and width are in very good agreement with the well-established $K^*(1680)$ state, 
which is observed here in the $\phi K$ decay mode for the first time, and 
fits the $\nslj{1}{3}{D}{1}$ interpretation.
When allowing an extra $1^-$ state (candidate for $\nslj{3}{3}{S}{1}$), 
its significance is $2.6\sigma$ with a mass of $1853\pm5 \mev$, but with a width of only $33\pm11 \mev$,
which cannot be accommodated in the $\bar{s} u$ quark model. 
When limiting the width to be $100 \mev$ or more,
the significance drops to $1.4\sigma$. 
We do not include it in the default model, but consider its 
inclusion as a systematic variation.
We also include 
among the considered variations the effect of an insignificant
($<0.3\sigma$) tail from the $K^*(1410)$ resonance.

There is a significant ($5.4\sigma$) $2^+$ contribution, which we describe with one 
very broad resonance, consistent with the claims of a $K^*_2(1980)$ state seen in other decays 
and also consistent with a broad enhancement seen in $K^-p\to\phi \overline{K}^0n$ scattering
data \cite{Kwon:1993xb}. It can be interpreted as the $\nslj{2}{3}{P}{2}$ state predicted in this mass range.
An extra $2^+$ state added to the model,
as suggested \eg by the possibility that the $\nslj{1}{3}{F}{2}$ state is in the probed mass range, 
is less than $0.7\sigma$ significant and is
considered among the systematic variations.

There is also $3.5\sigma$ evidence for a $0^-$ contribution, consistent with the 
previously unconfirmed $K(1830)$ state 
seen in $K^-p\to\phi K^-p$ scattering data \cite{Armstrong:1982tw}.
It could be a $\nslj{3}{1}{S}{0}$ state. 
An extra $0^-$ state added to the model (\eg $\nslj{4}{1}{S}{0}$) is less than $0.2\sigma$ significant and is
considered among the systematic variations.

We consider among the systematic variations the inclusion of several 
further states that are found not to be significant in the fit.  These 
are a $3^+$ state (\eg $\nlj{1}{F}{3}$, $<1.8\sigma$),
a $4^+$ state (\eg $\nslj{1}{3}{F}{4}$, $<2.0\sigma$ or $<0.6\sigma$ if fixed to the $K^*_4(2045)$ parameters \cite{PDG2014})
or a $3^-$ state (\eg $\nslj{1}{3}{D}{3}$, $<2.0\sigma$ if fixed to the $K^*_3(1780)$ parameters).

Overall, the $K^{*+}$ composition of our data is in good agreement 
with the expectations for the $\bar{s} u$ states,
and also in agreement with previous experimental results on $K^*$ states
in this mass range. 
These results add significantly to the knowledge of $K^*$ spectroscopy.

A near-threshold $\jpsi\phi$ structure in our data is the most significant 
($8.4\sigma$)
exotic contribution to our model. We determine its quantum numbers to be 
$J^{PC}=1^{++}$ at $5.7\sigma$ significance (Appendix~\ref{sec:spinana}).
When fitted as a resonance, its mass (\hbox{$4146.5\pm4.5\,^{+4.6}_{-2.8} \mev$}) is in excellent
agreement with previous measurements for the $X(4140)$ state,
although the width ($83\pm21\,^{+21}_{-14} \mev$) is substantially larger. 
The upper limit which we previously set for production of a narrow ($\Gamma=15.3 \mev$) $X(4140)$ 
state based on a small subset of our present data \cite{LHCb-PAPER-2011-033}
does not apply to such a  broad resonance, \ie the present results are 
consistent with our previous analysis.
The statistical power of the present data sample is not sufficient 
to study its phase motion \cite{LHCb-PAPER-2014-014}. 
A model-dependent study discussed in Appendix~\ref{sec:cusp} suggests that 
the $X(4140)$ structure may be affected by the nearby
$D_s^{\pm}D_s^{*\mp}$ coupled-channel threshold. 
However, larger data samples will be required to resolve this issue.
 
We establish the existence of the $\Xtwo$ structure
with statistical significance of $6.0\sigma$, at a mass of
\hbox{$4273.3\pm8.3\,^{+17.2}_{-\9 3.6} \mev$} and
a width of \hbox{$56.2\pm10.9\,^{+\9 8.4}_{-11.1} \mev$}.
Its quantum numbers are also $1^{++}$ at $5.8\sigma$ significance. 
Due to interference effects, the data peak above the pole mass,
underlining the importance of proper amplitude analysis. 

The high $J/\psi\phi$ mass region also shows 
structures that cannot be described in a model containing 
only $K^{*+}$ states. These features are best described in our model by two
$J^{PC}=0^{++}$ resonances at $4506\pm11\,^{+12}_{-15}$ \mev
and $4704\pm10\,^{+14}_{-24} \mev$, 
with widths of $92\pm21\,^{+21}_{-20} \mev$ and $120\pm31\,^{+42}_{-33} \mev$,
and significances of $6.1\sigma$ and $5.6\sigma$, respectively.
The resonances interfere with a nonresonant $0^{++}$ $\jpsi\phi$ contribution 
that is also significant ($6.4\sigma$).
The significances of the quantum number determinations for the 
high mass states are   
$4.0\sigma$ and $4.5\sigma$, respectively.

Additional $X$ resonances of any $J^P$ value ($J\le2$) added to our model
have less than $2\sigma$ significance.
A modest improvement in fit quality can be achieved by adding $Z^+\to\jpsi K^+$
resonances to our model, however the significance of such contributions
is too small to justify introducing an exotic hadron contribution 
(at most $3.1\sigma$ without accounting for systematic uncertainty).
The parameters obtained for the default model components stay within their
systematic uncertainties when such extra $X$ or $Z^+$ contributions are introduced. 

\begin{figure}[bthp] 
  \begin{center}
    \includegraphics*[width=\figsize]{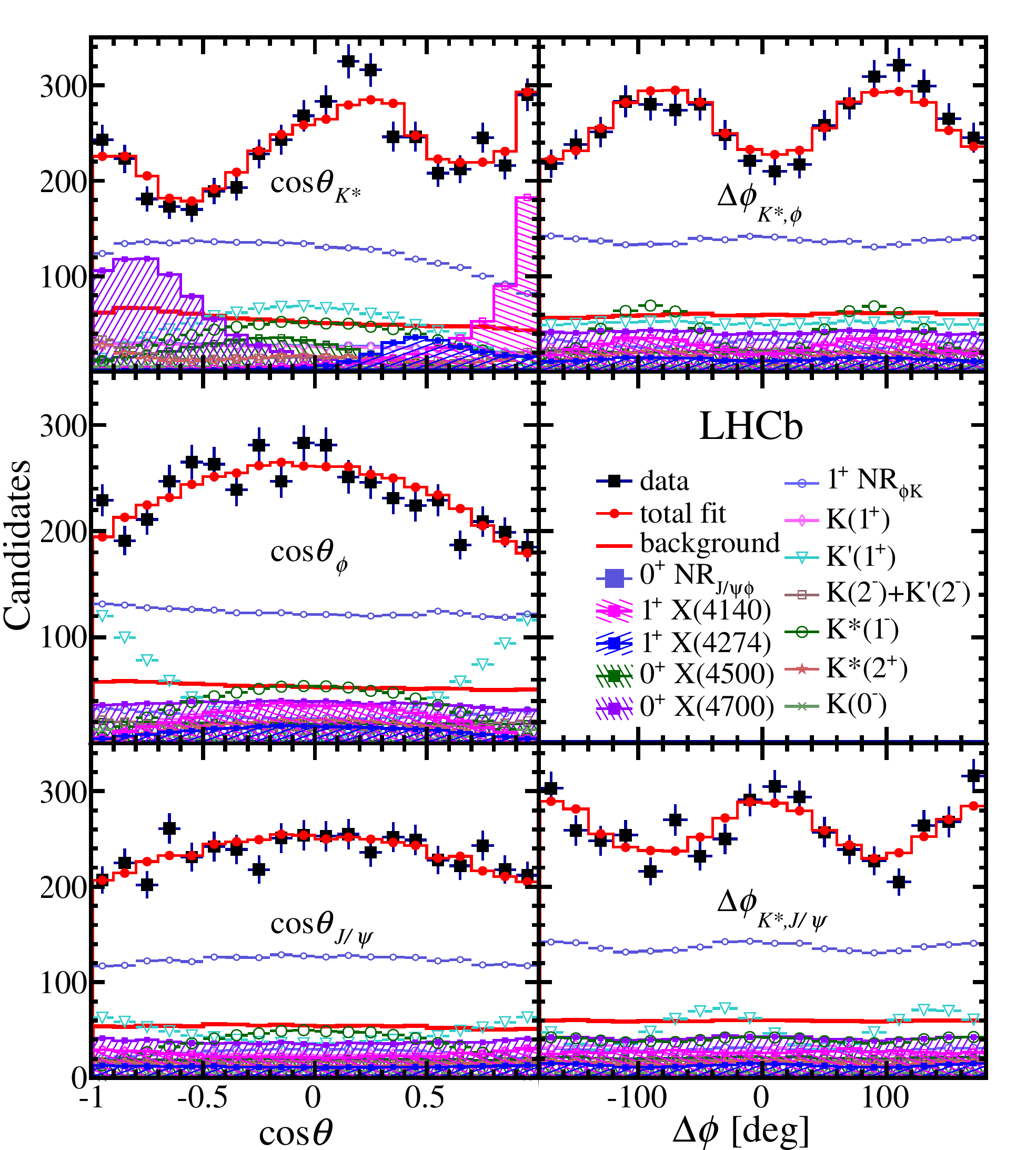} \\
  \end{center}
  \vskip-0.3cm\caption{\small
    Distributions of the fitted decay angles from the $K^{*+}$ decay chain together with the display of the
    default fit model described in the text.    
  \label{fig:defksangles}
  }
\end{figure}

\begin{figure}[bthp] 
  \begin{center}
    \includegraphics*[width=\figsize]{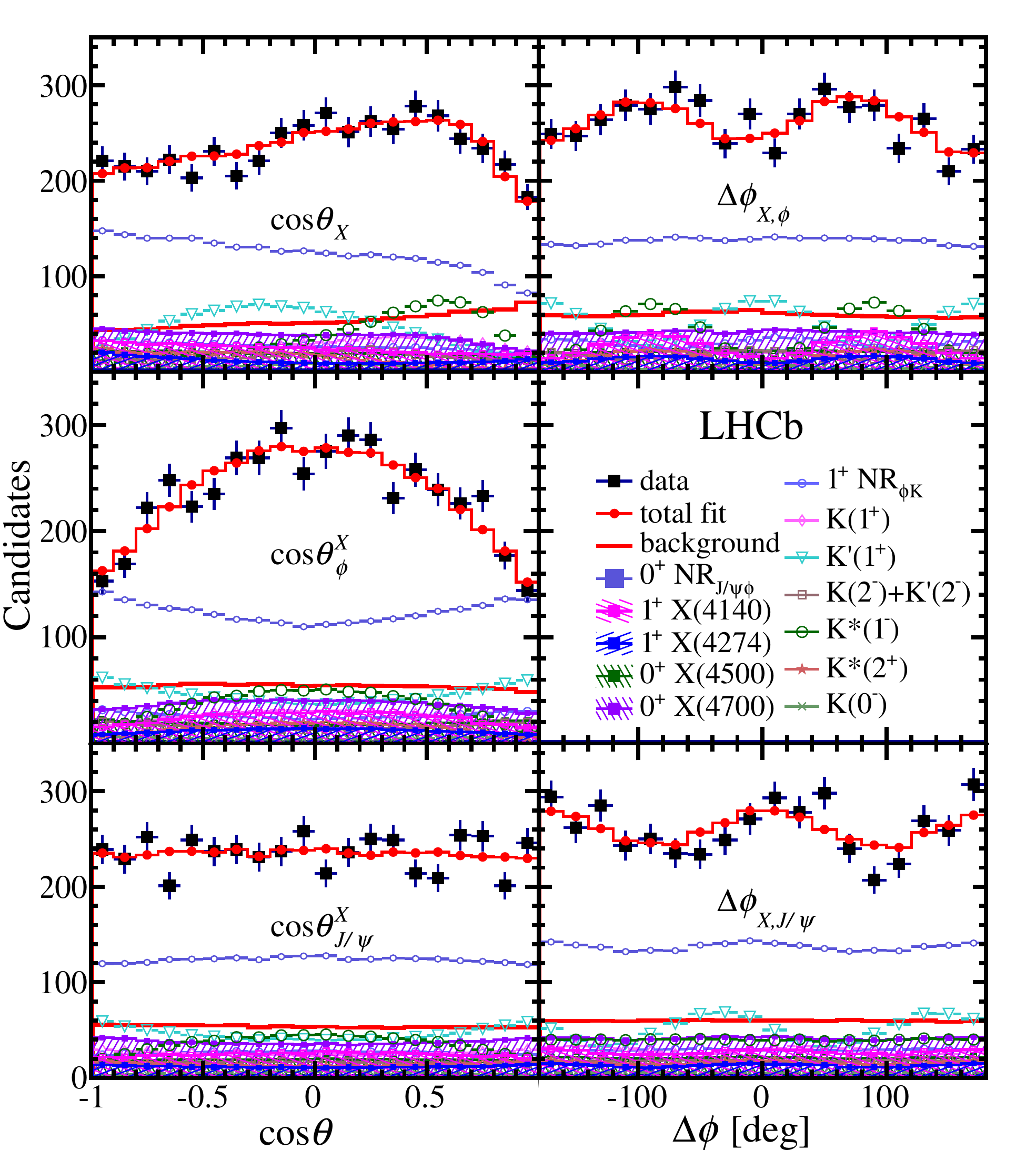} \\
  \end{center}
  \vskip-0.3cm\caption{\small
    Distributions of the fitted decay angles from the $X$ decay chain together with the display of the
    default fit model described in the text.    
  \label{fig:defxangles}
  }
\end{figure}

\begin{figure}[bthp] 
  \begin{center}
    \includegraphics*[width=\figsize]{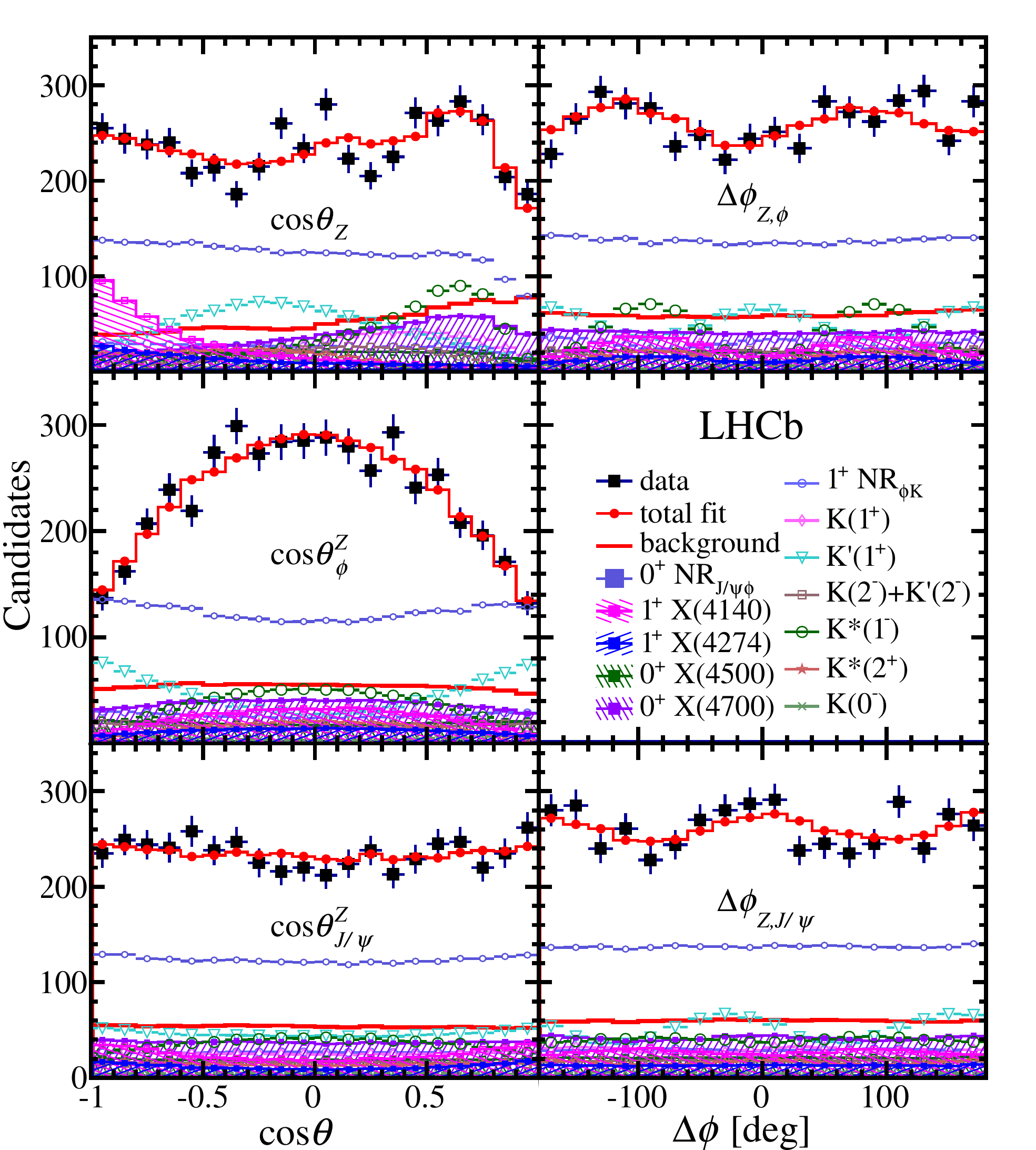} \\
  \end{center}
  \vskip-0.3cm\caption{\small
    Distributions of the fitted decay angles from the $Z$ decay chain together with the display of the
    default fit model described in the text.    
  \label{fig:defzangles}
  }
\end{figure}

\ifthenelse{\boolean{prl}}{
\begin{figure}[p]
  \begin{center}
\includegraphics*[width=\figsize]{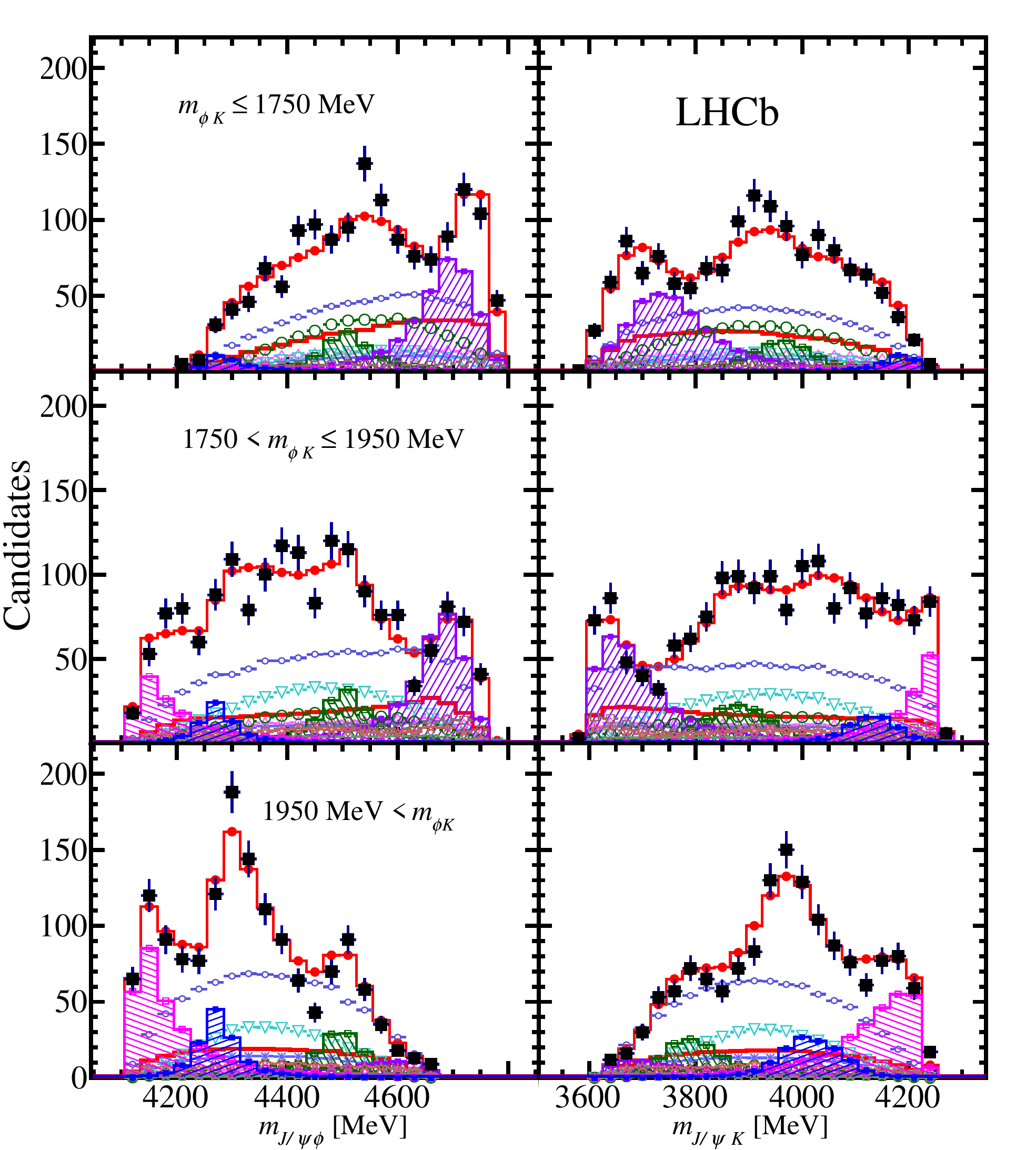}
  \end{center}
  \vskip-0.3cm\caption{\small
    Distribution of (left) $m_{\jpsi\phi}$ and (right) $m_{\jpsi K}$  
    in three slices of $m_{\phiz\kaon}$: $<1750 \mev$, $1750-1950 \mev$, and $>1950 \mev$ from top to bottom,
    together with the projections of the default amplitude model. 
    See the legend in Fig.~\ref{fig:defmasses} for a description of the components.
  \label{fig:msliced}
  }
\end{figure}
}{
\begin{figure}[p]
  \begin{center}
\includegraphics*[width=1.1\figsize]{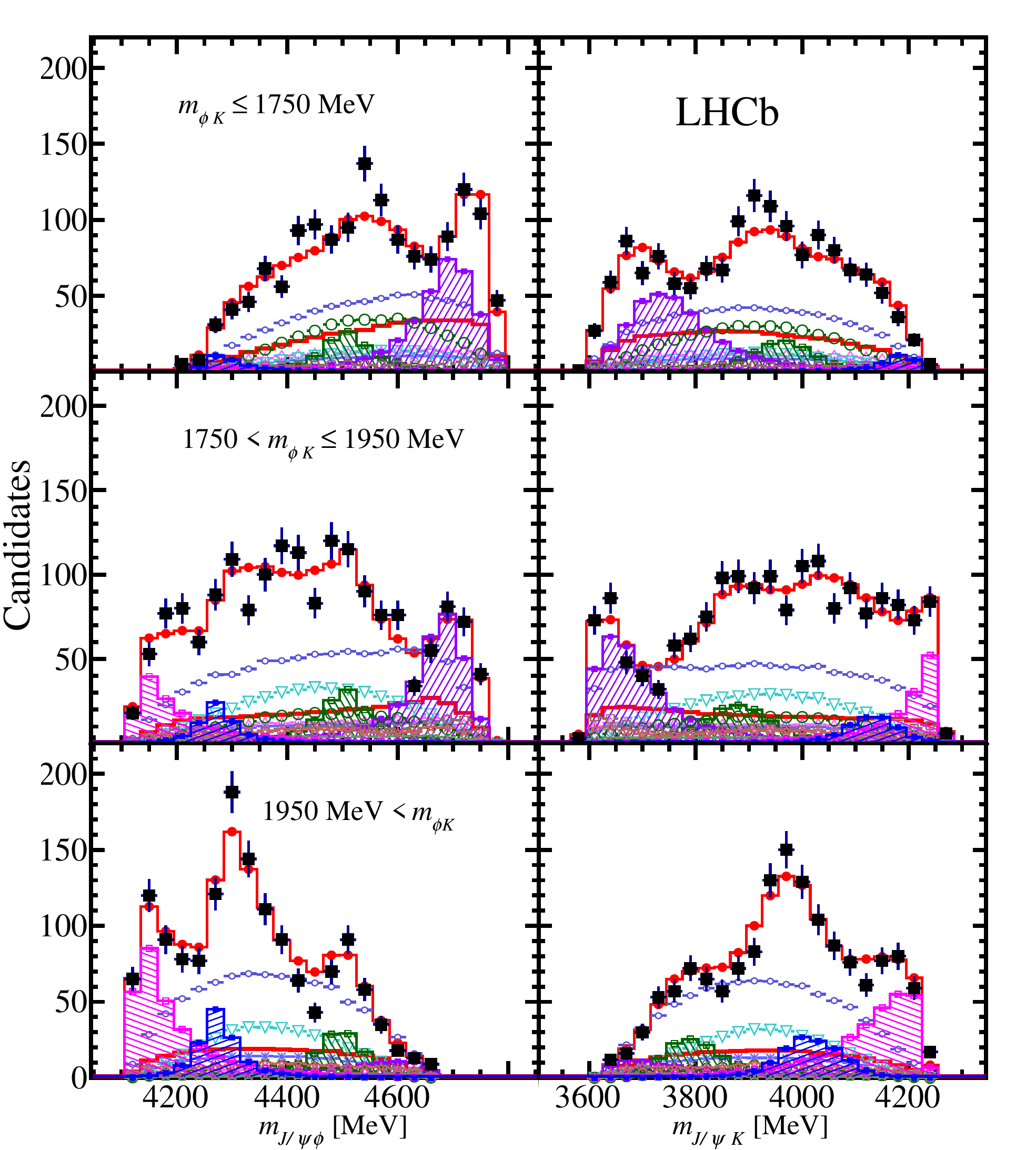}
  \end{center}
  \vskip-0.3cm\caption{\small
    Distribution of (left) $m_{\jpsi\phi}$ and (right) $m_{\jpsi K}$ 
    in three slices of $m_{\phiz\kaon}$: $<1750 \mev$, $1750-1950 \mev$, and $>1950 \mev$ from the top to the bottom,
    together with the projections of the default amplitude model. 
    See the legend in Fig.~\ref{fig:defmasses} for a description of the components.
  \label{fig:msliced}
  }
\end{figure}
}

%\FloatBarrier

\ifthenelse{\boolean{prl}}{\FloatBarrier}{\FloatBarrier}%

\section{Summary}

In summary, we have performed the first amplitude analysis of $\bujphik$ decays.
We have obtained a good description of data in the 6D phase space 
including invariant masses and decay angles.

  Even though no peaking structures are observed in the $\phi K^+$ mass distributions,
  correlations in the decay angles reveal a rich spectrum of $K^{*+}$ resonances.
  In addition to the angular information contained in the $K^{*+}$ and 
  $\phi$ decays, the $\jpsi$ decay also helps to probe these resonances, as the 
  helicity states of the $K^{*+}$ and $\jpsi$ mesons coming from the $B^+$ decay 
  must be equal. 
  Unlike the earlier scattering experiments
  investigating $K^{*}\to\phi K$ decays, we have good sensitivity to states with 
  both natural and unnatural $J^P$ combinations. 

  The dominant $1^{+}$ partial wave ($\FiFr=42\pm8\,^{+5}_{-9}\%$)
  has a substantial nonresonant component, 
  and at least one resonance that is $7.6\sigma$ significant. There is also $2\sigma$
  evidence that this structure can be better described with two resonances 
  matching the expectations for the two $\nlj{2}{P}{1}$ excitations of the kaon. 
  
  Also prominent is the $2^{-}$ partial wave
  ($\FiFr=10.8\pm2.8\,^{+1.5}_{-4.6}\%$).
  It contains at least one resonance at $5.0\sigma$ significance. 
  This structure is also better described with two
  resonances at $3.0\sigma$ significance. 
  Their masses and widths
  are in good agreement with the well established $K_2(1770)$ and $K_2(1820)$
  states, matching the predictions for the two $\nlj{1}{D}{2}$ kaon excitations.
  
  The $1^-$ partial wave
  ($\FiFr=6.7\pm1.9\,^{+3.2}_{-3.9}\%$)
  exhibits $8.5\sigma$ evidence for a resonance
  which matches the $K^*(1680)$ state, which was well established 
  in other decay modes, and matches the expectations for the $\nslj{1}{3}{D}{1}$ kaon
  excitation.  This is the first observation of its decay to 
  the $\phi K$ final state. 
   
  The $2^+$ partial wave has a smaller intensity
  ($\FiFr=2.9\pm0.8\,^{+1.7}_{-0.7}\%$),
  but provides $5.4\sigma$ evidence for a broad structure that 
  is consistent with the $K_2^*(1980)$ state observed previously
  in other decay modes and matching the expectations for the $\nslj{2}{3}{P}{2}$ state.

  We also confirm the $K(1830)$ state ($\nslj{3}{1}{S}{0}$ candidate) 
  at $3.5\sigma$ significance
  ($\FiFr=2.6\pm1.1\,^{+2.3}_{-1.8}\%$),
  earlier observed in the $\phi K$ decay by the $K^-p$ scattering experiment. 
  We determine its mass and width with properly evaluated uncertainties for the first
  time. %: $1874\pm43\,^{+\9 59}_{-115} \mev$ and $168\pm90\,^{+281}_{-104} \mev$. 

  Overall, our $K^{*+}\to\phi K^+$ results show excellent consistency with
  the states observed in other experiments, often in other decay modes,
  and fit the mass spectrum predicted for the kaon excitations by the Godfrey-Isgur model.
  Most of the $K^{*+}$ structures we observe were previously observed or hinted at 
  by the $Kp\to\phi K(p\,\,{\rm or}\,\,n)$ experiments, which were,
  however, sometimes inconsistent with each other.

Our data cannot be well described without several $\jpsi\phi$ contributions.
The significance of the near-threshold $X(4140)$ structure is $8.4\sigma$
with \FiFr$=13.0\pm3.2^{+4.8}_{-2.0}\%$.
Its width is substantially larger than previously
determined. 
We determine the $J^{PC}$ quantum numbers of
this structure to be $1^{++}$ at $5.7\sigma$. 
This has a large impact on its possible
interpretations, in particular ruling out the $0^{++}$ or $2^{++}$ 
$D_s^{*+}D_s^{*-}$ 
molecular models
\cite{Liu:2009ei,Branz:2009yt,Albuquerque:2009ak,Ding:2009vd,Zhang:2009st,Chen:2015fdn}. 
The below-$\jpsi\phi$-threshold $D_s^{\pm}D_{s}^{*\mp}$ 
cusp \cite{Swanson:2014tra,Karliner:2016ith} 
may have an impact on the $X(4140)$ structure, 
but more data will be required to address this issue.
The existence of the $X(4274)$ structure is established ($6\sigma$)
with \FiFr$=7.1\pm2.5^{+3.5}_{-2.4}\%$\ and 
its quantum numbers are determined to be $1^{++}$ ($5.8\sigma$). 
Together, these two $J^{PC}=1^{++}$ contributions 
have a fit fraction of $16\pm3\,^{+6}_{-2}\%$.
Molecular bound-states or cusps cannot account for the $X(4274)$ $J^{PC}$ values. 
A hybrid charmonium state would have $1^{-+}$ \cite{Mahajan:2009pj,Wang:2009ue}. 
Some tetraquark models expected $0^{-+}$, $1^{-+}$ \cite{Drenska:2009cd}
or $0^{++}$, $2^{++}$ \cite{Wang:2015pea} state(s) in this mass range.  
A tetraquark model implemented by Stancu \cite{Stancu:2009ka}
not only correctly assigned $1^{++}$ to $X(4140)$, 
but also predicted a second $1^{++}$ state at mass
not much higher than the $X(4274)$ mass. 
Calculations by Anisovich \etal \cite{Anisovich:2015caa} 
based on the diquark tetraquark model 
predicted only one $1^{++}$ state 
at a somewhat higher mass.
Lebed--Polosa \cite{Lebed:2016yvr} predicted the $X(4140)$ peak to be a $1^{++}$ tetraquark,
although they expected the $X(4274)$ peak to be a $0^{-+}$ state in the same model.  
A lattice QCD calculation with diquark operators found no evidence for 
a $1^{++}$ tetraquark below $4.2\gev$ \cite{Padmanath:2015era}. 

The high $\jpsi\phi$ mass region
is investigated with good sensitivity for the first time 
and shows very significant structures, which can be described as $0^{++}$
contributions ($\FiFr=28\pm5\pm7\%$) with a nonresonant term plus two 
new resonances: $X(4500)$ ($6.1\sigma$ significant) 
and $X(4700)$ ($5.6\sigma$).  
The quantum numbers of these states are determined with 
significances of more than $4\sigma$.
The work of Wang \etal \cite{Wang:2009ry} predicted a virtual 
$D_s^{*+}D_s^{*-}$ state at $4.48\pm0.17 \gev$.
None of the observed $\jpsi\phi$ states is consistent with the
state seen in two-photon collisions
by the Belle collaboration \cite{Shen:2009vs}.

\quad\newline
\input{acknowledgements}

\newpage
\ifthenelse{\boolean{prl}}{\clearpage}{} 

\appendix

\FloatBarrier

\section{Calculation of decay angles}
\label{sec:angles}

The decay angles are calculated in a way analogous 
to that documented in Appendix IX of Ref.~\cite{Chilikin:2013tch}.
The five angles for each decay chain are: three helicity angles of $\jpsi$, $\phi$ 
and of the resonance in question (\eg $\Kst$) and two angles between 
the decay plane of the resonance and the decay plane of either $\jpsi$ or $\phi$.  
In addition, a rotation is needed to align the muon helicity frames of the $X$ and $Z^+$ decay chains 
to that of the $\Kst$ in order to properly describe the interferences.  
The choice of $\Kst$ as the reference decay chain is arbitrary.  
The cosine of a helicity angle of particle $P$, 
produced in two-body decay $A\to P\,B$, 
and decaying to two particles $P\to C\,D$ is calculated from (Eq.~(16) in Ref.\cite{Chilikin:2013tch})
\begin{equation}
\cos\theta_P = - \frac{\vec{p}_B\cdot\vec{p}_C}{\vert\vec{p}_B\vert\,\vert\vec{p}_C\vert},
\label{eq:cosRes}
\end{equation}
where the momentum vectors are in the rest frame of the particle $P$. 

For the $B^+\to\jpsi K^{*+}$ decay, the angle between the $\jpsi\to\mu^+\mu^-$ 
and the $K^{*+}\to\phi K^+$ decay planes
is calculated from\footnote{The function atan2$(x,y)$ is the $\tan^{-1}(y/x)$ function with two arguments. The purpose of using two arguments
instead of one is to gather information on the signs of the inputs in order to return the appropriate
quadrant of the computed angle.} 
(Eqs.~(14)--(15) in Ref.\cite{Chilikin:2013tch})
\begin{align}
\phikspsi & = {\rm atan2}(\sin\phikspsi\,,\,\cos\phikspsi) \\
\cos\phikspsi & = \frac{\vec{a}_{K^+}\cdot\vec{a}_{\mu^+}}{
\vert\vec{a}_{K^+}\vert\,\vert\vec{a}_{\mu^+}\vert} \\
\sin\phikspsi & = \frac{[\vec{p}_{\jpsi}\times\vec{a}_{K^+}]\cdot\vec{a}_{\mu^+}}{
\vert\vec{p}_{\jpsi}\vert\,\vert\vec{a}_{K^+}\vert\,\vert\vec{a}_{\mu^+}\vert} \\
\vec{a}_{K^+} & =\vec{p}_{K^+}-\frac{\vec{p}_{K^+}\cdot\vec{p}_{K^{*+}}}{
\vert \vec{p}_{K^{*+}} \vert^2}\,\vec{p}_{K^{*+}}\\
\vec{a}_{\mu^+} & =\vec{p}_{\mu^+}-\frac{\vec{p}_{\mu^+}\cdot\vec{p}_{\jpsi}}{
\vert \vec{p}_{\jpsi} \vert^2}\,\vec{p}_{\jpsi},
\end{align}
with all vectors being in the $B^+$ rest frame. 
For the $B^+\to Z^+ \phi$ decay, the angle between the $Z^+\to\jpsi K^+$ 
and the $\phi\to K^+K^-$ decay planes, $\phizphi$, can be calculated in the same way
with $\jpsi\to\phi$, $\mu^+\to K^+$ (the $K^+$ from the $\phi$ decay) and
the accompanying $K^+$ staying the same.

The angle between the decay planes of two sequential decays,
\eg between the $Z^+\to\jpsi K^+$ and $\jpsi\to\mu^+\mu^-$ decay planes
after the $B^+\to Z^+\phi$ decay, 
is calculated from  
(Eqs.~(18)--(19) in Ref.\cite{Chilikin:2013tch})
\begin{align}
\phizpsi & = {\rm atan2}(\sin\phizpsi \,,\,\cos\phizpsi ) \\
\cos\phizpsi & = \frac{\vec{a}_{\phi}\cdot\vec{a}_{\mu^+}}{
\vert\vec{a}_{\phi}\vert\,\vert\vec{a}_{\mu^+}\vert} \\
\sin\phizpsi & = \frac{-[\vec{p}_{K^+}\times\vec{a}_{\phi}]\cdot\vec{a}_{\mu^+}}{
\vert\vec{p}_{K^+}\vert\,\vert\vec{a}_{\phi}\vert\,\vert\vec{a}_{\mu^+}\vert} \\
\vec{a}_{\phi} & =\vec{p}_{\phi}-\frac{\vec{p}_{\phi}\cdot\vec{p}_{K^+}}{
\vert \vec{p}_{K^+} \vert^2}\,\vec{p}_{K^+}\\
\vec{a}_{\mu^+} & =\vec{p}_{\mu^+}-\frac{\vec{p}_{\mu^+}\cdot\vec{p}_{K^+}}{
\vert \vec{p}_{K^+} \vert^2}\,\vec{p}_{K^+},
\end{align}
with all vectors being in the $\jpsi$ rest frame. 
The other angles of this type are calculated in the same way, with appropriate 
substitutions. 
For example, $\phiksphi$ between the $K^{*+}\to\phi K^+$ and $\phi\to K^+K^-$
decay planes after $B^+\to K^{*+}\jpsi$ decay, is calculated substituting 
$\phi\longrightarrow\jpsi$,  $\mu^+\longrightarrow K^+$ ($K^+$ from the $\phi$ decay), 
and the accompanying $K^+$ staying the same (all vectors are in the $\phi$ rest frame here).

The angle aligning the muon helicity frames between the $K^{*+}$  
and $Z^+$ decay chains is calculated from 
(Eqs.~(20)--(21) in Ref.\cite{Chilikin:2013tch})
\begin{align}
\alpha^Z & = {\rm atan2}(\sin\alpha^Z \,,\,\cos\alpha^Z ) \\
\cos\alpha^Z & = \frac{\vec{a}_{K^+}\cdot\vec{a}_{K^{*+}}}{
\vert\vec{a}_{K^+}\vert\,\vert\vec{a}_{K^{*+}}\vert} \\
\sin\alpha^Z & = \frac{-[\vec{p}_{\mu^+}\times\vec{a}_{K^+}]\cdot\vec{a}_{K^{*+}}}{
\vert\vec{p}_{\mu^+}\vert\,\vert\vec{a}_{K^+}\vert\,\vert\vec{a}_{K^{*+}}\vert} \\
\vec{a}_{K^{*+}} & =\vec{p}_{K^{*+}}-\frac{\vec{p}_{K^{*+}}\cdot\vec{p}_{\mu^+}}{
\vert \vec{p}_{\mu^+} \vert^2}\,\vec{p}_{\mu^+}\\
\vec{a}_{K^+} & =\vec{p}_{K^+}-\frac{\vec{p}_{K^+}\cdot\vec{p}_{\mu^+}}{
\vert \vec{p}_{\mu^+} \vert^2}\,\vec{p}_{\mu^+},
\end{align}
where the $K^+$ is the accompanying kaon and all vectors are in the $\jpsi$ rest frame.
Similarly, $\alpha^X$ is obtained from the above equations with the
$K^+\longrightarrow\phi$ substitution.
 
For the charge-conjugate $B^-\to\jpsi\phi K^-$ decays, the same formulae apply with 
the accompanying kaon being $K^-$,
$\mu^+$ replaced by $\mu^-$ and $K^+$ from the $\phi$ decay 
replaced by the $K^-$ from the $\phi$ decay.  
All azimuthal angles ($\Delta\phi$ and $\alpha$) have their signs flipped after
applying the formulae above (see the bottom of Appendix IX in Ref.~\cite{Chilikin:2013tch}). 

\section{Systematic uncertainty}
\label{sec:sys}

Individual systematic uncertainties on masses, widths
and fit fractions are presented
for $K^{*+}$ contributions in 
Tables~\ref{tab:ksysA}--\ref{tab:ksysB},
and for $X$ contributions in Table~\ref{tab:xsys}.  
Positive and negative deviations are summed in quadrature separately
for total systematic uncertainties.
The statistical uncertainties are included for comparison.

In many instances, the uncertainty in the $K^{*+}$ model composition is the dominant
systematic uncertainty. The $K^{*+}$ model variations include adding the 
following contributions (one-by-one) to the default amplitude model:
second $0^-$, $1^-$ or $2^+$ states, 
a third $2^-$ state, 
the $3^-$ $K_3^*(1780)$ state, 
a $3^+$ state, 
the $4^+$ $K_4^*(2045)$ state, and
the below threshold $1^-$ $K^*(1410)$ state.
The variations also include omitting the second $1^+$ or $2^-$ states.
The observed deviations in the fit parameters are added in quadrature
and then listed in Tables~\ref{tab:ksysA}--\ref{tab:xsys}.

The other sizable source of systematic uncertainty is due to the $L_B$ and $L_{K^*}$ (or $L_X$) dependence of the 
Breit--Wigner amplitude in the numerator of Eq.~(\ref{eq:resshapeBW}) via Blatt-Weisskopf factors. 
Helicity states correspond to mixtures of allowed $L$ values, 
but we assume the lowest $L$ values in Eq.~(\ref{eq:resshapeBW}) in the default fit.
We increase $L_B$ values by $1$ for all the components (one-by-one).
Values of $L_{K^*}$ or $L_X$ can only differ by an even number because of parity conservation in
strong decays. We performed such variations for states in which the higher value is allowed, except 
for the $X$ states, since the fit results indicate that the higher $L_X$ amplitudes are insignificant.  
Again, the observed deviations in the fit parameters are added in quadrature
and then listed in Tables~\ref{tab:ksysA}--\ref{tab:xsys}.

The energy release in the $B^+\to\jpsi\phi K^+$ decay is small ($\sim13\%$ on $M_B$), and the 
phase space is very limited, not offering much range for nonresonant interactions to change. 
In the default model the nonresonant terms are represented by constant amplitudes.
When allowing them to change exponentially with mass-squared, $\exp(-\alpha\,m^2)$,
the slope parameters, $\alpha$, are consistent with zero. 
The observed deviations in the measured parameters are included among the systematic contributions.

Replacing the Breit--Wigner amplitude for the $X(4140)$ structure with a 
$D_s^{\pm}D_s^{*\mp}$ cusp in one particular model (see Appendix~\ref{sec:cusp}) is 
included among the systematic model variations.

The dependence on mass of the total resonance width (Eq.~(\ref{eq:mwidth})) used in the default fit assumes 
that it is dominated by the observed decay mode.
All $K^{*+}$ states are expected to have sizable widths to the other decay
modes, $K\pi$, $K\rho$, $K^*(892)\pi$ etc. 
However, ratios of these partial widths to the $\phi K$ partial width are unknown.
To check the related systematic uncertainty, we perform an alternative fit 
(marked $\Gamma_{\rm tot}$ in the tables) in which the mass dependence of the width is set by 
the lightest possible decay mode allowed:
$K\pi$ for natural spin-parity resonances and $K\omega$ for the others.
This includes changing the $L_{K^*}$ value.

The Blatt-Weisskopf factors contain the $d$ parameter for the effective hadron size \cite{Aston:1987ir} (Eq.~(\ref{eq:blattw})), 
which we set to $3.0 \gev^{-1}$ in the default fit. 
As a systematic variation we change its value between $1.5$ and $5.0 \gev^{-1}$.  

To address the systematic uncertainty in the background parameterization, 
we perform amplitude fits with either the left or right $B^+$ mass sideband only.
The default fit uses both. 

We perform fits to $m_{\jpsi \phi K}$ with alternative signal and background
parameterizations to determine the systematic uncertainty on the background fraction 
in the signal region ($\beta)$. The largest deviation in its value 
($\Delta\beta/\beta=+25\%$) is then used in the alternative amplitude fit to
the data.

In the default fit, the simulated events used for the efficiency corrections
are weighted to improve the agreement between the data and the simulation.
The total Monte Carlo event weight ($w^{MC}$) is a product of weights
determined as the ratios between the signal distributions in the data and in the
simulated sample (generated according to the preliminary amplitude model) 
as functions of $\pt(B^+)$, number of charged tracks in the event, and 
each kaon momentum. These weights are intended to correct for any inaccuracies 
in simulation of $pp$ collisions, of $B^+$ production kinematics and in kaon
identification. To account for the uncertainty associated with the efficiency modelling 
we include among the systematic 
variations a fit in which the weights are not applied.

To check the uncertainty related to non-$\phi$ background, we reduce its fraction 
by narrowing the $\phi\to K^+K^-$ mass selection window by a factor of two. 
This also accounts for any uncertainty related to averaging over this mass in the amplitude fit.

\begin{figure}[tbhp]
  \begin{center}
    \includegraphics*[width=\figsize]{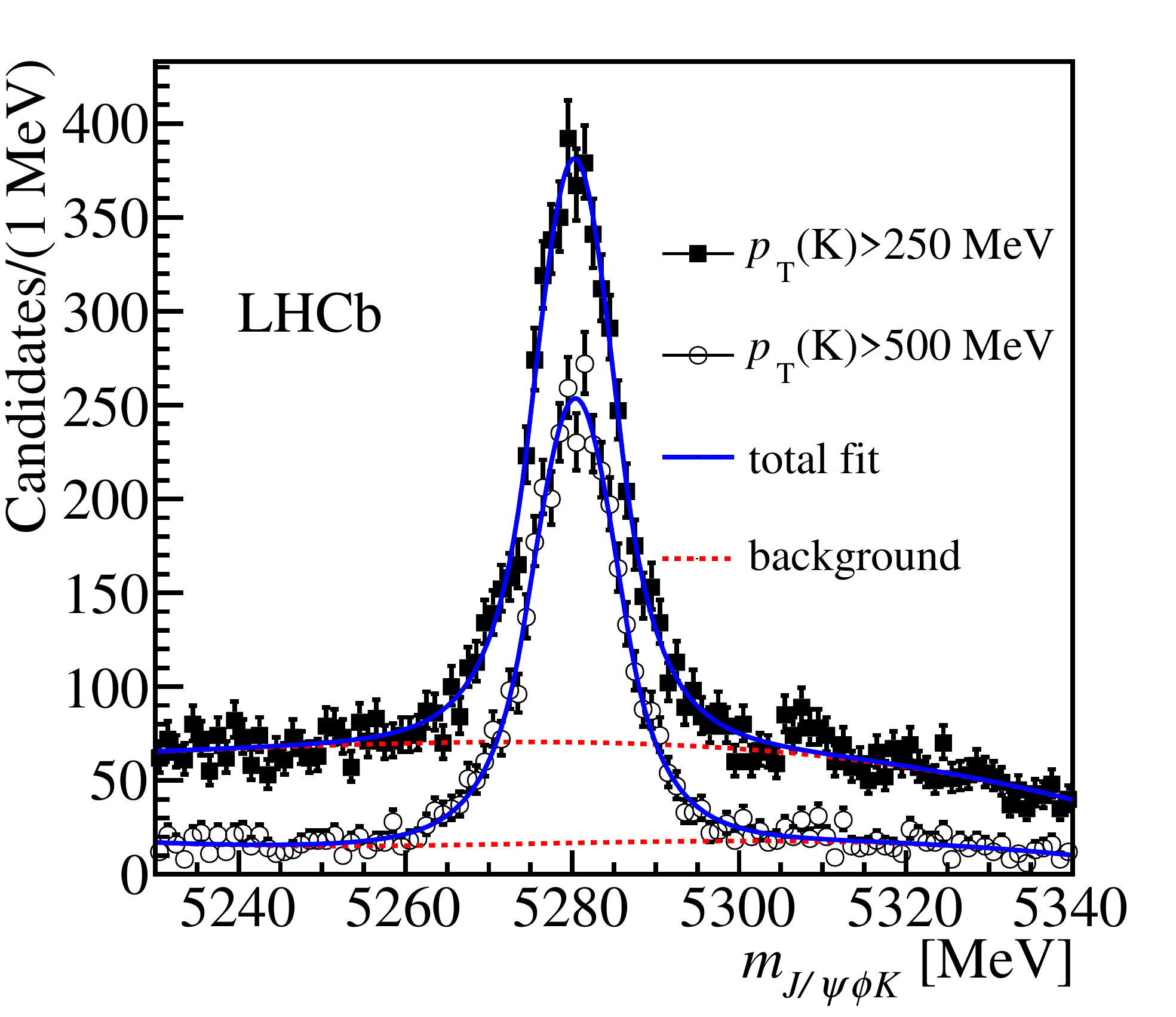}
  \end{center}
  \vskip-0.3cm\caption{\small
    Mass of $B^+\to\jpsi\phi K^+$ candidates in the data 
    with the $\pt(K)>250 \mev$ (default) and $\pt(K)>500 \mev$ selection requirements.
  \label{fig:KPT500mjpsiphik}
  }
\end{figure}

As a cross-check on both the background subtraction and the efficiency corrections
the minimal value of $\pt$ for kaon candidates 
is changed from $0.25 \gev$ to $0.5 \gev$, 
which reduces the background fraction by 
54\%\ ($\beta=10.4\%$) and the signal efficiency by 20\%, as illustrated in Fig.~\ref{fig:KPT500mjpsiphik}.
The mass projections of the fit are shown in 
Fig.~\ref{fig:KPT500masses}.
The fit results are within the assigned total uncertainties as shown at the bottom
rows of Tables~\ref{tab:ksysA}--\ref{tab:xsys}. 

More details on the systematic error evaluations can be found in Ref.~\cite{ThomasBritton:2016}.

\ifthenelse{\boolean{prl}}{ 
\begin{figure}[bthp] 
  \begin{center}
    \includegraphics*[width=\figsize]{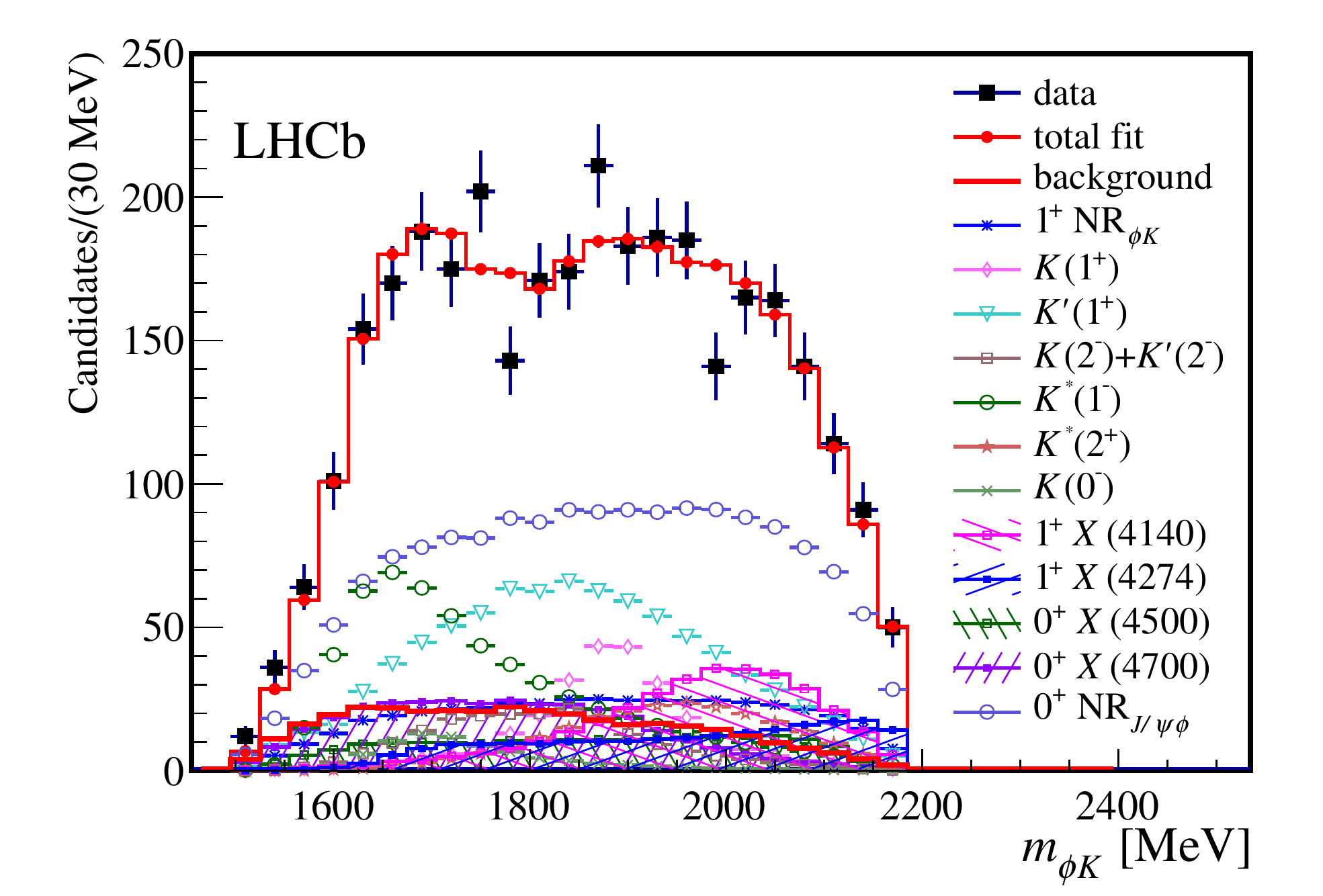} \\[-0.2cm]
    \includegraphics*[width=\figsize]{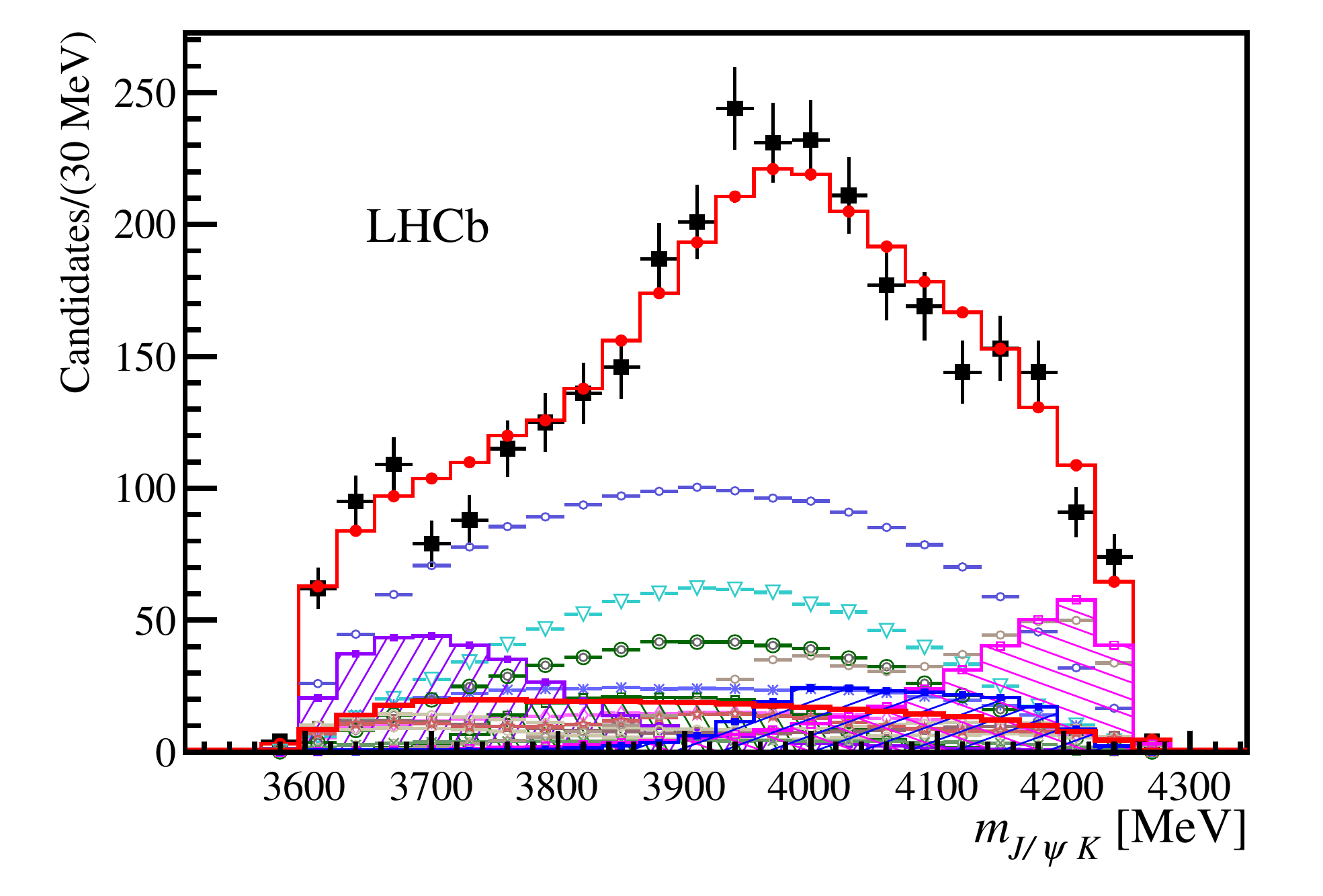} \\[-0.2cm]
    \includegraphics*[width=\figsize]{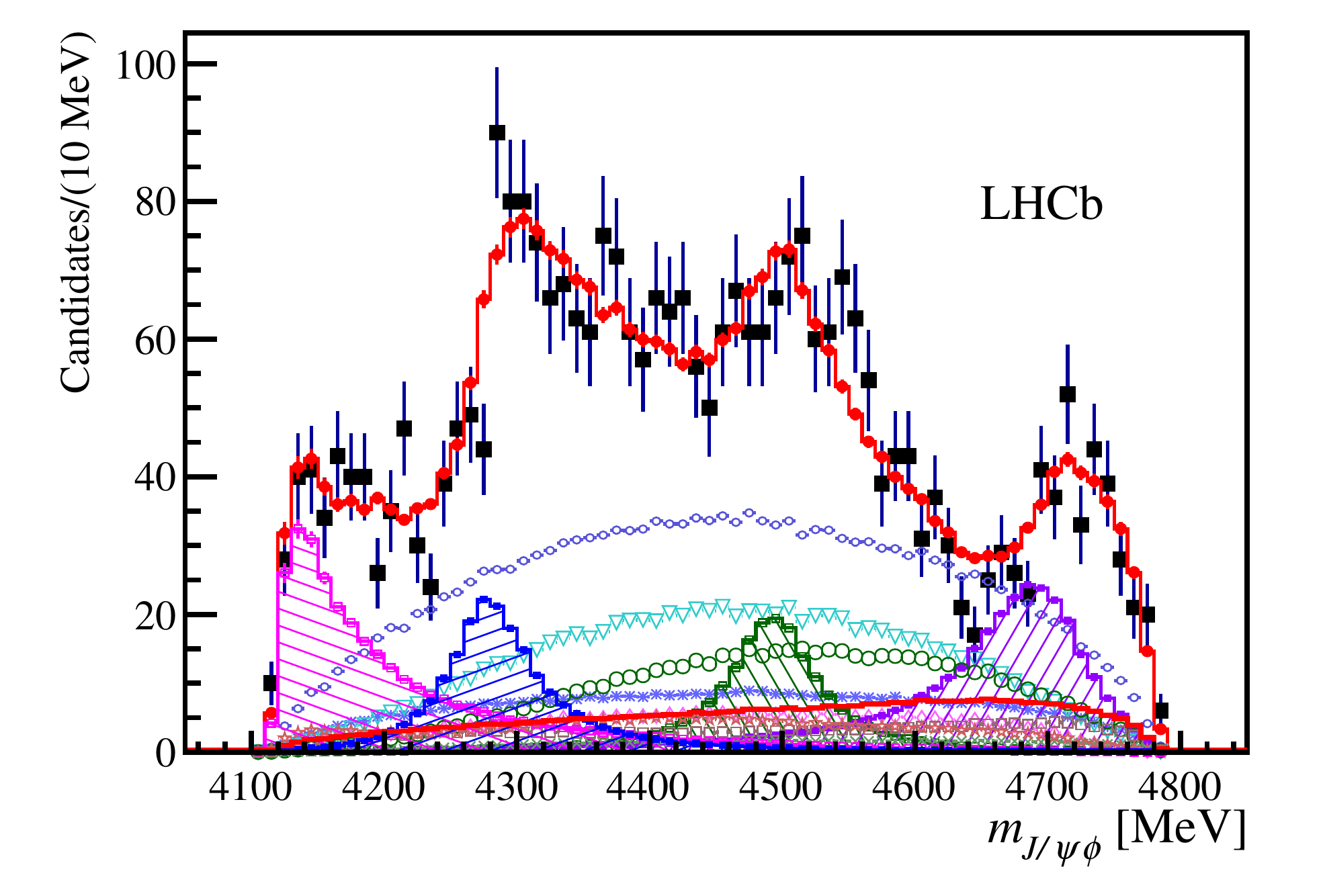} 
  \end{center}
  \vskip-0.3cm\caption{\small
    Distributions of (top) $\phiz K^+$, (middle) $\jpsi K^+$ and (bottom) $\jpsi\phi$ 
    invariant masses for the $\bujphik$ data 
    after changing the $\pt(K)>0.25 \gev$ requirement to $\pt(K)>0.5 \gev$,
    together with the fit projections.
    Compare to Fig.~\ref{fig:defmasses}. 
  \label{fig:KPT500masses}
  }
\end{figure} 
}{
\begin{figure}[bthp]  
  \begin{center}
    \hbox{\hskip-2.9cm \includegraphics*[width=0.75\figsize]{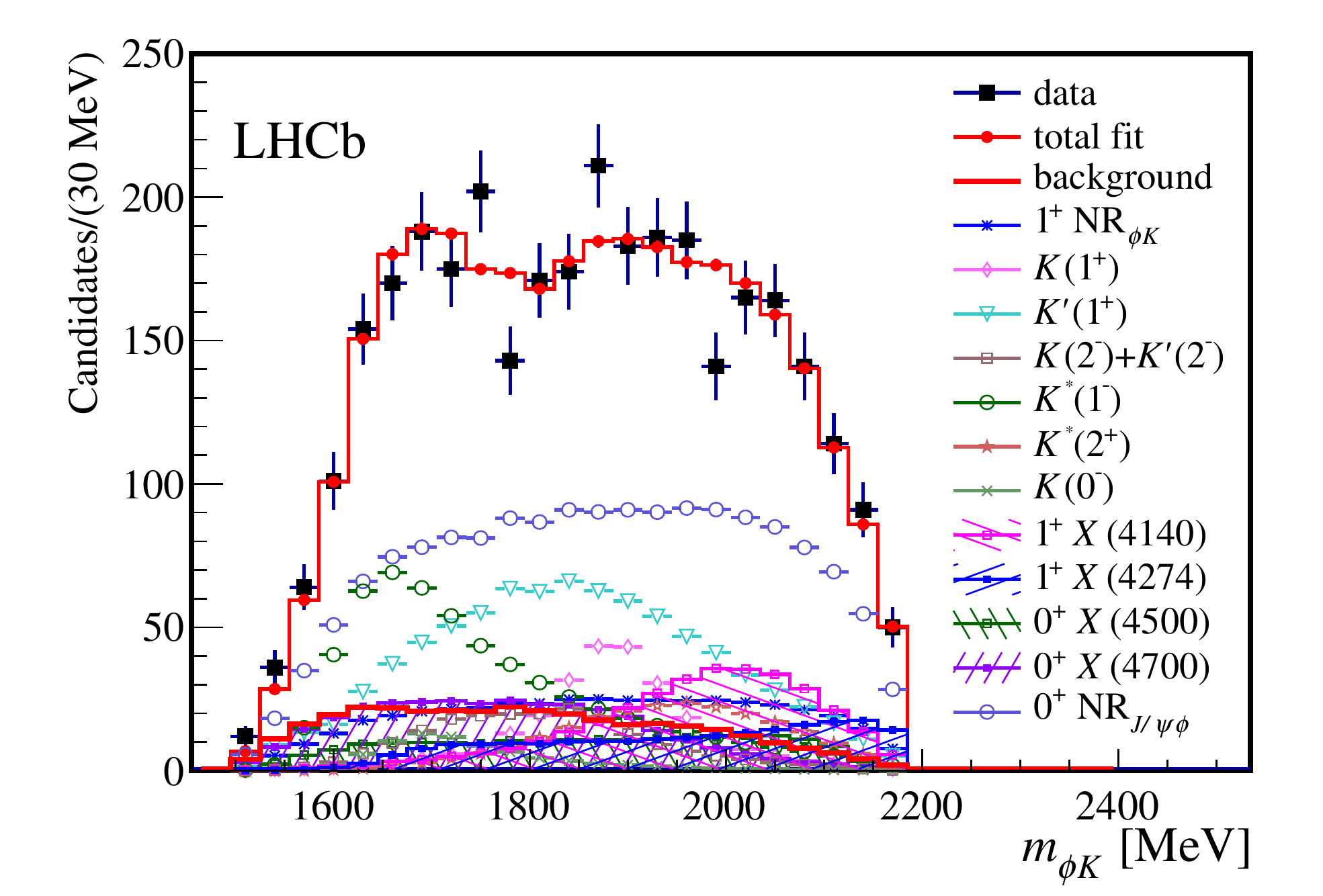} \hskip-0.6cm 
    \includegraphics*[width=0.75\figsize]{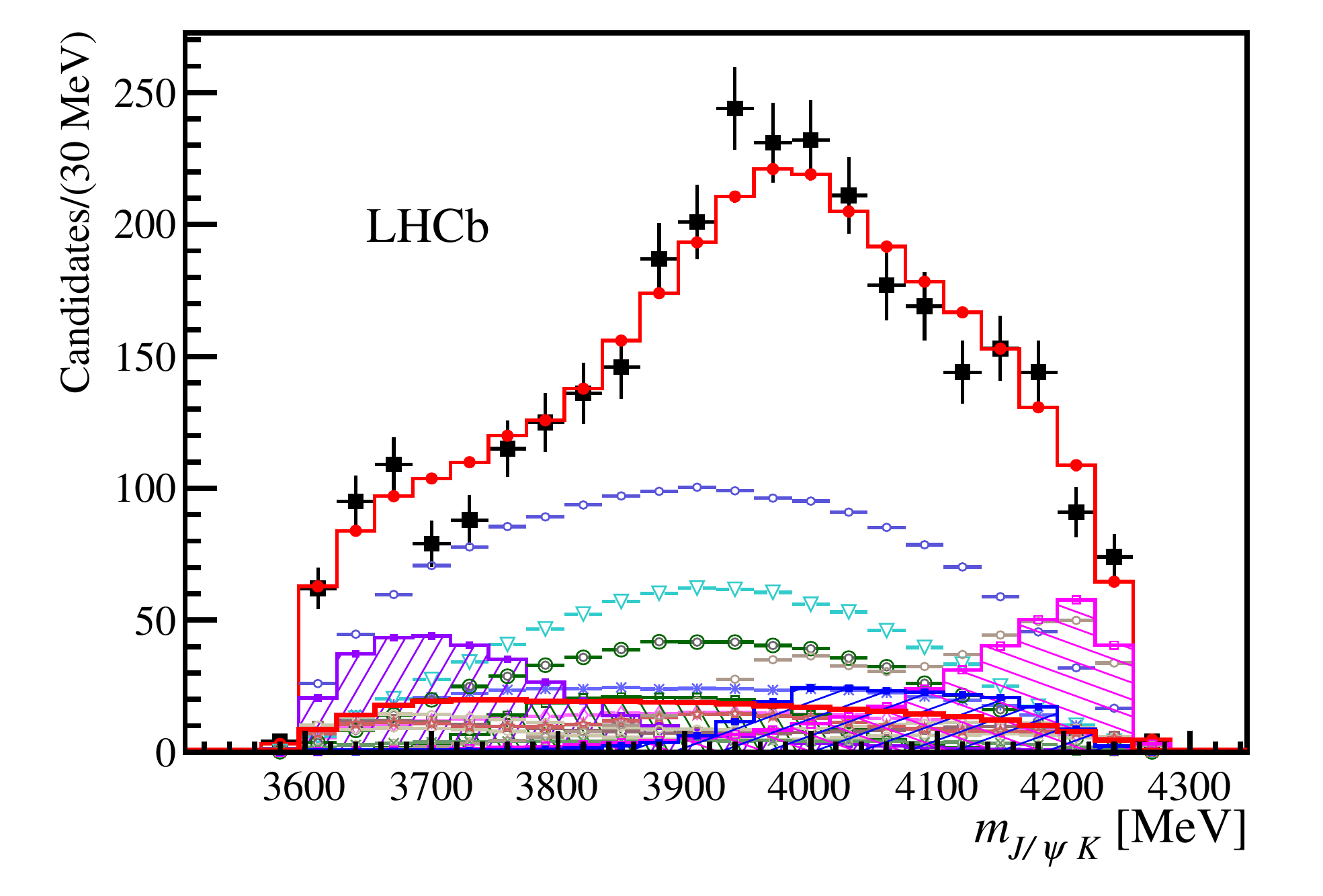} \hskip-2.1cm}\quad \\
    \includegraphics*[width=\figsize]{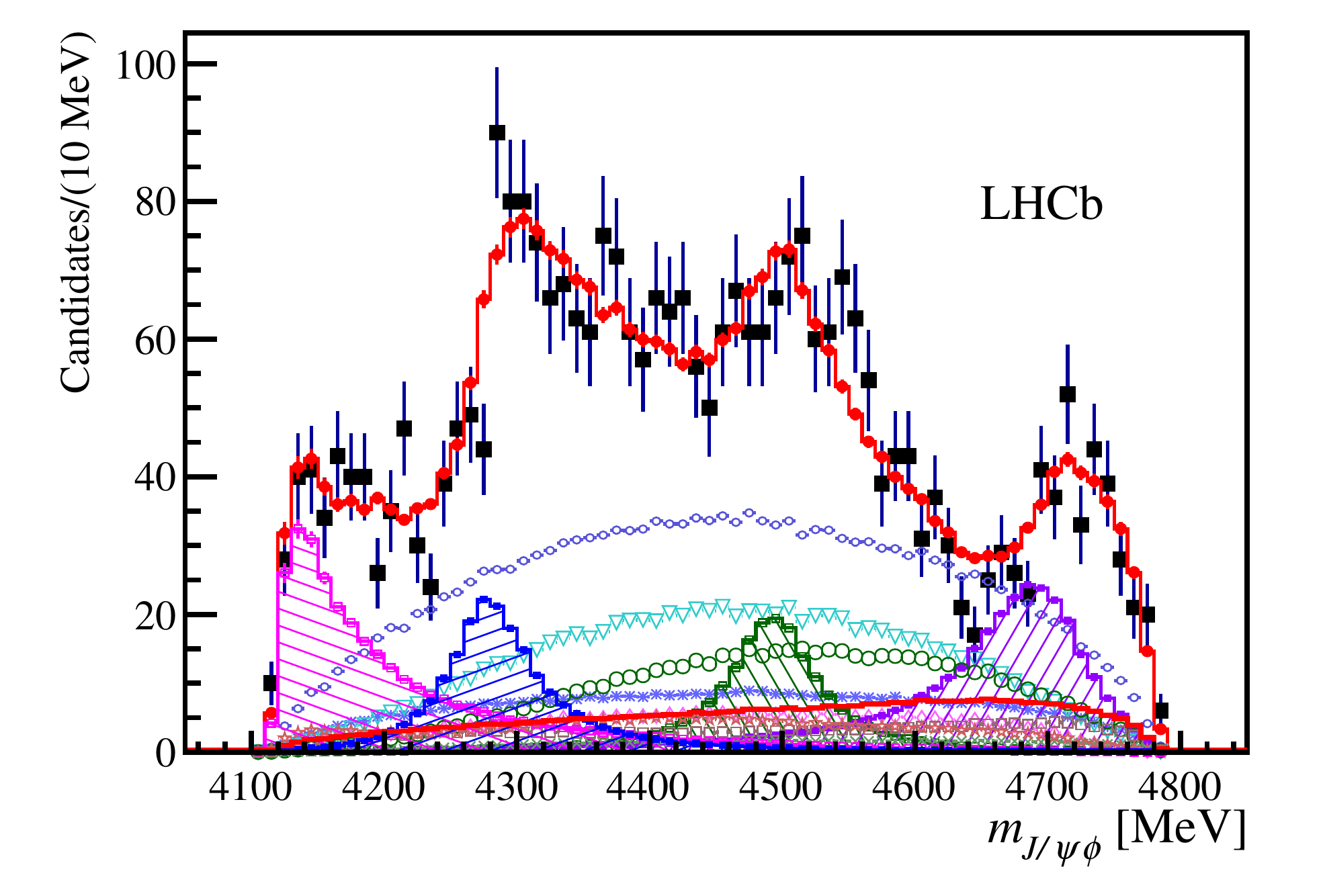} 
  \end{center}
  \vskip-0.3cm\caption{\small
    Distributions of (top) $\phiz K^+$, (middle) $\jpsi K^+$ and (bottom) $\jpsi\phi$ 
    invariant masses for the $\bujphik$ data 
    after changing the $\pt(K)>0.25 \gev$ requirement to $\pt(K)>0.5 \gev$,
    together with the fit projections.
    Compare to Fig.~\ref{fig:defmasses}. 
      \label{fig:KPT500masses}
  }
\end{figure} 
}

\def\ftiny{}

\begin{table*}[bthp]
\caption{\small Summary of the systematic uncertainties on the parameters of the $K^{*+}\to\phi K^+$ states with 
         $J^P=2^-$ and $1^+$. The kaon $\pt$ cross-check results are given at the bottom.
         All numbers for masses and widths are in \mev and fit fractions in \%.}
\label{tab:ksysA}
\def\1{\ifthenelse{\boolean{prl}}{}{\hbox{\quad\hskip-0.65cm}\quad}}
\def\2{\ifthenelse{\boolean{prl}}{}{\hbox{\quad\hskip-0.95cm}\quad}}
\def\3{\ifthenelse{\boolean{prl}}{}{\!\!}}
\def\6{\phantom{-}}
\ifthenelse{\boolean{prl}}{}{\begin{footnotesize}}
\renewcommand{\arraystretch}{1.2}
\resizebox{\textwidth}{!}{
%\begin{tabular}{|l||r|rrr|rrr||r|rrr|rrr|r|}
\begin{tabular} { l  r rrr rrr| r rrr rrr r}
\hline
{\3\ftiny sys.} & $2^-$ & \multicolumn{3}{c}{$K(2^-)$} & \multicolumn{3}{c|}{$K'(2^-)$} 
           & $1^+$ & \multicolumn{3}{c}{$K(1^+)$} & \multicolumn{3}{c}{$K'(1^+)$} & NR \\
\cline{2-16}
{\3\ftiny var.}  \1&\2  FF \1&\2 $M_0$ \1&\2 $\Gamma_0$ \1&\2 FF \1&\2 $M_0$ \1&\2 $\Gamma_0$ \1&\2 FF 
           \1&\2  FF \1&\2 $M_0$ \1&\2 $\Gamma_0$ \1&\2 FF \1&\2 $M_0$ \1&\2 $\Gamma_0$ \1&\2 FF \1&\2  FF \\
\hline
\hline

{\3\ftiny $K^*$} \1&\2 +1.2 \1&\2 +118.1 \1&\2 +194.8 \1&\2 +4.0 \1&\2 +16.2 \1&\2 +53.8 \1&\2 +4.4 \1&\2 +3.9 \1&\2 +150.8 \1&\2 +122.4 \1&\2 +15.6 \1&\2 +49.0 \1&\2 +159.5 \1&\2 +28.5 \1&\2 +34.4\\ 
{\3\ftiny model} \1&\2 $-$4.1 \1&\2 $-$22.3 \1&\2 $-$71.0 \1&\2 $-$8.6 \1&\2 $-$14.9 \1&\2 $-$38.5 \1&\2 $-$5.5 \1&\2 $-$7.5 \1&\2 $-$79.2 \1&\2 $-$196.2 \1&\2 $-$6.1 \1&\2 $-$53.8 \1&\2 $-$143.2 \1&\2 $-$27.2 \1&\2 $-$5.1\\

\hline

{\3\ftiny $L$} \1&\2 +0.7 \1&\2 +8.6 \1&\2 +54.1 \1&\2 +3.7  \1&\2 +5.5 \1&\2 +14.0 \1&\2 +3.5 \1&\2 +0.8 \1&\2 +22.3 \1&\2 +20.4 \1&\2 +3.4 \1&\2 +47.5 \1&\2 +37.7 \1&\2 +4.8 \1&\2 +5.0\\
{\3\ftiny var.} \1&\2 $-$1.5 \1&\2 $-$63.3 \1&\2 $-$127.9 \1&\2 $-$9.3 \1&\2 $-$31.2 \1&\2 $-$59.5 \1&\2 $-$8.6  \1&\2 $-$2.2 \1&\2 $-$48.6 \1&\2 $-$70.3 \1&\2 $-$0.9 \1&\2 $-$159.9 \1&\2 $-$72.5 \1&\2 $-$8.7 \1&\2 $-$2.2\\

\hline

{\3\ftiny NR exp.} \1&\2 +0.5 \1&\2 $-$4.8 \1&\2 $-$13.5 \1&\2 +0.4 \1&\2 $-$0.6 \1&\2 +8.6 \1&\2 +1.8 \1&\2 $-$2.2 \1&\2 $-$5.9 \1&\2 $-$3.7 \1&\2 +0.7 \1&\2 $-$21.4 \1&\2 $-$45.4 \1&\2 +0.8 \1&\2 +0.3\\

%\hline

{\3\ftiny $X$ cusp}   \1&\2 +0.0 \1&\2 +24.6 \1&\2 +42.2 \1&\2 +5.4 \1&\2 $-$0.8 \1&\2 +10.8 \1&\2 +3.8 \1&\2 +1.8 \1&\2 +4.5 \1&\2 +5.5 \1&\2 +4.4 \1&\2 $-$12.0 \1&\2 +40.6 \1&\2 +8.4 \1&\2 $-$0.3\\

%\hline
{\3\ftiny $\Gamma_{\rm tot}$} \1&\2 $-$0.2 \1&\2 +0.8 \1&\2 +38.7 \1&\2 $-$1.6 \1&\2 $-$1.9 \1&\2 $-$12.6 \1&\2 $-$2.4 \1&\2 +0.6 \1&\2 $-$29.5 \1&\2 +17.2 \1&\2 +0.9 \1&\2 $-$0.1 \1&\2 +7.1 \1&\2 $-$2.3 \1&\2 +2.2\\

%\hline
{\3\ftiny d=1.5} \1&\2 +0.1 \1&\2 +18.2 \1&\2 +67.2 \1&\2 $-$0.6 \1&\2 +2.7 \1&\2 +6.0 \1&\2 $-$1.5 \1&\2 +0.7 \1&\2 $-$17.4 \1&\2 $-$5.6 \1&\2 $-$1.0 \1&\2 +8.2 \1&\2 +13.9 \1&\2 $-$2.1 \1&\2 +1.7\\

{\3\ftiny d=5.0} \1&\2 +0.2 \1&\2 $-$7.2 \1&\2 $-$25.8 \1&\2$-$0.1  \1&\2 $-$1.0 \1&\2 $-$0.5 \1&\2 +1.3 \1&\2 $-$1.5 \1&\2 +12.2 \1&\2 $-$6.9 \1&\2 +0.5 \1&\2 $-$8.4 \1&\2 $-$42.8 \1&\2 $-$1.0 \1&\2 $-$1.5\\

%\hline
{\3\ftiny Left s.} \1&\2 +0.1 \1&\2 $-$4.2 \1&\2 $-$9.5 \1&\2 $-$0.2  \1&\2 $-$1.1 \1&\2 +2.0 \1&\2 +0.9 \1&\2 $-$1.0 \1&\2 +0.9 \1&\2 +0.2 \1&\2 +1.1 \1&\2 $-$8.7 \1&\2 $-$30.2 \1&\2 +0.9 \1&\2 +0.3\\

{\3\ftiny Right s.} \1&\2 $-$0.1 \1&\2 +3.2 \1&\2 +5.0 \1&\2 $-$0.4  \1&\2 +3.8 \1&\2 +0.1 \1&\2 $-$1.2 \1&\2 +1.2 \1&\2 $-$1.3 \1&\2 +12.5 \1&\2 $-$0.4 \1&\2 +11.6 \1&\2 +36.5 \1&\2 $-$1.5 \1&\2 $-$0.1\\

%\hline
$\beta$    \1&\2 +0.2 \1&\2 $-$8.1 \1&\2 $-$35.4 \1&\2 +1.7 \1&\2 $-$9.3 \1&\2 $-$6.7 \1&\2 +2.6 \1&\2 $-$2.7 \1&\2 +28.0 \1&\2 $-$8.2 \1&\2 +4.0 \1&\2 $-$23.4 \1&\2 $-$63.0 \1&\2 +4.8 \1&\2 $-$0.8\\

%\hline
{\3\ftiny No $w^{MC}$} \1&\2 $-$0.8 \1&\2 $-$0.2 \1&\2 +0.4 \1&\2 $-$1.1 \1&\2 +0.0 \1&\2 $-$0.5 \1&\2 $-$1.5 \1&\2 $-$0.8 \1&\2 +1.9 \1&\2 +1.2 \1&\2 +0.1 \1&\2 +0.6 \1&\2 +1.8 \1&\2 $-$0.7 \1&\2 +0.7\\

%\hline
{\3\ftiny $\phi$ window}  \1&\2 $-$1.0 \1&\2 $-$25.0 \1&\2 $-$27.2 \1&\2 $-$2.6 \1&\2 $-$1.1 \1&\2 +41.2 \1&\2 $-$1.4 \1&\2 $-$2.7 \1&\2 $-$11.3 \1&\2 $-$36.5 \1&\2 +0.0 \1&\2 $-$15.2 \1&\2 $-$23.1 \1&\2 +6.0 \1&\2 $-$1.9\\
%\hline
\hline
{\3\ftiny Total} \1&\2 +1.5 \1&\2 +122.3 \1&\2 +220.7 \1&\2 +7.7  \1&\2 +17.7 \1&\2 +82.0 \1&\2 +7.2  \1&\2 +4.7 \1&\2 +153.0 \1&\2 +138.0 \1&\2 +16.7 \1&\2 +69.7  \1&\2 +173.5 \1&\2 +31.3 \1&\2 +34.5\\
{\3\ftiny sys.}  \1&\2 $-$4.6 \1&\2 $-$76.5  \1&\2 $-$154.3 \1&\2 $-$13.3 \1&\2 $-$34.7 \1&\2 $-$72.0 \1&\2 $-$10.9 \1&\2 $-$9.2 \1&\2 $-$100.5 \1&\2 $-$214.8 \1&\2 $-$6.3  \1&\2 $-$172.3 \1&\2 $-$177.9 \1&\2 $-$28.8 \1&\2 $-$6.4\\
\hline
{\3\ftiny Stat.}     \1&\2  {\6}2.8     \1&\2  {\6}34.9 \1&\2 {\6}116.3 \1&\2 {\6}11.0 \1&\2   {\6}26.6\1&\2 {\6}58.1\1&\2 {\6}11.2      
                \1&\2   {\6}8.1     \1&\2  {\6}59.0 \1&\2 {\6}157.0\1&\2  {\6}10.3 \1&\2     {\6}65.0 \1&\2  {\6}170.3\1&\2 {\6}20.4 \1&\2    {\6}13.1       \\
\hline
\hline
{\3\ftiny $\pt^K\!\!\ftiny >$500} \1&\2 $-$2.7 \1&\2 $-$0.4 \1&\2 +4.9 \1&\2 $-$3.7 \1&\2 $-$10.1 \1&\2 $-$67.0 \1&\2 $-$5.7 \1&\2 +6.4 \1&\2 +95.2 \1&\2 $-$238.7 \1&\2 $-$3.7 \1&\2 $-$87.7 \1&\2 +33.6 \1&\2 $-$3.8 \1&\2 +4.7\\
\hline
\end{tabular}
}
\ifthenelse{\boolean{prl}}{}{\end{footnotesize}}
\end{table*}

\begin{table*}[bthp]
\caption{\small Summary of the systematic uncertainties on the parameters of the $K^{*+}\to\phi K^+$ states with
$J^P=0^-$, $1^-$ and $2^+$. The kaon $\pt$ cross-check results are given at the bottom.
All numbers for masses and widths are in \mev and fit fractions in \%.}
\label{tab:ksysB}
\def\1{\ifthenelse{\boolean{prl}}{}{\hbox{\quad\hskip-0.6cm}\quad}}
\def\2{\ifthenelse{\boolean{prl}}{}{\hbox{\quad\hskip-0.95cm}\quad}}
\def\3{\ifthenelse{\boolean{prl}}{}{\!\!}}
\begin{center}
\ifthenelse{\boolean{prl}}{}{\begin{footnotesize}}
\renewcommand{\arraystretch}{1.2}
%\begin{tabular}{|l||rrr||rrr||rrr|}
\begin{tabular}{ l rrr | rrr | rrr}
\hline
sys. &  \multicolumn{3}{c|}{$K^*(1^-)$}   & \multicolumn{3}{c|}{$K(0^-)$} 
           &  \multicolumn{3}{c}{$K^*(2^+)$} \\
\cline{2-10}
var.  \1&\2   $M_0$ \1&\2 $\Gamma_0$ \1&\2 FF \1&\2 $M_0$ \1&\2 $\Gamma_0$ \1&\2 FF 
           \1&\2   $M_0$ \1&\2 $\Gamma_0$ \1&\2 FF \\
\hline
\hline

$\Kst$ \1&\2 +19.9 \1&\2 +31.4 \1&\2 +2.6 \1&\2 +54.8 \1&\2 +236.9 \1&\2 +1.7 \1&\2 +214.3 \1&\2 +805.2 \1&\2 +1.6\\ 
model \1&\2 $-$33.1 \1&\2 $-$141.0 \1&\2 $-$2.7 \1&\2 $-$90.2 \1&\2 $-$96.3 \1&\2 $-$1.7 \1&\2 $-$66.9 \1&\2 $-$223.8 \1&\2 $-$0.6\\

\hline
%\hline
$L$      \1&\2 +14.2 \1&\2 +59.3 \1&\2 +1.8 \1&\2 +12.8 \1&\2 +51.6 \1&\2 +0.7 \1&\2 +52.0 \1&\2 +172.3 \1&\2 +0.3\\
var.     \1&\2 $-$17.7 \1&\2 $-$44.7 \1&\2 $-$0.2 \1&\2 $-$44.4 \1&\2 $-$31.1 \1&\2 $-$0.2 \1&\2 $-$19.1 \1&\2 $-$107.4 \1&\2 $-$0.3\\
\hline
%\hline
NR exp. \1&\2 +3.3 \1&\2 +11.5 \1&\2 +0.2 \1&\2 $-$22.9 \1&\2 +36.3 \1&\2 +0.4 \1&\2 $-$13.7 \1&\2 $-$65.1 \1&\2 +0.0\\

%\hline
%\hline
$X$ cusp \1&\2 +4.5 \1&\2 +5.5 \1&\2 $-$1.2 \1&\2 +7.8 \1&\2 +11.4 \1&\2 +0.1 \1&\2 +26.5 \1&\2 +6.1 \1&\2 $-$0.2\\

%\hline
%\hline
$\Gamma_{\rm tot}$ \1&\2 $-$101.5 \1&\2 $-$93.1 \1&\2 +0.2 \1&\2 $-$2.8 \1&\2 $-$6.2 \1&\2 $-$0.1 \1&\2 $-$167.6 \1&\2 $-$230.0 \1&\2 +0.3\\
%\hline
%\hline
d=1.5 \1&\2 +21.1 \1&\2 +121.7 \1&\2 +0.0 \1&\2 +12.1 \1&\2 +2.5 \1&\2 $-$0.1 \1&\2 +102.2 \1&\2 +806.2 \1&\2 +0.0\\
%\hline
d=5.0 \1&\2 $-$4.9 \1&\2 $-$21.0 \1&\2 +0.0 \1&\2 $-$10.3 \1&\2 +6.3 \1&\2 +0.2 \1&\2 $-$72.0 \1&\2 $-$242.5 \1&\2 +0.0\\
%\hline
%\hline
Left s. \1&\2 +2.7 \1&\2 +7.7 \1&\2 +0.0 \1&\2 $-$12.6 \1&\2 20.1 \1&\2 +0.2 \1&\2 $-$17.9 \1&\2 $-$28.8 \1&\2 +0.2\\
%\hline
Right s. \1&\2 $-$3.0 \1&\2 +7.7 \1&\2 +0.0 \1&\2 +10.0 \1&\2 $-$23.5 \1&\2 $-$0.2 \1&\2 +19.2 \1&\2 +24.7 \1&\2 $-$0.2\\
%\hline
%\hline
$\beta$  \1&\2 +2.2 \1&\2 $-$4.1 \1&\2 +0.1 \1&\2 $-$43.0 \1&\2 +32.2 \1&\2 +0.5 \1&\2 $-$18.5 \1&\2 +1.1 \1&\2 +0.4\\
%\hline
%\hline
No $w^{MC}$ \1&\2 +0.2 \1&\2 $-$0.4 \1&\2 +0.1 \1&\2 +1.0 \1&\2 $-$2.4 \1&\2 $-$0.4 \1&\2 $-$0.4 \1&\2 $-$3.1 \1&\2 $-$0.2\\
%\hline
%\hline
$\phi$ window \1&\2 +0.5 \1&\2 $-$28.9 \1&\2 $-$1.8 \1&\2 $-$33.6 \1&\2 +94.5 \1&\2 +0.9 \1&\2 $-$97.0 \1&\2 $-$258.9 \1&\2 +0.2\\
\hline
%\hline

Total \1&\2 +32.9 \1&\2 +139.8 \1&\2 +3.2 \1&\2 +59.0 \1&\2 +280.2 \1&\2 +2.3 \1&\2 +245.2 \1&\2 +1152.7 \1&\2 +1.7\\
sys. \1&\2 $-$108.4 \1&\2 $-$180.7 \1&\2 $-$3.9 \1&\2 $-$114.8 \1&\2 $-$104.1 \1&\2 $-$1.8 \1&\2 $-$239.7 \1&\2 $-$559.0 \1&\2 $-$0.7\\

\hline
%\hline
Stat.       \1&\2   19.9  \1&\2  74.7 \1&\2  1.9     \1&\2   43.2 \1&\2  90.4 \1&\2  1.1
                  \1&\2   94.2  \1&\2 310.6 \1&\2  0.8    \\
\hline
\hline
$\pt^K\!\!>$500 \1&\2 $-$15.6 \1&\2 $-$47.1 \1&\2 $-$0.2 \1&\2 $-$161.9 \1&\2 $-$2.4 \1&\2 $-$0.2 \1&\2 $-$10.1 \1&\2 $-$102.2 \1&\2 $-$0.1\\
\hline
\end{tabular}
\ifthenelse{\boolean{prl}}{}{\end{footnotesize}}
\end{center}
\end{table*}

\begin{table*}[bthp]
\caption{\small
Summary of the systematic uncertainties on the parameters of the $X\to\jpsi\phi$ states.
The kaon $\pt$ cross-check results are given at the bottom.
All numbers for masses and widths are in \mev and fit fractions in \%.}
\label{tab:xsys}
\def\1{\ifthenelse{\boolean{prl}}{}{\hbox{\quad\hskip-0.6cm}\quad}}
\def\2{\ifthenelse{\boolean{prl}}{}{\hbox{\quad\hskip-0.95cm}\quad}}
\def\3{\ifthenelse{\boolean{prl}}{}{\!\!}}
\ifthenelse{\boolean{prl}}{}{\begin{footnotesize}}
\renewcommand{\arraystretch}{1.0}
\resizebox{\textwidth}{!}{
%\begin{tabular}{|l||r|rrr|rrr||r|rrr|rrr|r|}
\begin{tabular}{ l  r rrr rrr |r rrr rrr r }
\hline
{\3  sys.} & $1^+$ & \multicolumn{3}{c}{$\Xone$} & \multicolumn{3}{c|}{$\Xtwo  $} 
           & $0^+$ & \multicolumn{3}{c}{$\Xthree$} & \multicolumn{3}{c}{$\Xfour $} & NR \\
\cline{2-16}
{\3  var.}  \1&\2  FF \1&\2 $M_0$ \1&\2 $\Gamma_0$ \1&\2 FF \1&\2 $M_0$ \1&\2 $\Gamma_0$ \1&\2 FF 
           \1&\2  FF \1&\2 $M_0$ \1&\2 $\Gamma_0$ \1&\2 FF \1&\2 $M_0$ \1&\2 $\Gamma_0$ \1&\2 FF \1&\2  FF \\
\hline
\hline

{\3  $\Kst$}  \1&\2 +2.0 \1&\2 +3.6 \1&\2 +17.1 \1&\2 +2.2 \1&\2 +11.2 \1&\2 +7.9 \1&\2 +1.4 \1&\2 +1.8 \1&\2 +9.3 \1&\2 +13.8 \1&\2 +2.0 \1&\2 +7.5 \1&\2 +38.6 \1&\2 +6.7 \1&\2 +8.0\\
{\3  model} \1&\2 $-$1.7 \1&\2 $-$2.6 \1&\2 $-$11.7 \1&\2 $-$1.9 \1&\2 $-$2.5 \1&\2 $-$8.5 \1&\2 $-$1.5 \1&\2 $-$11.0 \1&\2 $-$8.6 \1&\2 $-$16.6 \1&\2 $-$1.7 \1&\2 $-$18.9 \1&\2 $-$13.5 \1&\2 $-$4.8 \1&\2 $-$16.6\\

%\hline
\hline

{\3  $L$}  \1&\2 +3.2 \1&\2 +2.2 \1&\2 +7.3 \1&\2 +2.1 \1&\2 +10.6 \1&\2 +1.4 \1&\2 +1.0 \1&\2 +0.3 \1&\2 +1.3 \1&\2 +10.8 \1&\2 +1.7 \1&\2 +9.0 \1&\2 +12.4 \1&\2 +1.5 \1&\2 +1.2\\
 var.      \1&\2 +0.0 \1&\2 $-$1.2 \1&\2 $-$6.2 \1&\2 $-$0.5 \1&\2 $-$0.8 \1&\2 $-$4.6 \1&\2 $-$1.2 \1&\2 $-$4.7 \1&\2 $-$9.6 \1&\2 $-$11.2 \1&\2 $-$1.6 \1&\2 $-$6.8 \1&\2 $-$24.9 \1&\2 $-$0.8 \1&\2 $-$8.5\\

%\hline
\hline

{\3  NR exp.} \1&\2 +0.4 \1&\2 $-$0.2 \1&\2 $-$0.1 \1&\2 +0.4 \1&\2 $-$0.2 \1&\2 +0.6 \1&\2 +0.8 \1&\2 $-$1.7 \1&\2 +6.3 \1&\2 +0.3 \1&\2 +0.2 \1&\2 +7.1 \1&\2 $-$15.7 \1&\2 $-$1.7 \1&\2 $-$9.1\\

%\hline
%\hline

{\3 $X$ cusp} \1&\2 +2.2 \1&\2       \1&\2      \1&\2 +0.9 \1&\2 +6.4 \1&\2 $-$5.4 \1&\2 $-$1.4 \1&\2  $-$1.2  \1&\2 +0.0 \1&\2 +1.2 \1&\2 +0.2 \1&\2 +1.9 \1&\2 $-$2.5 \1&\2 0.5 \1&\2 $-$1.6 \\

%\hline
%\hline

{\3  $\Gamma_{\rm tot}$} \1&\2 $-$0.6 \1&\2 +0.2 \1&\2 +1.5 \1&\2 $-$0.4 \1&\2 +3.2 \1&\2 +0.2 \1&\2 $-$0.3 \1&\2 +0.1 \1&\2 +0.8 \1&\2 $-$0.1 \1&\2 $-$0.3 \1&\2 +0.9 \1&\2 $-$5.8 \1&\2 $-$0.9 \1&\2 $-$1.1\\
%\hline
%\hline

{\3  d=1.5}  \1&\2 $-$0.9 \1&\2 +1.1 \1&\2 +5.3 \1&\2 $-$0.5 \1&\2 +2.2 \1&\2 +0.8 \1&\2 $-$0.4 \1&\2 +0.5 \1&\2 +1.7 \1&\2 +3.2 \1&\2 +0.1 \1&\2 $-$0.1 \1&\2 +1.7 \1&\2 +0.0 \1&\2 +1.1 \\
%\hline
{\3  d=5.0}   \1&\2 +1.1 \1&\2 $-$0.2 \1&\2 $-$2.0 \1&\2 +0.6 \1&\2 +0.2 \1&\2 $-$0.8 \1&\2 +0.3 \1&\2 $-$0.5 \1&\2 $-$1.0 \1&\2 $-$3.1 \1&\2 $-$0.1 \1&\2 $-$1.2 \1&\2 $-$3.2 \1&\2 $-$0.7 \1&\2 $-$2.5\\

%\hline
%\hline

{\3  Left s.}  \1&\2 +0.1 \1&\2 $-$0.4 \1&\2 $-$2.0 \1&\2 +0.1 \1&\2 +0.4 \1&\2 $-$0.8 \1&\2 +0.1 \1&\2 $-$0.5 \1&\2 $-$2.4 \1&\2 $-$2.6 \1&\2 $-$0.2 \1&\2 $-$1.5 \1&\2 $-$3.1 \1&\2 $-$0.7 \1&\2 $-$1.2\\

%\hline
{\3  Right s.} \1&\2 $-$0.3 \1&\2 +0.3 \1&\2 +2.6 \1&\2 $-$0.2 \1&\2 $-$0.6 \1&\2 +1.0 \1&\2 +0.0 \1&\2 +0.5 \1&\2 +3.7 \1&\2 +3.4 \1&\2 +0.4 \1&\2 +1.2 \1&\2 +7.0 \1&\2 +0.8 \1&\2 +1.6\\

%\hline
%\hline

$\beta$   \1&\2 +1.2 \1&\2 $-$0.6 \1&\2 $-$3.6 \1&\2 +1.2 \1&\2 +1.7 \1&\2 $-$0.7 \1&\2 +0.9 \1&\2 $-$2.5 \1&\2 $-$4.6 \1&\2 $-$11.1 \1&\2 $-$0.5 \1&\2 $-$3.9 \1&\2 $-$6.1 \1&\2 $-$1.4 \1&\2 $-$1.4 \\

%\hline
%\hline

{\3  No $w^{MC}$} \1&\2 +1.6 \1&\2 +0.0 \1&\2 +0.0 \1&\2 +0.1 \1&\2 +0.0 \1&\2 +0.0 \1&\2 +1.4 \1&\2 +1.7 \1&\2 +0.0 \1&\2 +0.2 \1&\2 +0.2 \1&\2 +0.1 \1&\2 +0.0 \1&\2 +1.2 \1&\2 +2.7\\

%\hline
%\hline

{\3  $\phi$ window} \1&\2 +2.5 \1&\2 +1.1 \1&\2 +4.7 \1&\2 +2.4 \1&\2 $-$1.6 \1&\2 +1.4 \1&\2 +1.8 \1&\2 +4.2 \1&\2 $-$4.3 \1&\2 +7.1 \1&\2 +1.2 \1&\2 $-$9.3 \1&\2 +5.8 \1&\2 +0.7 \1&\2 +4.7\\

%\hline
\hline

{\3 Total} \1&\2 +5.9 \1&\2 +4.6 \1&\2 +20.7 \1&\2 +4.7 \1&\2 +17.2 \1&\2 +8.4 \1&\2 +3.5 \1&\2 +6.5 \1&\2 +12.0 \1&\2 +20.8 \1&\2 +3.2 \1&\2 +13.9 \1&\2 +42.0 \1&\2 +7.2 \1&\2 +11.0\\
{\3 sys.}   \1&\2 $-$2.1 \1&\2 $-$2.8 \1&\2 $-$13.5 \1&\2 $-$2.0 \1&\2 $-$3.6 \1&\2 $-$11.1 \1&\2 $-$2.4 \1&\2 $-$6.7 \1&\2 $-$14.5 \1&\2 $-$20.4 \1&\2 $-$2.3 \1&\2 $-$24.1 \1&\2 $-$33.3 \1&\2 $-$5.3 \1&\2 $-$21.0\\

%\hline
\hline
{\3  Stat.}       \1&\2   2.8      \1&\2  4.5  \1&\2 20.7 \1&\2   3.2    \1&\2   8.3 \1&\2 10.9  \1&\2   2.5  \1&\2   5.1     \1&\2  11.1 \1&\2 21.2 \1&\2  2.4 \1&\2 10.1  \1&\2  30.7     \1&\2  4.9   \1&\2    10.7       \\
\hline
\hline
{\!\!  $\pt^K\!\!>$500} \1&\2 $-$1.3 \1&\2 +1.6 \1&\2 +1.7 \1&\2 $-$2.7 \1&\2 +7.8 \1&\2 +12.2 \1&\2 +0.2 \1&\2 $-$9.6 \1&\2 $-$10.9 \1&\2 $-$18.6 \1&\2 $-$3.2 \1&\2 $-$4.7 \1&\2 $-$12.7 \1&\2 $-$6.6 \1&\2 $-$17.1\\
\hline
\end{tabular}
}
%}
%}
\ifthenelse{\boolean{prl}}{}{\end{footnotesize}}
\end{table*} 

\FloatBarrier

\section{Spin analysis for the $X\to\jpsi\phi$ states}
\label{sec:spinana}

To determine the quantum numbers of each $X$ state, fits are done under 
alternative $J^{PC}$ hypotheses.
The likelihood-ratio test is used to quantify rejection of these hypotheses.
Since different spin-parity assignments are represented by different functions 
in the angular part of the fit PDF, they represent separate hypotheses.
For two models representing separate hypotheses,
assuming a $\chi^2$ distribution with one degree of freedom 
for $\dll$ under the disfavored $J^{PC}$ hypothesis 
gives a lower limit on the significance of its 
rejection \cite{james2006statistical}.
The results for the default fit approach are shown in Table~\ref{tab:defaultSpinAnalysis}.
The $J^{PC}$ values of the $\Xone$ and $\Xtwo$ states are both determined to be $1^{++}$ with $7.6\sigma$ and $6.4\sigma$ significance, respectively.
The quantum numbers of $\Xthree$ and of $\Xfour$ states are both established to be $0^{++}$ at $5.2\sigma$ and $4.9\sigma$ level,
respectively.

The separation from the alternative $J^{PC}$ hypothesis with 
likelihood closest to that for the favored  quantum numbers 
in the default fit 
is studied for each state under the 
fit variations which have dominant effects on the resonance parameters
as shown in Table~\ref{tab:sysSpin}.
The lowest values are taken for the final significances of the quantum number determinations:
$5.7\sigma$ for $\Xone$, $5.8\sigma$ for $\Xtwo$, $4.0\sigma$ for $\Xthree$ and $4.5\sigma$ for $\Xfour$.
 
\begin{table}[tbhp]
\caption{\small
         Statistical significance of $J^{PC}$ preference for the $X$ states in the default model.
         The lowest significance value for each state is highlighted. 
}
\label{tab:defaultSpinAnalysis}
\begin{center}
\begin{tabular}{crrrr}
\hline
 $J^{PC}$ & $\Xone$ & $\Xtwo$ & $\Xthree$ & $\Xfour$ \\
\hline\hline
$0^{++}$ &10.3$\sigma$ &7.8$\sigma$ &preferred & preferred\\
$0^{-+}$ &12.5$\sigma$ &7.0$\sigma$ &8.1$\sigma$ & 8.2$\sigma$\\
$1^{++}$ &preferred & preferred & $\mathbf{5.2\sigma}$ & $\mathbf{4.9\sigma}$\\
$1^{-+}$ &10.4$\sigma$ & $\mathbf{6.4\sigma}$ &6.5$\sigma$ & 8.3$\sigma$\\
$2^{++}$ &$\mathbf{7.6\sigma}$ 
                   &7.2$\sigma$ &  5.6$\sigma$ &6.8$\sigma$\\
$2^{-+}$ &9.6$\sigma$ & $\mathbf{6.4\sigma}$ & 6.5$\sigma$ &6.3$\sigma$\\
\hline
\end{tabular}
\end{center}
\end{table}

\begin{table}[bthp]
\caption{\small
         Significance, in standard deviations, of $J^{PC}$ preference for the $X$ states for dominant systematic variations
         of the fit model. The label ``$L+n$'' specifies which $L$ value in 
         Eq.~(\ref{eq:resshapeBW}) is increased relative to its minimal value and 
         by how much ($n$). 
         The lowest significance value for each state is highlighted.}
\label{tab:sysSpin} 
\renewcommand{\arraystretch}{1.2}
\begin{center}
\begin{tabular}{lccccc}
\hline
systematic variation &  \multicolumn{1}{c}{$\Xone$} 
           &  \multicolumn{2}{c}{$\Xtwo$} 
           &  \multicolumn{1}{c}{$\Xthree$}
           &  \multicolumn{1}{c}{$\Xfour$} \\
\multicolumn{1}{r}{alternative $J^{PC}$}  &  \multicolumn{1}{c}{$2^{++}$} 
           &  \multicolumn{1}{c}{$1^{-+}$} 
           &  \multicolumn{1}{c}{$2^{-+}$} 
           &  \multicolumn{1}{c}{$1^{++}$} 
           & \multicolumn{1}{c}{$1^{++}$} \\
\hline\hline
default fit   & 7.6  &  6.4 & 6.4 & 5.2 &  4.9 \\
\hline
$K'(1^+)$ $L_{\Kst}$ + 2 & {\!\!12.2} & 6.2 & 7.4 & 5.4 &  5.1\\
$K_2(2^-)$ $L_{\Kst}$ + 2 & {\bf 5.7} & 6.0 & {\bf 5.8} & 5.2  & 4.5\\
${K^*_3(1780)}$ included & 6.2 & 6.6 & 6.3 & 4.9  & {\bf 4.5}\\
extra $K^{*}(1^-)$ included & 6.8 & 6.1 & {\bf 5.8} & 5.8 &  4.7\\
extra $K_2(2^-)$ included & 6.9 & 6.7 & 6.2 & {\bf 4.0} &  4.8\\
NR exp & 7.5 & 6.5 & 6.1 & 8.9 &  4.7\\
\hline
\end{tabular}
\end{center}
\end{table}

%\FloatBarrier
\clearpage

\section{Is $X(4140)$ a $D_s^{\pm}D_s^{*\mp}$ cusp?}
\label{sec:cusp}

While our $1^{++}$ assignment to $X(4140)$ and its large width 
rule out an interpretation as a $0^{++}$ or $2^{++}$ $D_s^{*+}D_s^{*-}$ molecule 
(for which $1^{++}$ is not allowed \cite{Liu:2009ei})
with large $\sim83 \mev$ binding energy as suggested by many authors
\cite{Liu:2009ei,Branz:2009yt,Albuquerque:2009ak,Ding:2009vd,Zhang:2009st},
such a structure could be formed by molecular forces in a $D_s^{\pm}D_s^{*\mp}$ pair 
in S-wave \cite{Swanson:2014tra,Karliner:2016ith}.
Since the sum of $D_s^{\pm}$ and $D_s^{*\mp}$ masses ($4080 \mev$) is below the $\jpsi\phi$ mass threshold 
($4116$ \mev), such a contribution would not be described by the Breit--Wigner function 
with a pole above that threshold. 
The investigation of all possible parameterizations for such contributions, which are model 
dependent, goes beyond the scope of this analysis. 
However, we attempt a fit with a simple threshold cusp parameterization proposed by
Swanson (Ref.~\cite{Swanson:2015bsa} and private communications), 
in which the introduction of an exponential form factor, with a momentum scale 
($\beta_0$) characterizing the hadron size,  
makes the cusp peak slightly above the sum of masses of the 
rescattering mesons.
While controversial \cite{Guo:2014iya}, 
this model provided a successful description of
the $Z_c(3900)^+$ and $Z_c(4025)^+$ 
exotic meson candidates with masses peaking slightly above the molecular 
thresholds \cite{Swanson:2015bsa}.

\begin{figure}[tbhp]  
  \begin{center}
    \includegraphics*[width=1.0\figsize]{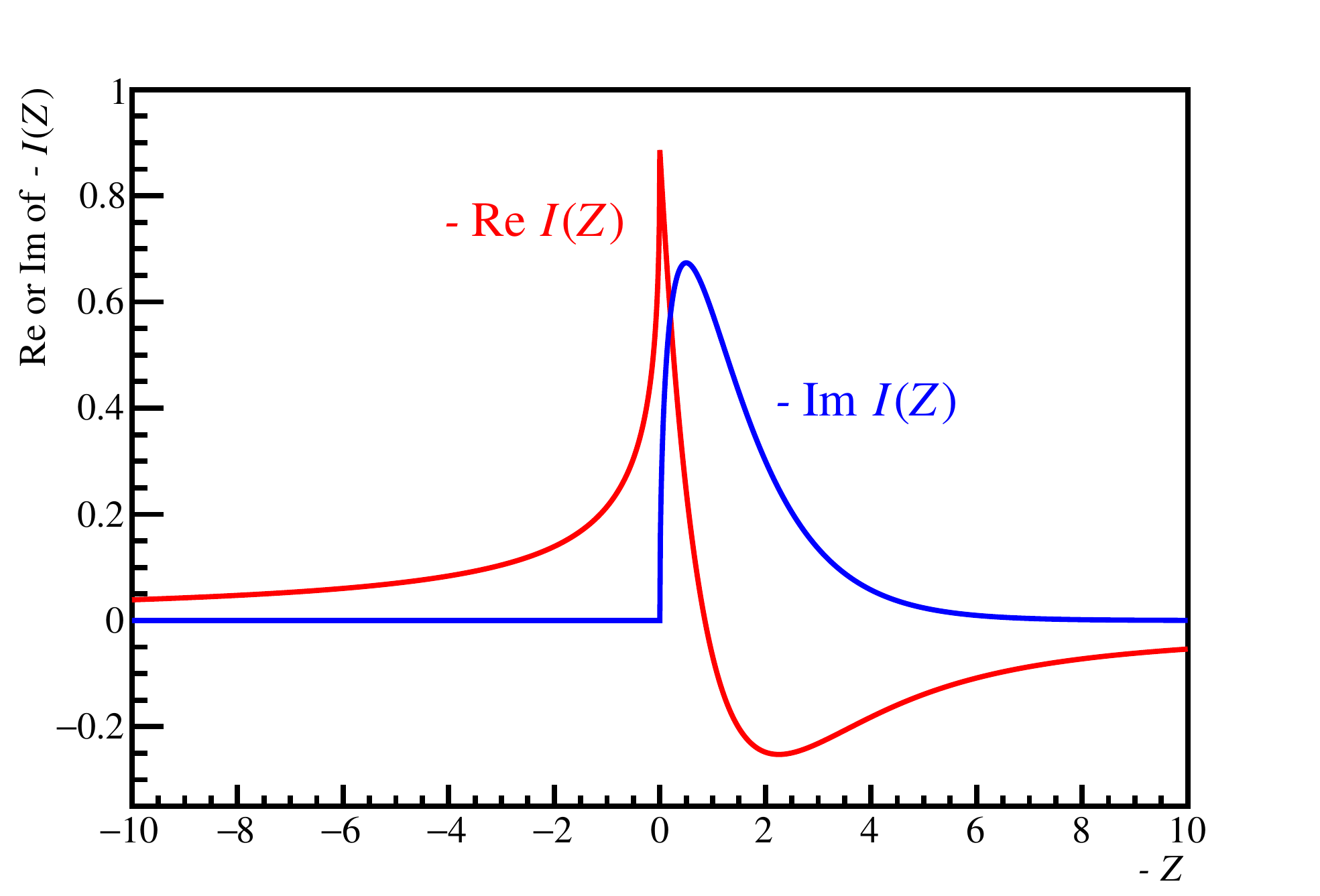}
  \end{center}
  \vskip-0.3cm\caption{\small
    Dependence of the real and imaginary parts of the cusp amplitude on the mass in Swanson's 
    model \cite{Swanson:2015bsa}.
    See the text for a more precise explanation.
  \label{fig:SwIZPlot}
  }
\end{figure}

\begin{figure}[bthp] 
  \begin{center}
    \includegraphics*[width=1.0\figsize]{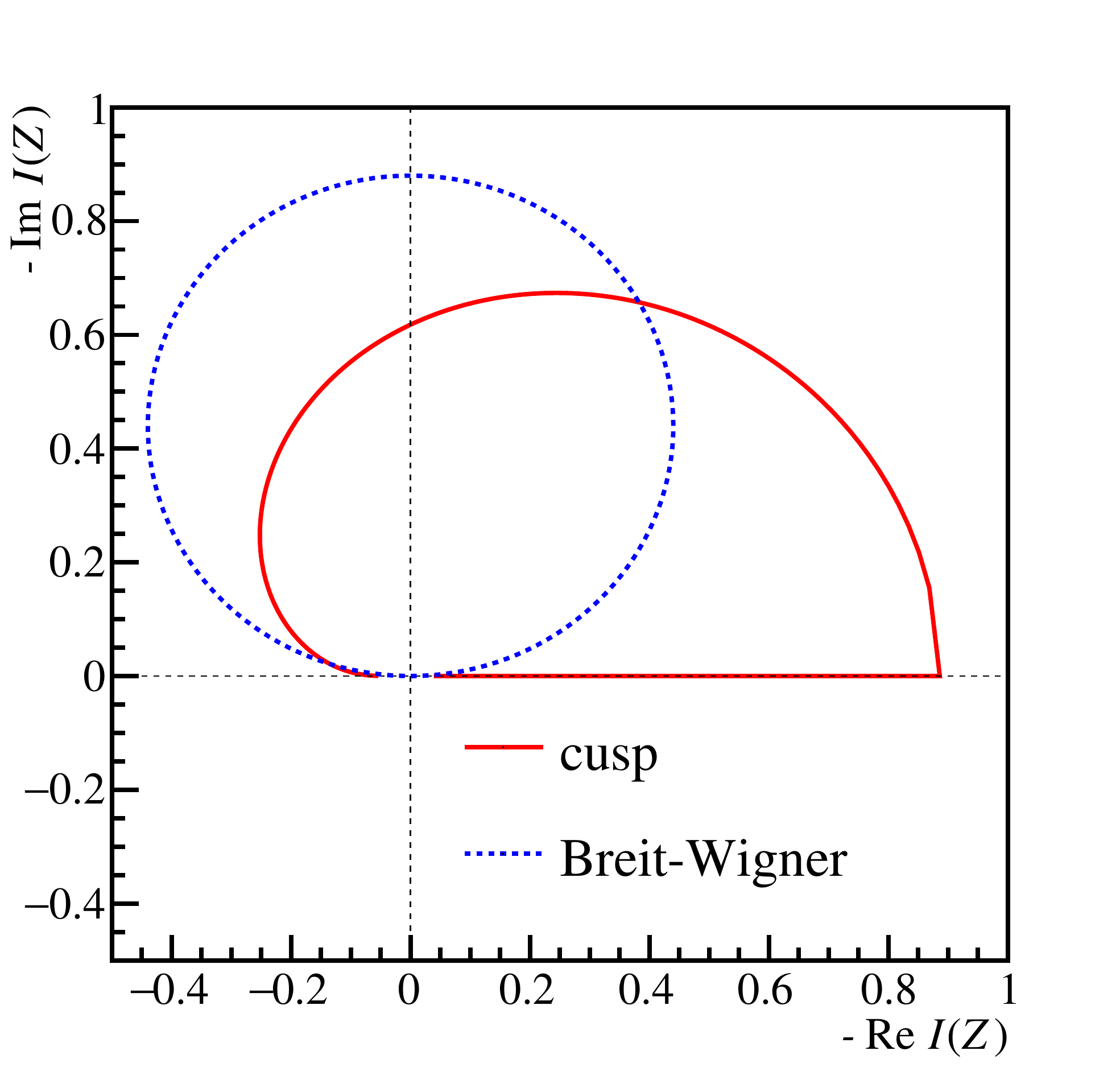}
  \end{center}
  \vskip-0.3cm\caption{\small
    The Argand diagram of the the cusp amplitude in Swanson's model \cite{Swanson:2015bsa}.
    Motion with the mass is counter-clockwise.
    The peak amplitude is reached at threshold 
    when the real part is maximal and the imaginary part is zero.  
    The Breit--Wigner amplitude gives circular phase motion,
    also with counter-clockwise mass evolution,
    with maximum magnitude when zero is crossed on the real axis.
  \label{fig:cuspArgand}
  }
\end{figure}

In Swanson's model a virtual loop with two mesons $A$ and $B$ inside 
(Fig.~1 left in Ref.~\cite{Swanson:2015bsa})
contributes, in the non-relativistic near-threshold approximation, 
the following amplitude,
\begin{equation}
\Pi(m)= \int\, \frac{d^3q}{(2\pi)^3}\,\frac{q^{2l}\,e^{-2q^2/\beta_{0}^2}}{
   m - M_A - M_B - \frac{q^2}{2\,\mu_{AB}} + i\,\epsilon},
\end{equation} 
where $m$ is $\jpsi\phi$ mass,
$\mu_{AB}=M_A\,M_B/(M_A+M_B)$ is the reduced mass of the pair,
$\beta_{0}$ is a hadronic scale of order of $\Lambda_{\rm QCD}$, 
(which can be $AB$ dependent),
$\epsilon$ is a very small number ($\epsilon\to0$), 
$l$ is the angular momentum between $A$ and $B$.
The lowest $l$ values are expected to dominate. 
The amplitude $\Pi(m)$ reflects coupled-channel kinematics. 
The above integral can be conveniently expressed as
\begin{align}
\Pi(m) & =-\frac{\mu_{AB}\,\beta_{0}}{\sqrt{2}\pi^2}\,I(Z)\\
Z & = \frac{4\mu_{AB}}{\beta_{0}^2}(M_A+M_B-m)\\
I(Z) & = \int_0^\infty\,dx\,\frac{x^{2+2l}\,e^{-x^2}}{x^2+Z-i\,\epsilon},
\end{align}
where $-Z$ is the scaled mass deviation from the $AB$ threshold. 
For $l=0$, the integral above evaluates to
\begin{equation}
I(Z) = \frac{1}{2}\,\sqrt{\pi}[1-\sqrt{\pi\,Z}\,e^Z\,{\rm erfc}(\sqrt{Z})].
\end{equation}
For masses below the $AB$ threshold $Z>0$ and $I(Z)$ (thus $\Pi(Z)$) has no imaginary part.
For masses above the threshold $Z<0$, $\sqrt{Z}$ is imaginary, 
which leads to both real and imaginary parts.
The real and imaginary parts of $-I(Z)$ as a function of $-Z$ 
are shown in
Fig.~\ref{fig:SwIZPlot}, while the corresponding Argand diagram is shown in 
Fig.~\ref{fig:cuspArgand} where it is compared to the 
phase motion of
the Breit--Wigner function.   

The function $\Pi(m)$ replaces the Breit--Wigner function 
$BW(m | M_0, \Gamma_0)$  in Eq.~(\ref{eq:breitwigner}).
The Blatt--Weisskopf functions in Eq.~(\ref{eq:resshapeBW})
still apply.
Thus, the functional form of this representation, 
has three free parameters to determine from the data 
($\beta_0$ and the complex S-wave helicity coupling). 
The value of $\beta_0$ obtained by the fit to the data, $297\pm20 \mev$,
is close to the value of $300 \mev$
with which Swanson was successful in describing the other 
near-threshold exotic meson candidates
\cite{Swanson:2015bsa}.
A fit with such parameterization (see Fig.~\ref{fig:cuspmasses} 
for mass distributions), has a better likelihood than the
Breit--Wigner fit by $1.6\sigma$ 
for the default model 
(8 free parameters in the $X(4140)$ 
Breit--Wigner parameterization), 
and better by $3\sigma$ 
when only S-wave couplings are allowed (4 free parameters),
providing an indication that the $X(4140)$ structure may not be a bound state
that can be described by the Breit--Wigner formula. 
Larger data samples will be required to obtain more insight.
We have included the $X(4140)$ 
cusp model among the systematic variations considered for
parameters of the other fit components.
The differences between the results obtained with 
the default amplitude model and the model in which the 
$X(4140)$ structure is represented by a cusp are given in 
Tables \ref{tab:ksysA}--\ref{tab:xsys}.   

\ifthenelse{\boolean{prl}}{ 
\begin{figure}[bthp]  
  \begin{center}
    \includegraphics*[width=\figsize]{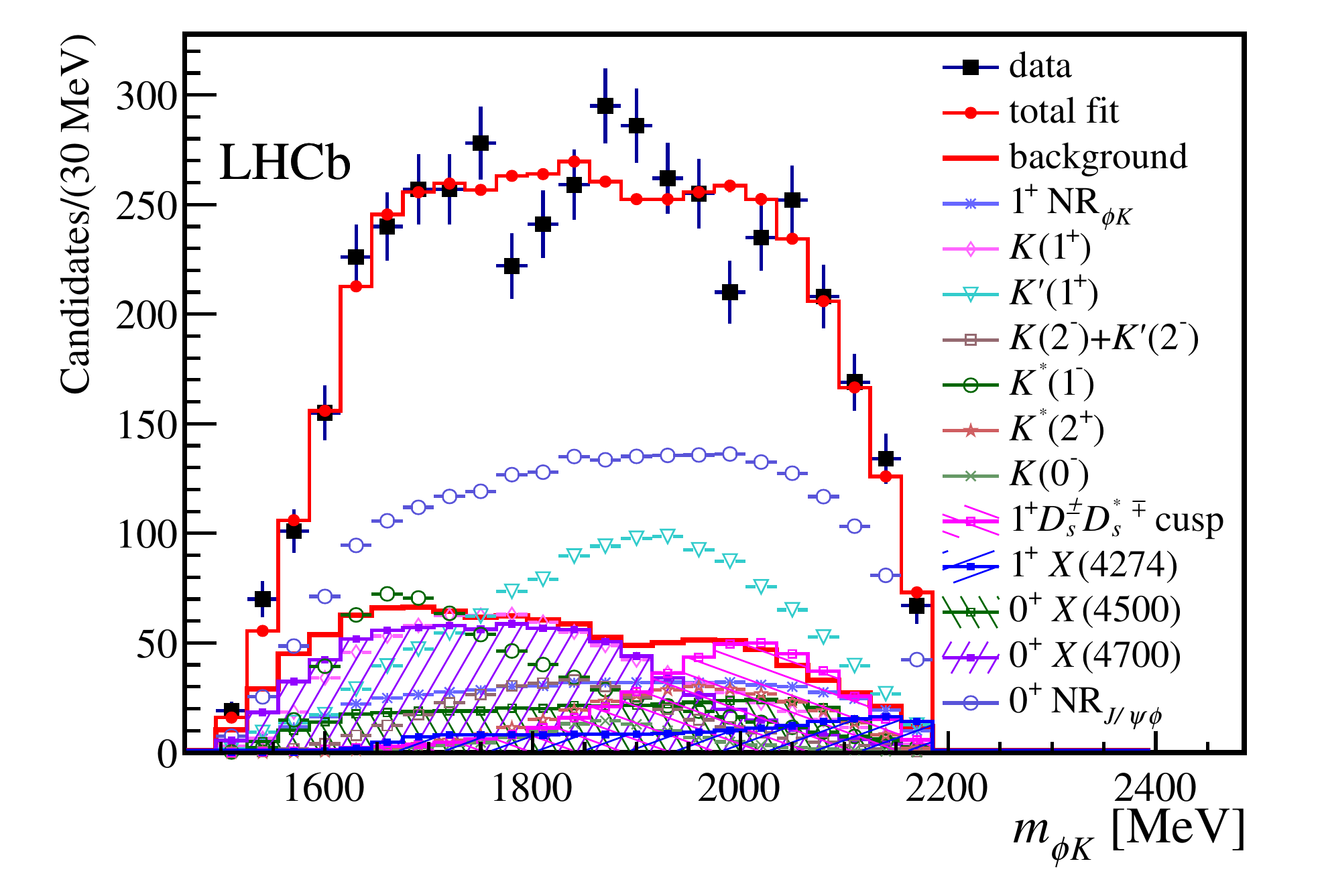} \\[-0.2cm]
    \includegraphics*[width=\figsize]{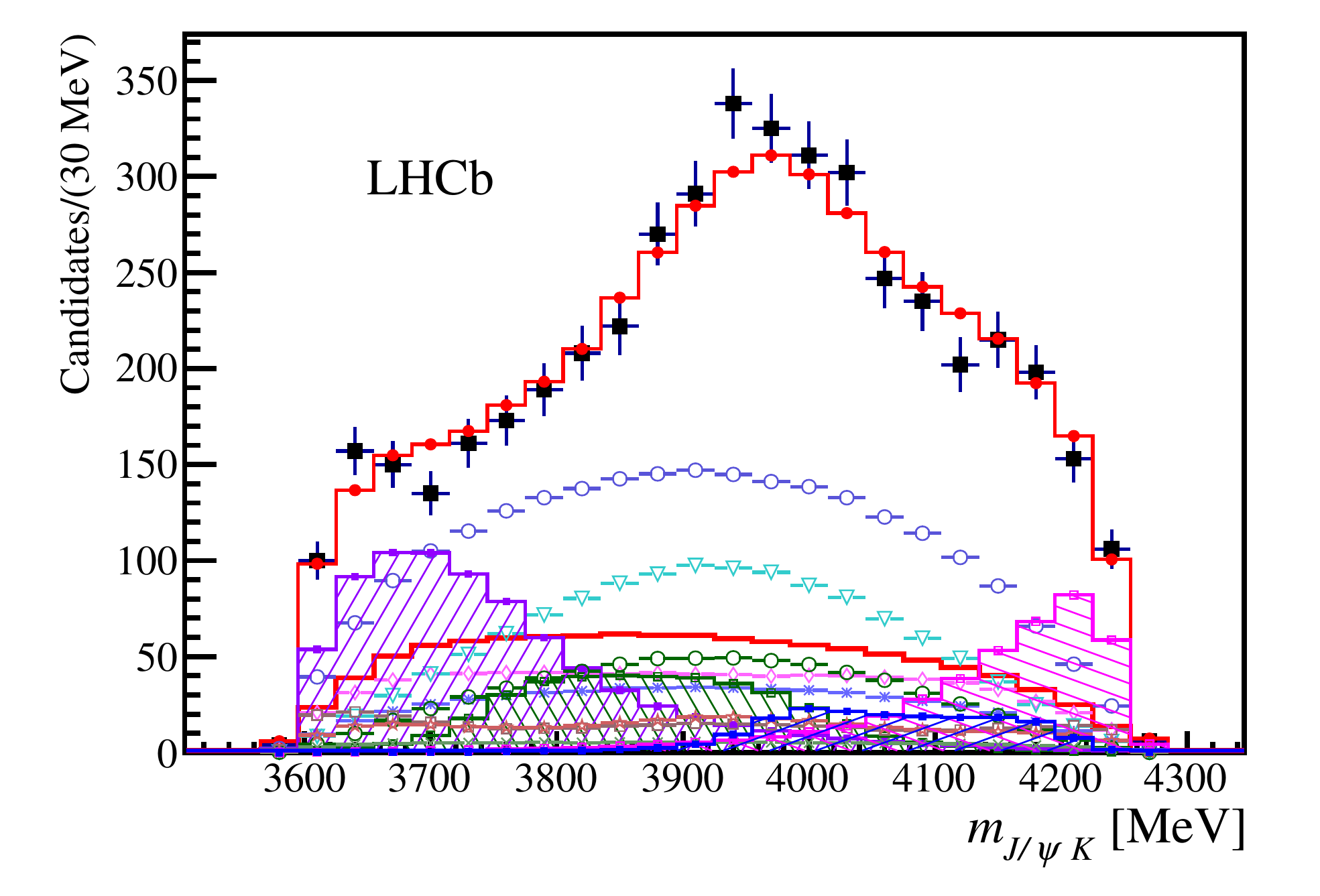} \\[-0.2cm]
    \includegraphics*[width=\figsize]{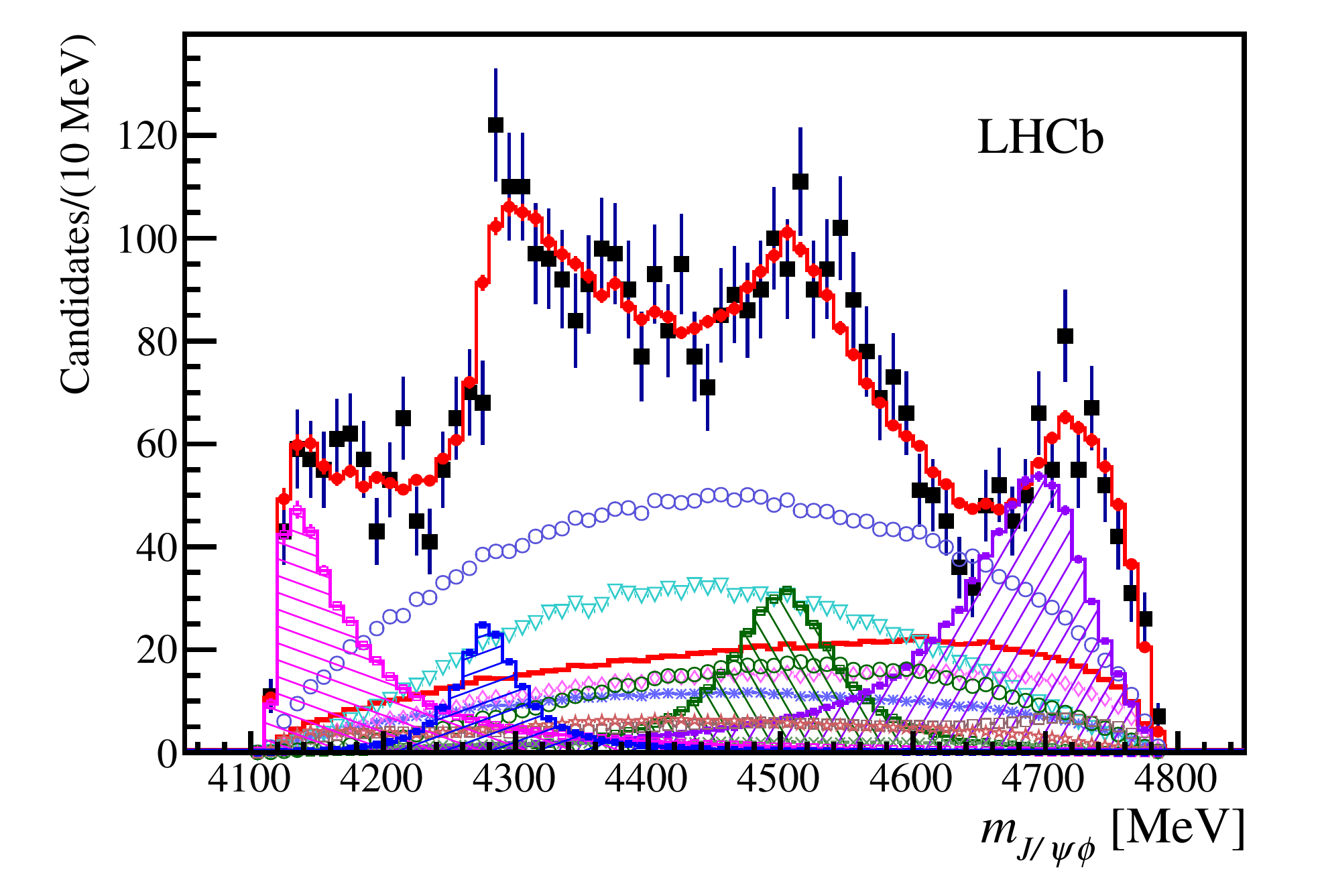} 
  \end{center}
  \vskip-0.3cm\caption{\small
    Distributions of (top) $\phiz K^+$, (middle) $\jpsi K^+$ and (bottom) $\jpsi\phi$ 
    invariant masses for the $\bujphik$ data (black data points) 
    compared with the results of the amplitude fit 
    containing $K^{*+}\to\phi K^+$ and $X\to\jpsi\phi$ contributions 
    in which $X(4140)$ is represented as a $J^{PC}=1^{++}$ $D_s^+D_s^{*-}$ cusp.
    The total fit is given by the red points with error bars. Individual fit components
    are also shown.
  \label{fig:cuspmasses}
  }
\end{figure}
}{
\begin{figure}[bthp]
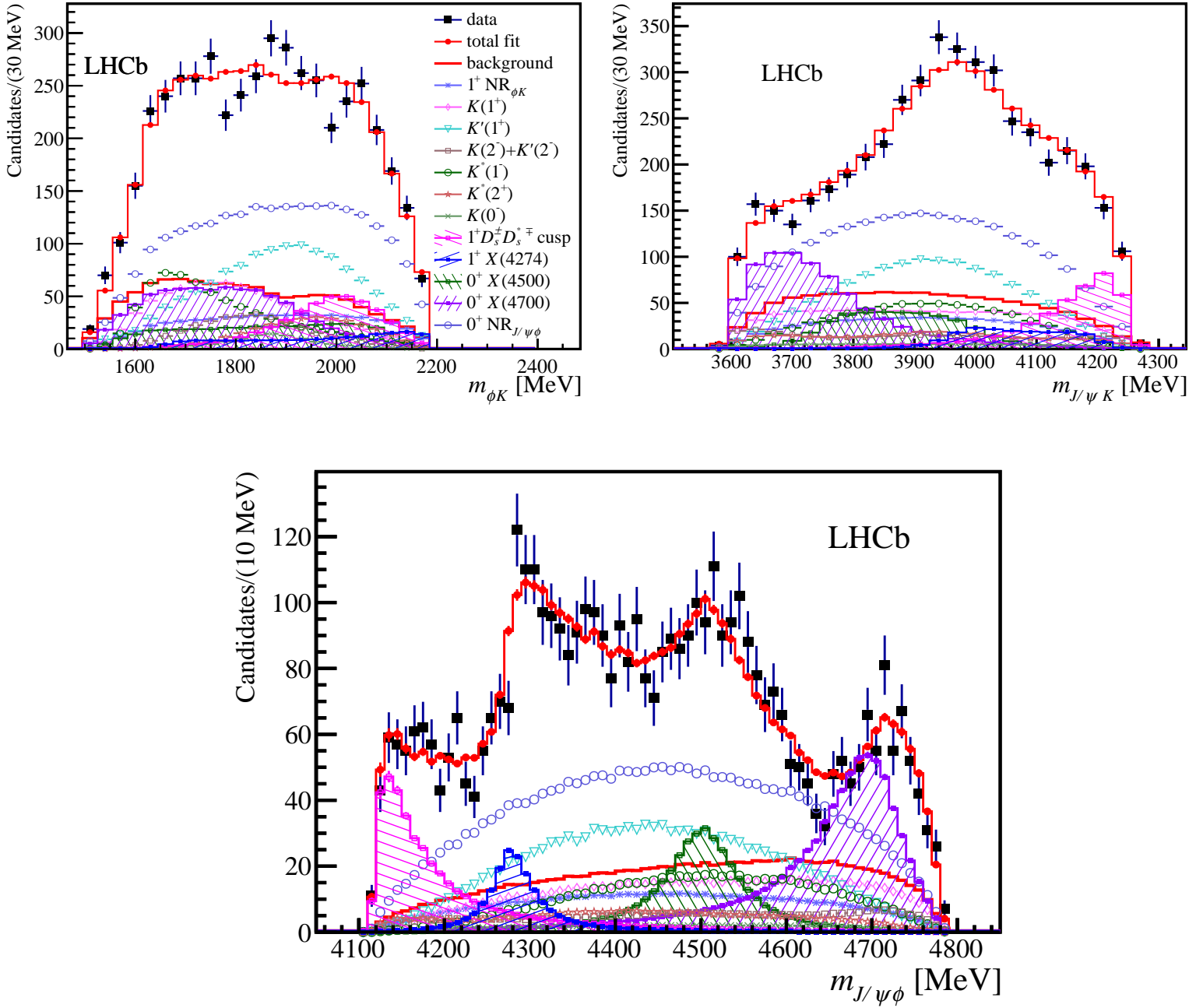
  
  \begin{center}
    \hbox{\hskip-2.9cm \includegraphics*[width=0.75\figsize]{xonecusp_PhiKh.pdf} 
     \hskip-0.7cm \includegraphics*[width=0.75\figsize]{xonecusp_JpsiKh.pdf} \hskip-2.1cm}\quad \\
    \includegraphics*[width=\figsize]{xonecusp_JpsiPhih.pdf} 
  \end{center}
  \vskip-0.3cm\caption{\small
    Distributions of (top left) $\phiz K^+$, (top right) $\jpsi K^+$ and (bottom) $\jpsi\phi$ 
    invariant masses for the $\bujphik$ data (black data points) 
    compared with the results of the amplitude fit 
    containing $K^{*+}\to\phi K^+$ and $X\to\jpsi\phi$ contributions 
    in which $X(4140)$ is represented as a $J^{PC}=1^{++}$ $D_s^+D_s^{*-}$ cusp.
    The total fit is given by the red points with error bars. Individual fit components
    are also shown.
  \label{fig:cuspmasses}
  }
\end{figure}
}

\begin{figure}[bthp]  
  \begin{center}
    \includegraphics*[width=\figsize]{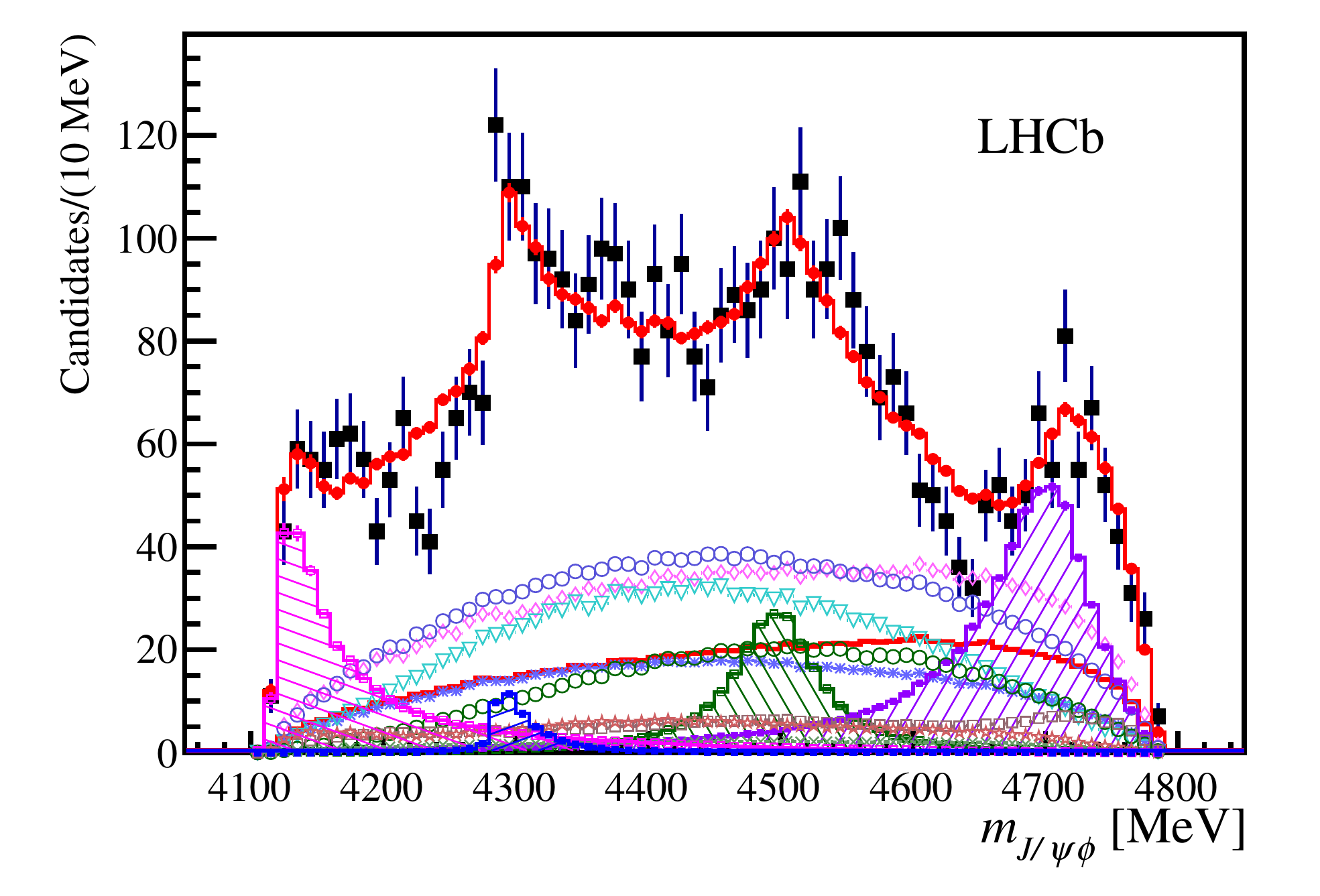} 
  \end{center}
  \vskip-0.3cm\caption{\small
    Distributions of $\jpsi\phi$  
    invariant mass for the $\bujphik$ data (black data points) 
    compared with the results of the amplitude fit 
    containing $K^{*+}\to\phi K^+$ and $X\to\jpsi\phi$ contributions 
    in which $X(4140)$ and $X(4274)$ are represented as $J^{PC}=1^{++}$ $D_s^\pm D_s^{*\mp}$
    and $0^{-+}$ $D_s^{\pm}D_{s0}^{*}(2317)^{\mp}$ cusps, respectively.
    The total fit is given by the red points with error bars. Individual fit components
    are also shown.
  \label{fig:twocuspmasses}
  }
\end{figure}

The $\Xtwo$ mass structure can be reasonably well described by the $0^{-+}$
cusp model for $D_s^{\pm}D_{s0}^{*}(2317)^{\mp}$ scattering (Fig.~\ref{fig:twocuspmasses}).
However, the multidimensional likelihood is substantially worse than for the 
default amplitude model ($6.6\sigma$).
The likelihood remains worse for the default fit even if $1^{++}$ quantum numbers are
assumed for such a cusp ($4.4\sigma$).
This particular cusp parameterization is not useful when trying to describe any
of the higher mass $\jpsi\phi$ structures. 

%\FloatBarrier
\ifthenelse{\boolean{prl}}{}{\FloatBarrier}%

\FloatBarrier

\bibliographystyle{LHCb}
\bibliography{main,LHCb-PAPER,LHCb-CONF,LHCb-DP}

\newpage

%\addcontentsline{toc}{section}{References}

%\end{document} 

%% file: acknowledgements.tex
\section*{Acknowledgements}

\noindent 
We thank Eric Swanson for discussions related to his cusp model.
We express our gratitude to our colleagues in the CERN
accelerator departments for the excellent performance of the LHC. We
thank the technical and administrative staff at the LHCb
institutes. We acknowledge support from CERN and from the national
agencies: CAPES, CNPq, FAPERJ and FINEP (Brazil); NSFC (China);
CNRS/IN2P3 (France); BMBF, DFG and MPG (Germany); INFN (Italy); 
FOM and NWO (The Netherlands); MNiSW and NCN (Poland); MEN/IFA (Romania); 
MinES and FASO (Russia); MinECo (Spain); SNSF and SER (Switzerland); 
NASU (Ukraine); STFC (United Kingdom); NSF (USA).
We acknowledge the computing resources that are provided by CERN, IN2P3 (France), KIT and DESY (Germany), INFN (Italy), SURF (The Netherlands), PIC (Spain), GridPP (United Kingdom), RRCKI and Yandex LLC (Russia), CSCS (Switzerland), IFIN-HH (Romania), CBPF (Brazil), PL-GRID (Poland) and OSC (USA). We are indebted to the communities behind the multiple open 
source software packages on which we depend.
Individual groups or members have received support from AvH Foundation (Germany),
EPLANET, Marie Sk\l{}odowska-Curie Actions and ERC (European Union), 
Conseil G\'{e}n\'{e}ral de Haute-Savoie, Labex ENIGMASS and OCEVU, 
R\'{e}gion Auvergne (France), RFBR and Yandex LLC (Russia), GVA, XuntaGal and GENCAT (Spain), Herchel Smith Fund, The Royal Society, Royal Commission for the Exhibition of 1851 and the Leverhulme Trust (United Kingdom).

%% file: LHCb_authorlist.tex
\centerline{\large\bf LHCb collaboration}
\begin{flushleft}
\small
R.~Aaij$^{39}$,
B.~Adeva$^{38}$,
M.~Adinolfi$^{47}$,
Z.~Ajaltouni$^{5}$,
S.~Akar$^{6}$,
J.~Albrecht$^{10}$,
F.~Alessio$^{39}$,
M.~Alexander$^{52}$,
S.~Ali$^{42}$,
G.~Alkhazov$^{31}$
% ...

{\footnotesize \it
% ...
$ ^{5}$Clermont Universit{\'e}, Universit{\'e} Blaise Pascal, CNRS/IN2P3, LPC, Clermont-Ferrand, France\\
$ ^{6}$CPPM, Aix-Marseille Universit{\'e}, CNRS/IN2P3, Marseille, France\\
% ...
$ ^{10}$Fakult{\"a}t Physik, Technische Universit{\"a}t Dortmund, Dortmund, Germany\\
% ...
$ ^{31}$Petersburg Nuclear Physics Institute (PNPI), Gatchina, Russia\\
% ...
$ ^{38}$Universidad de Santiago de Compostela, Santiago de Compostela, Spain\\
$ ^{39}$European Organization for Nuclear Research (CERN), Geneva, Switzerland\\
% ...
$ ^{42}$Nikhef National Institute for Subatomic Physics, Amsterdam, The Netherlands\\
% ...
$ ^{47}$H.H. Wills Physics Laboratory, University of Bristol, Bristol, United Kingdom\\
% ...
$ ^{52}$School of Physics and Astronomy, University of Glasgow, Glasgow, United Kingdom\\
% ...
}
\end{flushleft}